\documentclass[11pt,a4paper]{article}
\pdfoutput=1
\usepackage{jcappub, slashed}

\usepackage{bm,graphicx,amssymb}

\begin{document}

\title{Constraining particle dark matter using local galaxy distribution}
\author[a]{Shin'ichiro Ando}
\author[b]{and Koji Ishiwata}

\affiliation[a]{GRAPPA Institute, University of Amsterdam, 1098 XH
  Amsterdam, The Netherlands}
\affiliation[b]{Institute for Theoretical Physics, Kanazawa University,
  Kanazawa 920-1192, Japan}

\emailAdd{s.ando@uva.nl}
\emailAdd{ishiwata@hep.s.kanazawa-u.ac.jp}

\abstract{ It has been long discussed that cosmic rays may contain
  signals of dark matter. In the last couple of years an anomaly of
  cosmic-ray positrons has drawn a lot of attentions, and recently an
  excess in cosmic-ray anti-proton has been reported by AMS-02
  collaboration. Both excesses may indicate towards decaying or
  annihilating dark matter with a mass of around 1--10 TeV. In this
  article we study the gamma rays from dark matter and constraints
  from cross correlations with distribution of galaxies, particularly
  in a local volume. We find that gamma rays due to inverse-Compton
  process have large intensity, and hence they give stringent
  constraints on dark matter scenarios in the TeV scale mass
  regime. Taking the recent developments in modeling astrophysical
  gamma-ray sources as well as comprehensive possibilities of the
  final state products of dark matter decay or annihilation into
  account, we show that the parameter regions of decaying dark matter
  that are suggested to explain the excesses are excluded. We also
  discuss the constrains on annihilating scenarios.}

\maketitle
 
\section{Introduction}
\label{sec:intro}    
\setcounter{equation}{0} 

Discovery of the Higgs boson at the CERN Large Hadron Collider (LHC)
has confirmed the standard model of particle
physics~\cite{Aad:2012tfa,Chatrchyan:2012xdj}. To be precise the
standard model is an effective theory below TeV scale, meanwhile the
validity of the standard model above TeV scale has not been
unveiled. In cosmology, on the other hand, the observation of the
cosmic microwave background (CMB) strongly supports the cold dark
matter with a cosmological constant ($\Lambda$CDM model), which
indicates that about 26\% of the total energy density of the universe
is occupied by non-baryonic cold dark
matter~\cite{Hinshaw:2012aka,Ade:2013zuv,Ade:2015xua}. As well-known,
however, the standard model cannot explain the existence of dark
matter. While its energy density is precisely determined by the
analysis of the CMB, the other properties, such as the mass, lifetime,
spin, quantum numbers, etc., are still unknown. Weakly interacting
massive particles (WIMPs) are considered to be a good candidate for
dark matter. The reason is that the WIMPs are produced in thermal
plasma in the early universe, and then their relic abundance can
naturally fit to the observed density of dark matter if the mass is
around the weak scale.  In fact, one of the goals of the LHC
experiment has been the observation of dark matter. However, since no
signal of dark matter has been reported at the LHC so far, it may
indicate that the interaction of dark matter with ordinary particles
is extremely weak ({\it e.g.}, axion, very light gravitino or axino),
or simply its mass is larger than TeV.

The observation of cosmic rays has a potential for the search of such
classes of dark matter. Even if its mass is much larger than the
energy scale of collider experiments, dark matter may annihilate or
decay to produce lots of high-energy cosmic rays. Even in a case that
dark matter extremely weakly interacts with the standard model
particles, cosmic rays can still be produced if dark matter has finite
lifetime.  Gravitino or axino in supersymmetric models are
well-motivated examples. For gravitino, since its interaction with
ordinary particles is suppressed by the Planck scale, gravitino can be
enough long-lived to explain dark matter even under $R_p$
violation~\cite{Takayama:2000uz}. In annihilation or decay of dark
matter, variety of the cascading decay products could be observed as
anomalous fluxes among the astrophysical cosmic rays. In fact some
anomalous cosmic rays have been reported. The latest observations are
given by AMS-02. They have shown an excess in positron flux which has
a peak around 300\,GeV~\cite{Accardo:2014lma}. In addition, they have
also reported an excess in anti-proton flux in the same energy
range~\cite{AMS-02antiP}. If dark matter is the origin of these
anomalous cosmic rays, the observations indicate that the mass of dark
matter is 1--10\,TeV. On the other hand, however, it is not necessary
to attribute the excesses to dark matter. There are claims that it is
possible to account for the anomalous positrons by pulsars (see, {\it
  e.g.}, ref.\,\cite{Linden:2013mqa} for recent work) or anti-protons
by known-galactic sources~\cite{Giesen:2015ufa,Kappl:2015bqa}. These
claims, however, do not completely exclude the dark matter hypothesis
for the anomalous cosmic rays.

In this article, we mainly investigate the dark matter scenarios which
claim to explain the anomalous positron or anti-proton in light of
cross correlation between gamma rays from dark matter and the observed
galaxies. It was pointed out that the inverse-Compton (IC) process
gives large amount of gamma rays in decaying~\cite{Ishiwata:2009dk} or
annihilating~\cite{Profumo:2009uf} dark matter, and recently
ref.\,\cite{Ando:2015qda} has updated the constraints on decaying dark
matter by using the {\it spectral} data of 50-month extragalactic
gamma-ray background (EGRB) by Fermi-LAT~\cite{Ackermann:2014usa}.  On
the other hand, however, the IC process has not been taken into
account in the previous studies using the {\it cross-correlation} of
the gamma rays~\cite{Regis:2015zka,Cuoco:2015rfa}. We therefore
include the IC gamma rays in the cross correlation analysis, which is
shown to be essential for constraining decay lifetimes or annihilation
cross sections for heavy mass dark matter particularly for leptonic
channels.  We also investigate more final states than in the previous
works. Since the decay signal does not depend on the clustering of
dark matter, robust constraints are obtained since gamma rays from
dark matter can be computed with little uncertainties for each of
decaying dark matter scenarios.\footnote{The electroweak correction,
  which is not included in our study, gets important for TeV mass
  region and might give more stringent
  constraints~\cite{Cirelli:2010xx}.  Our main goal in the current
  study is to show the importance of the IC process.}  It will be
shown that the decaying dark matter scenarios to explain the
cosmic-ray positron or anti-proton excess are excluded. For the
annihilating scenario, on the other hand, we will show that the
constraints can be competitive with other probes such as dwarf
spheroidal galaxies~\cite{Ackermann:2015zua} by adopting reasonable
parameters for abundance of dark matter halos and subhalos.

In this paper, we use the following cosmological parameters: Hubble
constant $H_0=67.81~{\rm km}~{\rm s}^{-1}\,{\rm Mpc}^{-1}$ with
$h=0.6781$, density parameter of dark matter $\Omega_{\rm
  dm}h^2=0.1186$~\cite{Ade:2015xua}. The critical density is obtained
from Hubble parameter, $\rho_c=1.054\times 10^{-5}\,h^2~{\rm
  GeV}\,{\rm cm}^{-3}$.

\section{Extragalactic gamma rays  from dark matter}
\label{sec:EGRB_DM}
\setcounter{equation}{0}

In this section we explain how to compute the EGRB intensity from dark
matter. There are three components to determine gamma rays from dark
matter in a decaying or annihilating scenario. One is the intrinsic
gamma-ray energy spectrum per decay or annihilation. The others are
mass $m_{\rm dm}$ and lifetime $\tau_{\rm dm}$ of dark matter for the
former, while mass and annihilation cross section for the latter. (For
annihilation case, dark matter clustering is another factor to
determine the gamma-ray signals.) First, we mainly focus on decaying
scenarios. Sec.\,\ref{sec:scenario_DM} discusses possible final states
of decaying dark matter that can explain the AMS-02 excesses.  In
Sec.\,\ref{sec:gamma_DM} the EGRB from decaying dark matter is
given. We discuss the annihilating scenarios at the end of each
section.

\subsection{Scenarios of (decaying) dark matter}
\label{sec:scenario_DM}

Scenarios of decaying dark matter are classified by the final state of
the decay process, {\it i.e.}, the final state contains (a) hadrons,
(b) hadrons and leptons, and (c) leptons. (Here lepton means
electron/positron or muon.)  In the later discussion, we sometimes
refer to the scenarios (a), (b), and (c) as hadronic, hadroleptonic,
and leptonic scenarios (or final states), respectively.  Obviously the
final state needs to have leptons and hadrons in order to explain the
observed positron and anti-proton excesses, respectively.

\begin{table}[t]
 \begin{center}
  \begin{tabular}{lcccccc}
   \hline \hline
   Final state & AMS-02 $e^+$ region  & AMS-02 $\bar{p}$ region & $\gamma_{\rm pr}$ & $\gamma_{\rm fsr}$  &  $\gamma_{\rm ic}$ & Data \\
   \hline
   $b\bar{b}$, $W^+W^-$   & -- & excluded~\cite{Regis:2015zka,Cuoco:2015rfa}&$\checkmark$ & $\checkmark$ & no & \cite{Xia:2015wka} \\
   $W^{\pm}l^{\mp}_i$\,\cite{Hamaguchi:2015wga} & -- & partly
	   excluded~\cite{Ando:2015qda} & $\checkmark$ & no &
		       $\checkmark$ & \cite{Ackermann:2014usa}\\
   $\nu \tau^{\pm}l^{\mp}_k$\,\cite{Ibe:2014qya} & partly excluded~\cite{Ando:2015qda} & -- &$\checkmark$ & no& $\checkmark$& \cite{Ackermann:2014usa}\\
   $\nu \l^{\pm}_jl^{\mp}_k$\,\cite{Ibe:2014qya} & partly excluded~\cite{Ando:2015qda} & -- &-- &no & $\checkmark$& \cite{Ackermann:2014usa}\\
   $\mu^+\mu^-$ & excluded~\cite{Cirelli:2012ut} & -- & -- &no &
		       $\checkmark$& \cite{Abdo:2010nz}\\
   \hline \hline
  \end{tabular}
  \caption{\small Possible final states from decaying dark matter to
    explain AMS-02 positron or anti-proton and the constraints on them
    from the gamma ray observations. $i,k=1,2,3$ for
    $W^{\pm}l^{\mp}_i$, $\nu \tau^{\pm}l^{\mp}_k$ and $j,k$ are $1$ or
    $2$ for $\nu \l^{\pm}_jl^{\mp}_k$. (Neutrino flavor is irrelevant,
    thus it is omitted.) `AMS-02 $e^+$ ($\bar{p}$) region' indicates
    $m_{\rm dm}\simeq 1$--$10$ TeV and $\tau_{\rm dm}\simeq
    10^{26}$--$10^{27}$\,s, which is shown to explain the observed
    positron (anti-proton) excesses, and `--' in their column means
    there is no explicit study for the latest positron (anti-proton)
    data.  $\checkmark$ in $\gamma_{\rm pr}$, $\gamma_{\rm fsr}$,
    $\gamma_{\rm ic}$ (gamma rays produced primarily, via final state
    radiation, via inverse-Compton process, respectively) columns
    indicates gamma rays from dark matter which are computed to
    constrain the scenarios. For $b\bar{b}$, $W^+W^-$, for example,
    gamma rays produced primarily from the decay and final state
    radiation are taken into account, but gamma rays from
    inverse-Compton scattering are not. `--' in $\gamma_{\rm pr}$
    shows that there is no primary gamma rays. Final column shows the
    data used to give constrains in each study of
    refs.\,\cite{Regis:2015zka,Cuoco:2015rfa,Ando:2015qda,Cirelli:2012ut}.}
  \label{table:CurrentLimit}
 \end{center}
\end{table}

In decaying dark matter scenarios, $m_{\rm dm}\simeq 1$--$10$ TeV and
$\tau_{\rm dm}\simeq 10^{26}$--$10^{27}$\,s (roughly on $m_{\rm dm}
\tau_{\rm dm}\simeq 10^{27}\,{\rm TeV\,s}$ line) gives a good fit for
the anti-proton excess and the positron excess reported by AMS-02.  As
a concrete model, ref.\,\cite{Hamaguchi:2015wga} studied decaying
gravitino dark matter via $L_iH_u$-type $R_p$-violated supersymmetric
model for the anti-proton excess. It is known that the main decay mode
of gravitino decay under $L_iH_u$-type $R_p$-violation is $W^\pm
l^\mp$~\cite{Ishiwata:2008cu}, which corresponds to the hadroleptonic
case.  On the other hand, ref.\,\cite{Ibe:2014qya} considers decaying
wino dark matter in $L_iL_jE_k^c$-type $R_p$-violation to show that
the decay can explain the positron flux. In this case, the decay
products are leptons, {\it i.e.}, $\nu l^\pm l^{\mp}$, leptonic final
state (or hadroleptonic final state when $l=\tau$).

Table~\ref{table:CurrentLimit} summarizes the current status of
decaying dark matter that can explain the anomalous fluxes.  In
hadronic scenarios, such as $b\bar{b}$, $W^+W^-$, lots of high energy
gamma rays are produced primarily from the dark matter decay, which is
mainly from $\pi^0$ decay after hadronization. In addition, it has
been recently realized that the cross correlation between the
extragalactic gamma rays from dark matter and the observed galaxy
catalog gives severe constraints on both annihilating and decaying
dark matter scenarios. It is easy to read from
refs.\,\cite{Regis:2015zka,Cuoco:2015rfa,Hamaguchi:2015wga} that the
parameter region which is suitable for the positron or anti-proton
excesses is already excluded for final state $b\bar{b}$, $W^+W^-$. For
$\mu^+\mu^-$ final state, on the other hand, there is no primary gamma
rays from the final state.  However, it was pointed out that electrons
and positron from muons create high energy gamma rays by scattering
off the CMB photon, {\it i.e.} inverse-Compton scattering, even in
extragalactic region~\cite{Ishiwata:2009dk}. These gamma rays are
calculated with little theoretical uncertainty, thus it can give
robust constraints on decaying dark matter scenarios. As the result,
it was shown that leptonic scenarios, such as $\mu^+\mu^-$ final
state, to account for the positron excess were already
excluded~\cite{Cirelli:2012ut} by Fermi-LAT 2010
data~\cite{Abdo:2010nz}. This is basically due to the fact that the
spectrum of produced $e^\pm$ is very hard and yields large amount of
the IC gamma rays. Therefore, remaining possibility for the decaying
scenarios suitable for the cosmic ray excesses are hadroleptonic final
states, {\it e.g.}, $W^{\pm}l^{\mp}_i$, or three-body leptonic final
state, {\it e.g.}, $\nu l^{\pm}_jl^{\mp}_k$. ($i,j,k$ are flavor
indices and we omit flavor index for neutrino since it is irrelevant
in our discussion.  We do not distinguish neutrino and anti-neutrino
for the same reason.) In these cases, produced $e^\pm$ are softer
compared to $\mu^+\mu^-$ or $e^+e^-$ final states. Thus the
constraints from the gamma-ray observations are weaker.

In the present work we extend the past analysis to further constrain
the decaying dark matter scenario. We use cross correlation of gamma
rays from dark matter with galaxy catalogs to constrain the dark
matter scenarios. In the calculation of the gamma rays from dark
matter we take into account all contributions, {\it i.e.}, gamma rays
produced primarily, via final state radiation, and via IC
process. Namely, in the language of Table~\ref{table:CurrentLimit},
our strategy is expressed by all $\checkmark$ marks for $\gamma_{\rm
  pr}$, $\gamma_{\rm fsr}$, $\gamma_{\rm ic}$, and to adopt
cross-correlating technique using the galaxy catalog
data~\cite{Xia:2015wka}. Although our main target is to study the
final state such as $W^{\pm}l^{\mp}_i$ or $\nu l^{\pm}_jl^{\mp}_k$
motivated by the AMS-02 excesses, we analyze the other final states,
such as $b \bar{b}$, $W^+W^-$, $l^+_il^-_i$, to give constraints for
general use.

The final state does not depend on its origin, decay or
annihilation.\footnote{Of course, final states $W^{\pm}l^{\mp}_i$, $\nu
  l^{\pm}_jl^{\mp}_k$ are impossible for annihilation case. } Thus the
discussion here can also cover the annihilation case. Typical examples
are $b \bar{b}$, $W^+W^-$, and $l^+_il^-_i$.

\subsection{Gamma rays from dark matter}
\label{sec:gamma_DM}

As described at the beginning of this section, the spectrum of gamma
rays is determined by three components, the dark matter mass $m_{\rm
  dm}$ and lifetime $\tau_{\rm dm}$, and the decaying scenario, {\it
  i.e.}, final state of the decay process.  When the final states are
specified, the energy distributions $dN_I/dE$ of the decay products
$I=\gamma, e^{\pm}$, etc., are determined.

We derive the `window function', which is needed for the calculation
of both the gamma-ray intensity and the angular cross correlation
between gamma rays from dark matter and galaxy catalogs.  We start
with defining the gamma-ray intensity from decaying dark matter:
\begin{align}
\frac{d\Phi_\gamma^{\rm dm}}{d\chi}(E_\gamma,z) =
\frac{1}{4\pi}\frac{\Omega_{\rm dm}\rho_c}{m_{\rm dm}\tau_{\rm dm}} 
\frac{1}{1+z}Q_\gamma^{\rm dm}(E'_\gamma,z)
\, e^{-\tau(E'_\gamma,z)}\, .
\label{eq:dPhi_gamma}
\end{align}
$\chi$ is comoving distance, $E_\gamma$ is the energy of a gamma ray
which we observe today, and $E'_\gamma$ is the gamma-ray energy at the
redshift $z$, {\it i.e.},
$E'_\gamma=(1+z)E_\gamma$. $\tau(E'_\gamma,z)$ is the optical depth,
for which we will use data given in ref.\,\cite{Gilmore:2011ks}.
$Q_{\gamma}^{\rm dm}(E'_\gamma,z)$ is a gamma-ray source function due
to dark matter, given as
\begin{align}
Q_\gamma^{\rm dm}(E'_\gamma,z)=
Q_{\gamma_{\rm pr}}^{\rm dm}(E'_\gamma,z)+
Q_{\gamma_{\rm fsr}}^{\rm dm}(E'_\gamma,z)+
Q_{\gamma_{\rm ic}}^{\rm dm}(E'_\gamma,z)\,,
\end{align}
where
\begin{align}
&Q_{\gamma_{\rm pr}}^{\rm dm}(E'_\gamma,z)+
Q_{\gamma_{\rm fsr}}^{\rm dm}(E'_\gamma,z)
 =
(1+z)  \frac{dN_\gamma}{dE}(E'_\gamma)\,,
\\
&Q_{\gamma{\rm ic}}^{\rm dm}(E'_\gamma,z) =
c \int dE_e\, dE_{\gamma_{\rm BG}}(1+z) 
  \frac{d\sigma_{\rm IC}}{dE'_\gamma}(E'_\gamma,E_e,E_{\gamma_{\rm BG}})
  f_{\gamma}^{\rm BG} (E_{\gamma_{\rm BG}},z)
  \frac{Y_{e}(E_e)}{b_{\rm IC}(E_e,z)}\,.
\end{align}
$c$ is the speed of light. The former describes the gamma-ray source
primarily produced by the dark matter decay. $dN_{\gamma}/dE$ is the
energy distribution of primary gamma rays from single dark matter
decay, including final state radiation (FSR) photons. For the
computation, we use {\tt PYTHIA 6.4}~\cite{Sjostrand:2006za} (not
including the electroweak corrections).\footnote{The FSR is important
  especially for leptonic final state.  While we ignored the FSR in
  our previous study~\cite{Ando:2015qda}, we have checked that
  inclusion of FSR only give minor changes for hadronic and
  hadroleptonic scenarios. For leptonic case, on the other hand, we
  have found that the constraints become tighter for $m_{\rm
    dm}\lesssim 1~{\rm TeV}$. } The latter represents the IC
scattering between $e^{\pm}$ from dark matter and the background
photons.  $ d\sigma_{\rm IC}/dE'_\gamma$ is the differential cross
section of the IC process. $f_{\gamma}^{\rm BG}(E_{\gamma_{\rm
    BG}},z)$ is the energy spectrum of the background photon field (in
unit energy and volume). In our study we take into account both the
CMB and extragalactic background light (EBL) photons, {\it i.e.},
\begin{align}
f_{\gamma}^{\rm BG}(E_{\gamma_{\rm BG}},z) =
f_{\gamma}^{\rm CMB}(E_{\gamma_{\rm BG}},z) + f_{\gamma}^{\rm EBL}(E_{\gamma_{\rm BG}},z)
\,.
\end{align}
For the EBL, we will use the spectrum given in
ref.\,\cite{Gilmore:2011ks}.  $Y_e(E_e)$ is defined as
\begin{align}
Y_e(E_e) = \sum_{I=e^\pm}\int^\infty_{E_e} dE \frac{dN_I}{dE}(E)\,,
\end{align}
where $dN_{e^\pm}/dE$ is the $e^\pm$ energy distribution from dark
matter. Finally $b_{\rm IC}(E_e,z)$ is the energy loss rate (per
unit time) of $e^{\pm}$ with energy $E_e$, which is given by
\begin{align}
b_{\rm IC}(E_e,z)&=
\int dE'_{\gamma} dE_{\gamma_{\rm BG}} (E'_\gamma - E_{\gamma_{\rm BG}})
\frac{d\sigma_{\rm IC}}{dE'_\gamma}(E'_\gamma,E_e,E_{\gamma_{\rm BG}})
f_{\gamma}^{\rm BG} (E_{\gamma_{\rm BG}},z) \\
&\equiv b_{\rm IC}^{\rm CMB}(E_e,z)+b_{\rm IC}^{\rm EBL}(E_e,z)\,,
\end{align}
where the first and second terms of the right-hand side in the second
line are the energy loss rates due to the CMB and EBL, respectively.
Under the CMB, the energy loss rate well agrees with an analytic
expression $b_{\rm IC,T}^{\rm CMB}(E_e,z)=(1+z)^4(4/3)\sigma_T
(E_e/m_e)^2\rho_{\rm CMB}^{\rm (now)}$ ($\sigma_T$, $m_e$ are Thomson
scattering cross section, electron mass, respectively, and $\rho_{\rm
  CMB}^{\rm (now)}\simeq 0.260\,{\rm eV~cm}^{-3}$), especially for
$E_e\lesssim 1\,{\rm TeV}$.\footnote{Using the analytic expression for
  $f^{\rm CMB}_\gamma$ and $b^{\rm CMB}_{\rm IC}$ and neglecting the
  EBL, $Q_{\gamma,{\rm IC}}^{\rm dm}(E'_\gamma,z)$ agrees with
  $(1+z){\cal P}_{\rm ic}(E'_\gamma)$ in eq.\,(2.5) of
  ref.\,\cite{Ando:2015qda}. In this equation $d\sigma_{\rm
    IC}/dE'_\gamma (E'_{\gamma},E_e,E_{\gamma_{\rm CMB}})$ should be
  replaced by $d\sigma_{\rm IC}/dE'_\gamma
  (E'_{\gamma},E_e,(1+z)E_{\gamma_{\rm CMB}})$. Similarly,
  $\tau(z,E_{\gamma})$ should be $\tau(z,E'_{\gamma})$ in eq.\,(2.3)
  of ref.\,\cite{Ando:2015qda}.  Those are just typos, and the
  numerical calculation had been done in the correct expressions. }
However, for $e^\pm$ with $E_e\gtrsim 1\,{\rm TeV}$, the energy loss
rate deviates from the analytic expression. This is because it is
given in the Thomson limit where the background photon energy is much
smaller than the incident electron energy (or electron mass in the
electron rest frame).  (See Fig.\,\ref{fig:bic} in Appendix.)  In
addition, we have checked that the energy loss due to the EBL photon
is much smaller than one due to the CMB, {\it e.g.}, it accounts for
just around 5\% in the total energy loss rate for $z=0$ and smaller
for higher $z$. This is because the intensity of the EBL is much
smaller than the CMB. Thus, the energy loss rate of $e^{\pm}$ can be
computed with little theoretical uncertainty since it is determined
mainly by the CMB.

\begin{figure}
 \begin{center}
   \includegraphics[width=7cm]{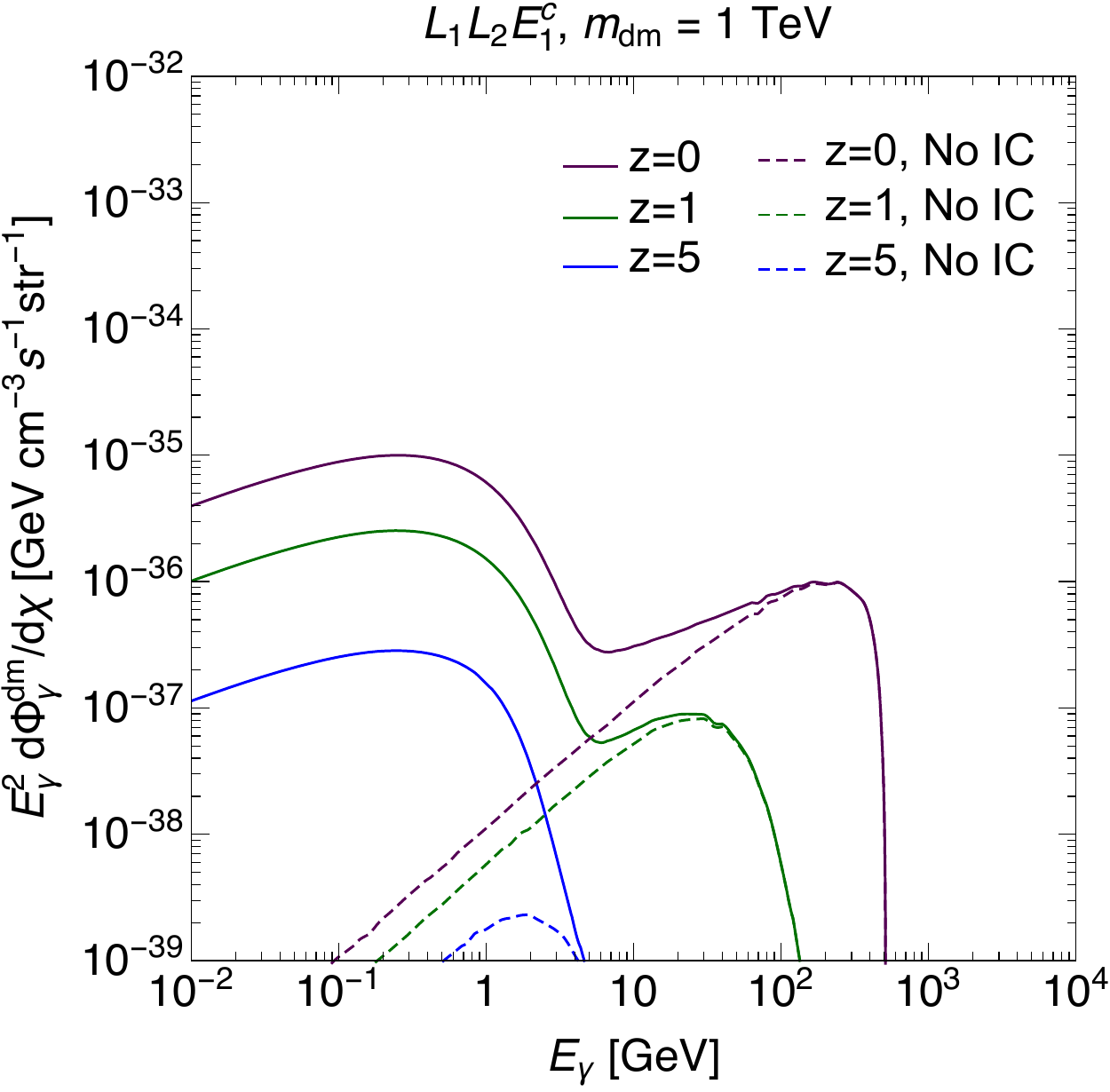}
   \includegraphics[width=7cm]{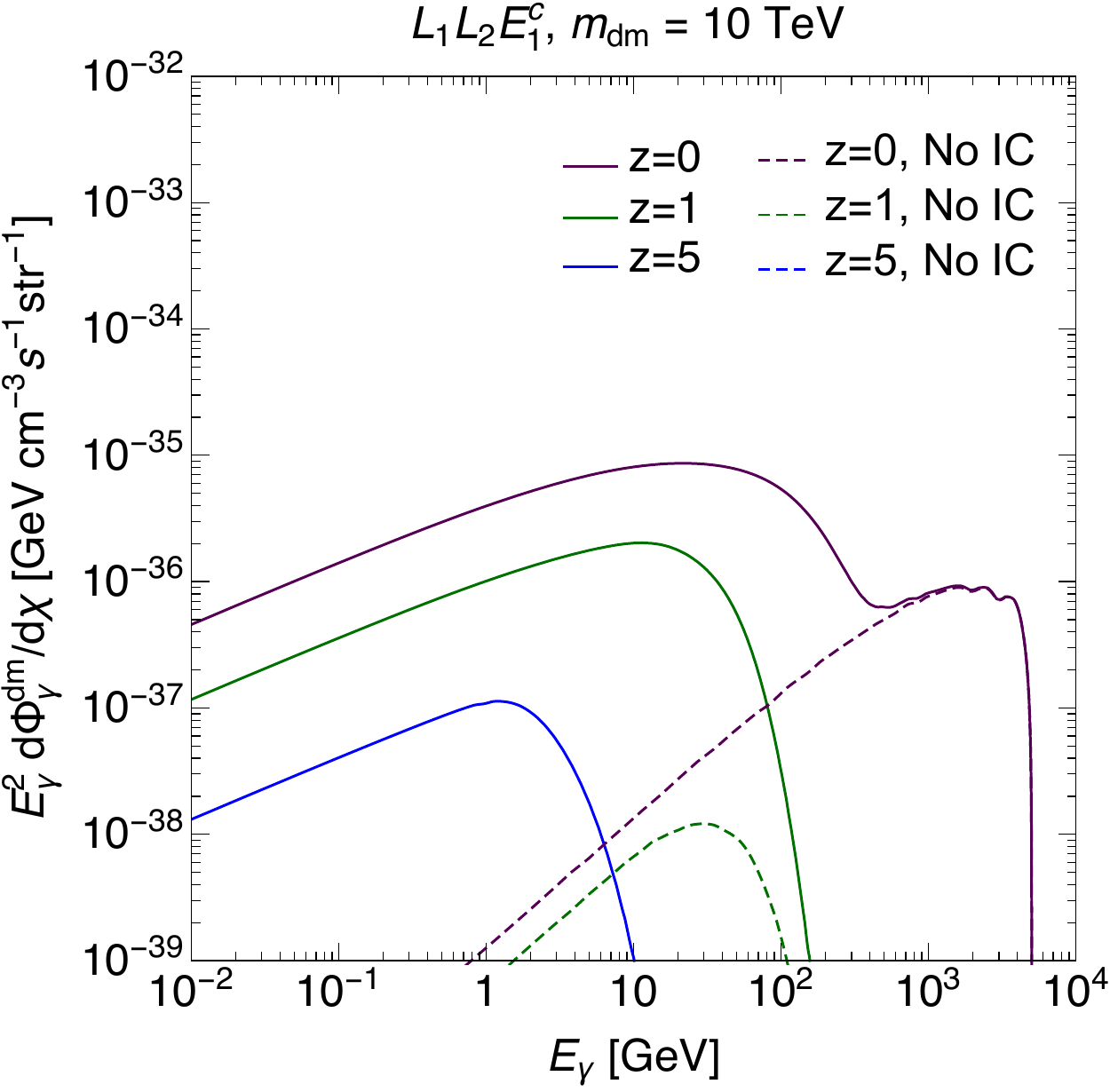}
   \includegraphics[width=7cm]{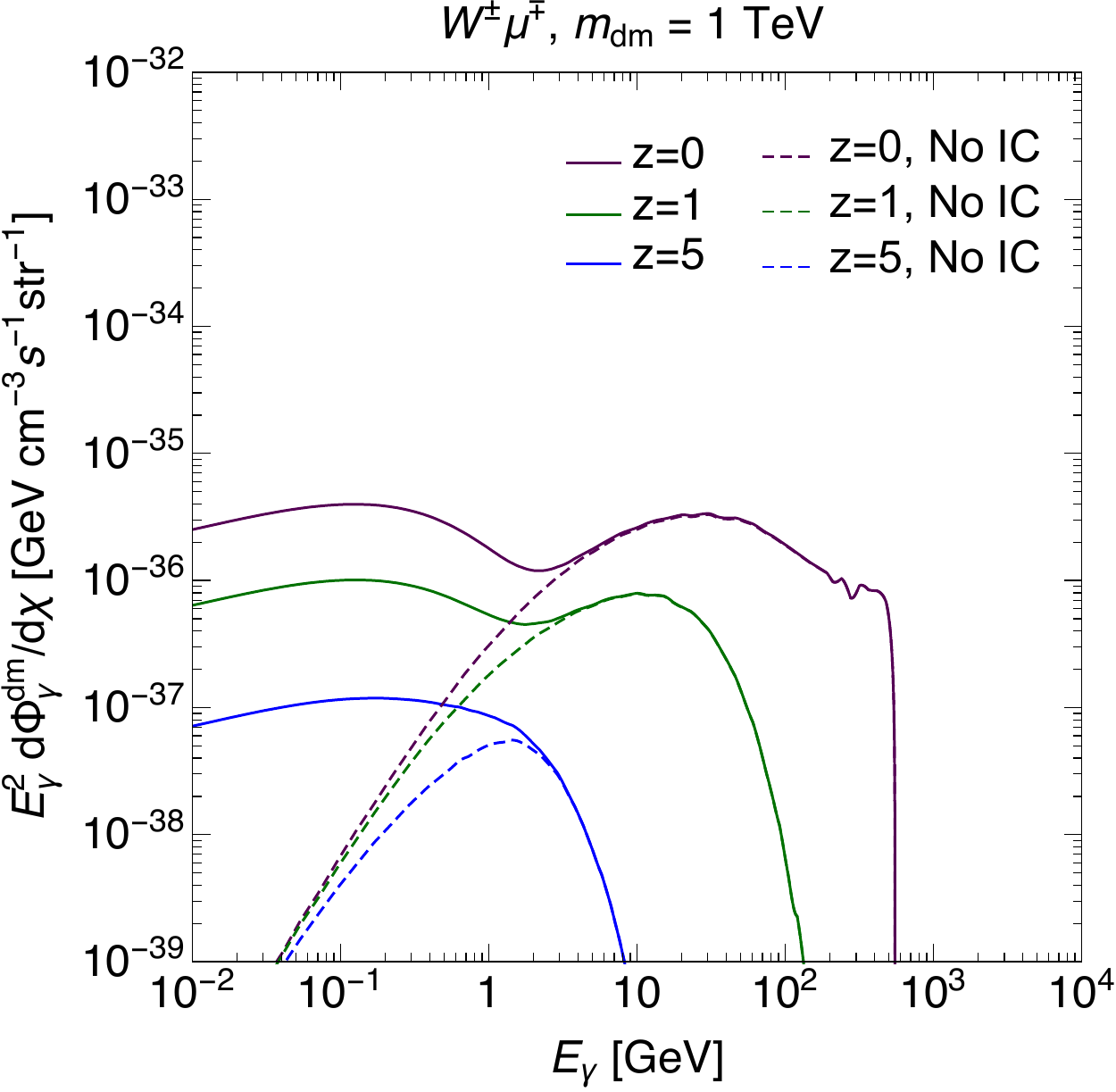}
   \includegraphics[width=7cm]{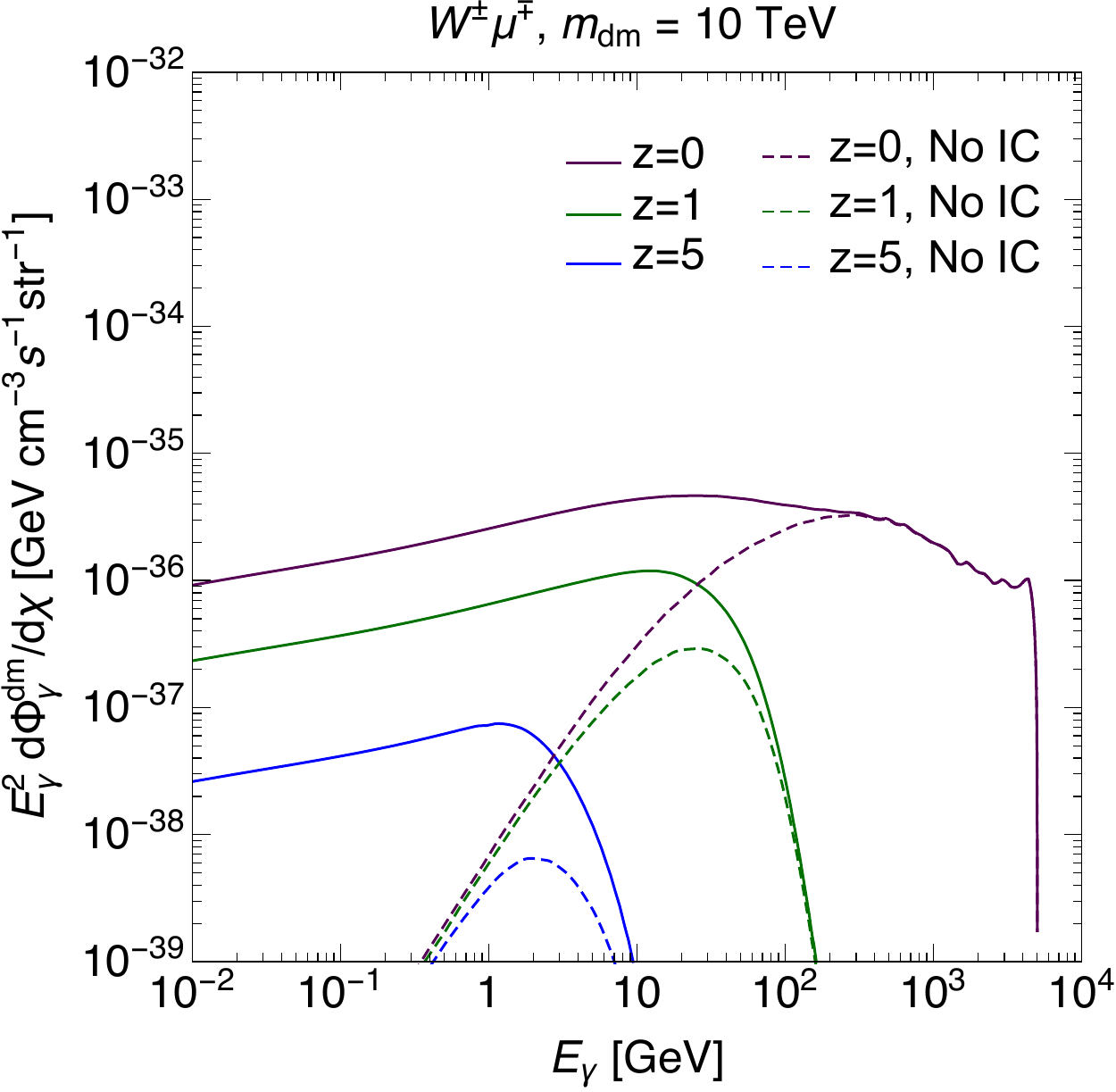}
  \caption{Gamma-ray intensity $d\Phi_\gamma^{\rm dm} /
    d\chi(E_\gamma,z)$ from decaying dark matter. Results are shown for
    $\nu \mu^{\pm}e^{\mp}\&\nu e^{\pm}e^{\mp}$ and $W^\pm \mu^\mp$
    final states. Total (without IC contribution) intensities for
    $z=0$, $1$, $5$ from top to bottom are plotted in solid (dashed)
    lines. Dark matter mass is taken as 1 (left) and 10 (right) TeV.}
  \label{fig:dPhi/dchi}
 \end{center}
\end{figure}

\begin{figure}
 \begin{center}
   \includegraphics[width=7cm]{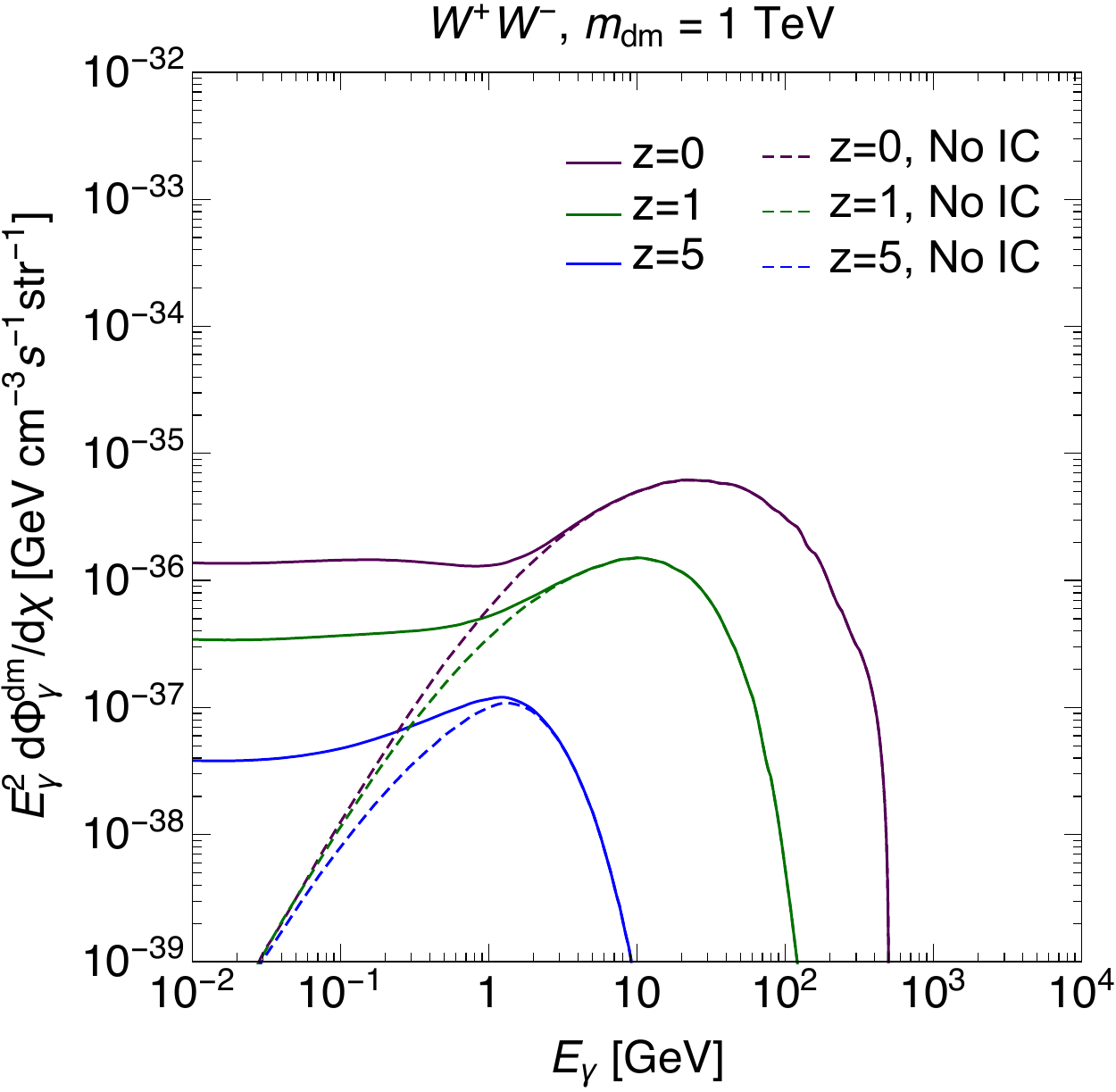}
   \includegraphics[width=7cm]{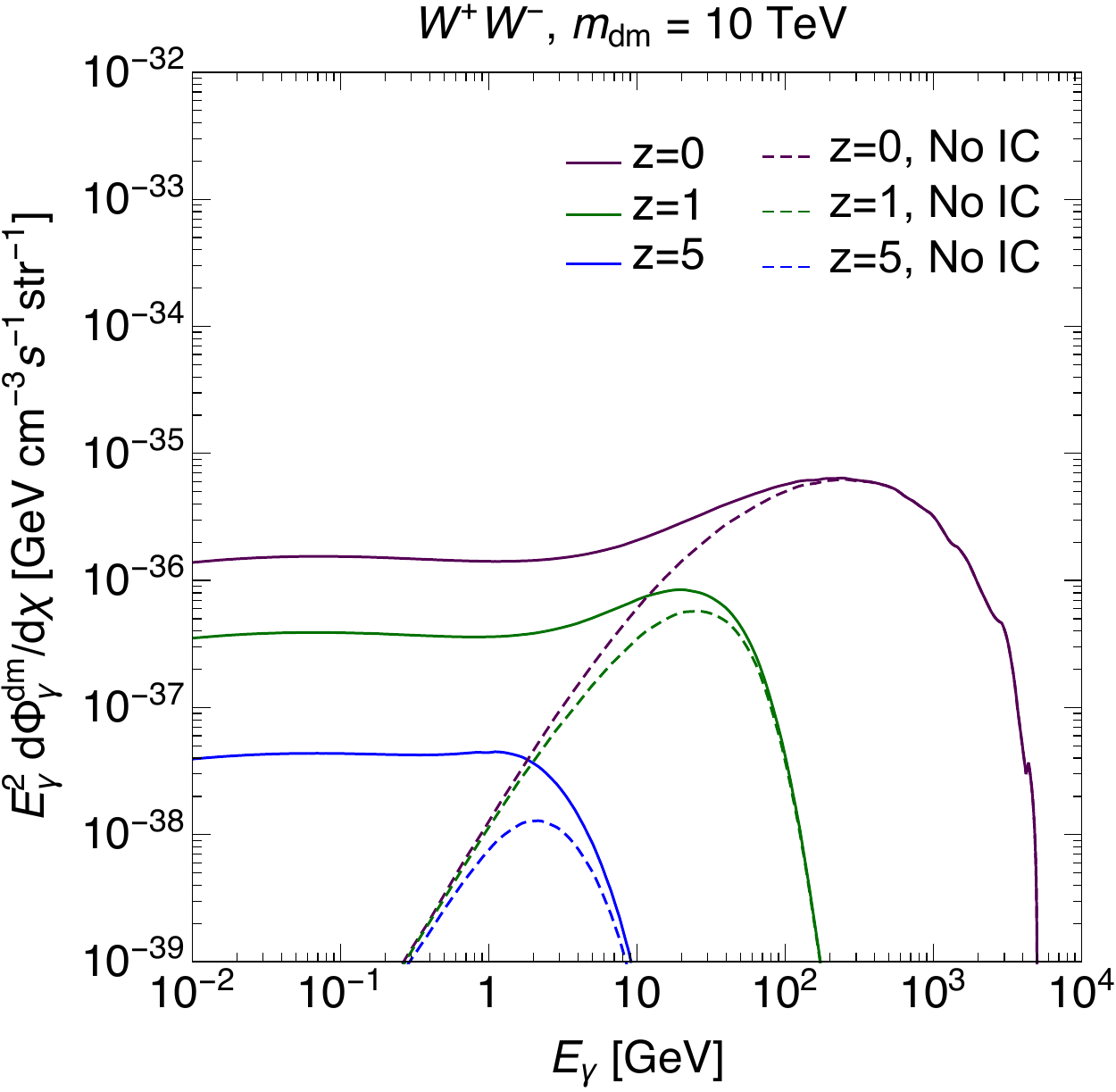}
   \includegraphics[width=7cm]{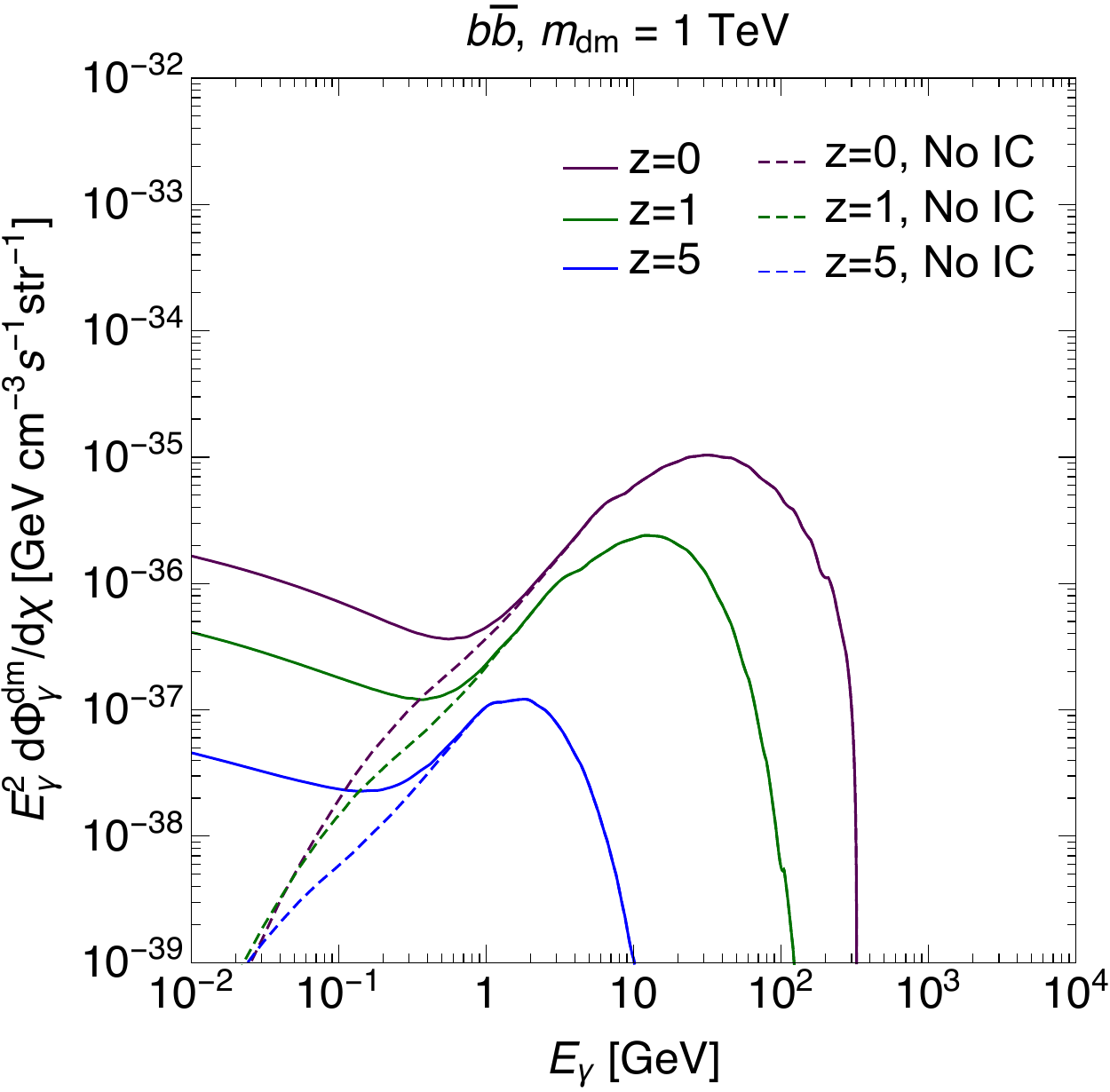}
   \includegraphics[width=7cm]{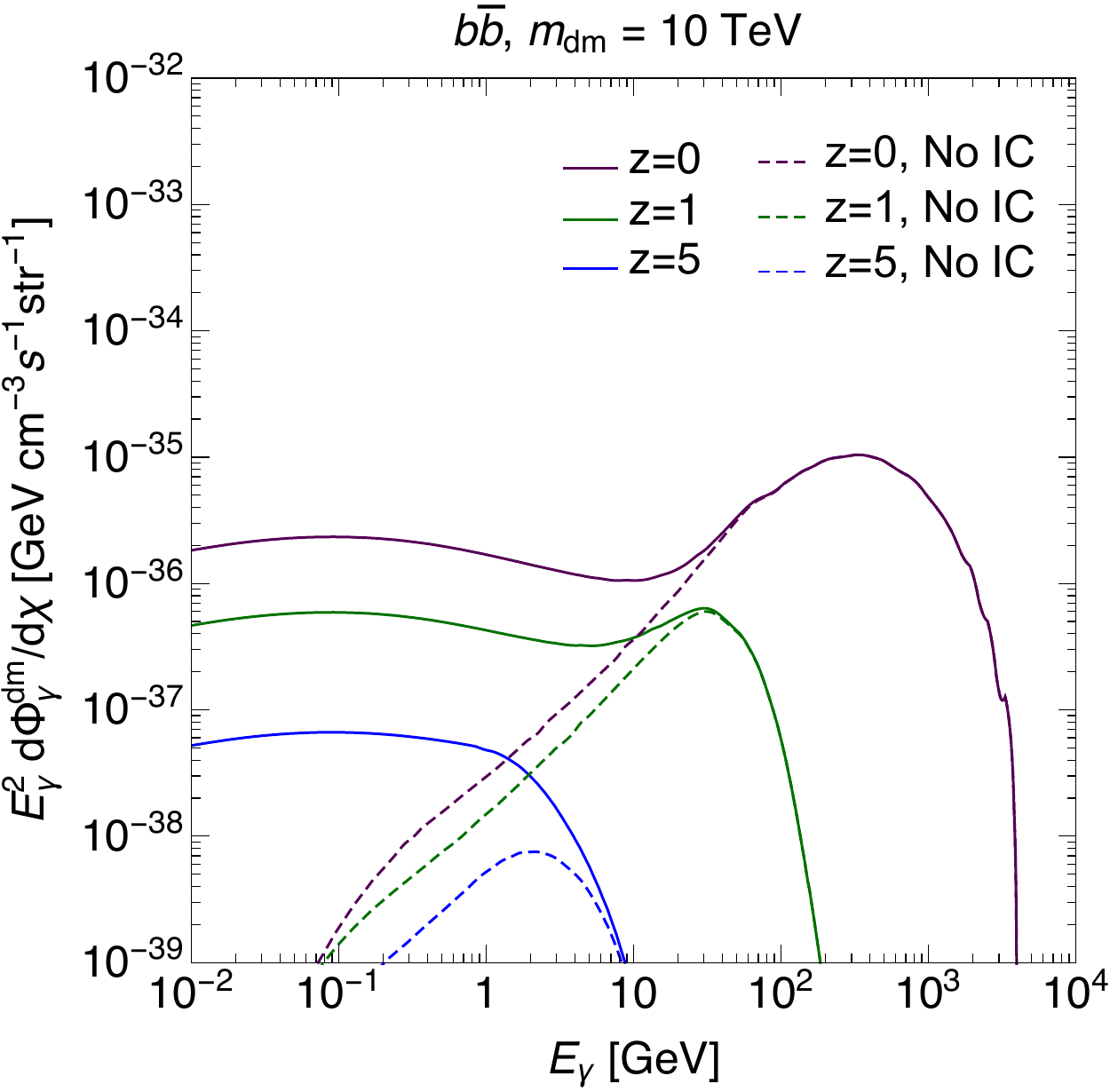}
  \caption{Same as Fig.\,\ref{fig:dPhi/dchi} but for $W^+W^-$ and
    $b\bar{b}$ final states.}
  \label{fig:dPhi/dchi_had}
 \end{center}
\end{figure}

In Fig.\,\ref{fig:dPhi/dchi}, we show the gamma-ray intensity
$d\Phi_\gamma^{\rm dm} / d\chi(E_\gamma,z)$ from decaying dark matter.
Here final states of the decay are $\nu \mu^{\pm}e^{\mp}\&\nu
e^{\pm}e^{\mp}$ (which corresponds to $L_1L_2E^c_1$-type $R_p$
violation) and $W^{\pm}\mu^{\mp}$, and the results for redshift $z=0$,
$1$ and $5$ are shown. For each $z$ the total gamma-ray intensity
(solid) and one without IC gamma ray (dashed) are plotted. The intensity
agrees with rough estimation from eq.\,\eqref{eq:dPhi_gamma}:
\begin{align}
E_\gamma^2\frac{d\Phi_\gamma^{\rm dm}}{d\chi}(E_\gamma,z) &\simeq
9.9 \times 10^{-36}\, {\rm GeV~cm}^{-3}\,{\rm s}^{-1}
\frac{10^{27}\,{\rm TeV~s}}{m_{\rm dm}\tau_{\rm dm}}
\left(\frac{E_{\gamma}}{1~{\rm GeV}}\right)^2
\nonumber \\ &~~~ \times
\frac{1}{1+z}~ \left[10~{\rm MeV}~ Q_\gamma^{\rm dm}(E'_\gamma,z)\right]
\, e^{-\tau(E'_\gamma,z)}\, .
\label{eq:dPhi_gamma_est}
\end{align}
Here we have estimated $Q^{\rm dm}_\gamma\sim Q^{\rm dm}_{\gamma_{\rm
    ic}}\sim 1/(10~{\rm MeV})$ in the Thomson limit. This is based on
the fact that in the Thomson limit the maximum energy of scattered
photon due to electron with an energy of $\sim$TeV in the CMB is
$\sim$GeV and the energy distribution of the scattered photon is
peaked at lower energy.  For $\nu \mu^{\pm}e^{\mp}\&\nu
e^{\pm}e^{\mp}$ final state, a peak in high-energy regions comes from
the FSR while another peak in lower energy corresponds to the IC gamma
rays. Although its energy range is lower, it is clear that the
intensity of the IC gamma rays is much larger than that of the
FSR. For $W^\pm\mu^\mp$ case, the first peak in high energy region is
mainly the contribution from the primary decay. The FSR is
subdominant, which can be seen by comparing the result of $W^+W^-$
final state in Fig.\,\ref{fig:dPhi/dchi_had}. (In the figure the same
results but for $b\bar{b}$ final states is shown too.)  The comparison
with $W^+W^-$ also gives us the importance of the IC gamma rays in
lower energy region. The IC gamma rays in $W^\pm\mu^\mp$ mainly
attributes to $e^\pm$ from $\mu^\pm$.

Eventually we get the window function 
\begin{align}
W^{\rm dm}_\gamma(z)=
\int^{E_{\rm max}}_{E_{\rm min}} dE_{\gamma}\ 
\frac{d\Phi_\gamma^{\rm dm}}{d\chi}(E_\gamma,z)\,.
\label{eq:Wdm}
\end{align}
$E_{\rm max}$ and $E_{\rm min}$ shows the energy region of gamma rays
we compute for the cross correlation with galaxy catalog. (See
Sec.\,\ref{sec:crosscorrelation}.) It is obvious from
eqs.\,\eqref{eq:dPhi_gamma}--\eqref{eq:Wdm} that the window function
can be obtained with little uncertainty if we specify $dN_I/dE$,
$m_{\rm dm}$ and $\tau_{\rm dm}$. This is not exactly the same for
annihilation.  In annihilation, although the window function is given
similarly to the decaying case, just by replacing the gamma-ray
intensity eq.\,\eqref{eq:dPhi_gamma} to
\begin{align}
\frac{d\Phi^{\rm dm}_\gamma}{d\chi}(E_\gamma,z)\bigr|_{\rm ann}&=
\frac{1}{8\pi}
\frac{\langle \sigma v \rangle(\Omega_{\rm dm}\rho_c)^2}{m_{\rm dm}^2} (1+z)^3
Q_\gamma^{\rm dm}(E'_\gamma,z)
\,\langle(1+\delta)^2\rangle \, e^{-\tau(E'_\gamma,z)}\, 
\\ &\simeq
6.2\times 10^{-43}\,{\rm GeV}^{-1}~{\rm cm}^{-3}\,{\rm s}^{-1}
\frac{\langle \sigma v \rangle}{10^{-25}\,{\rm cm}^3\,{\rm s}^{-1}}
\left(\frac{1~{\rm TeV}}{m_{\rm dm}}\right)^2
\nonumber \\ &~~~
\times (1+z)^3
 \left[10~{\rm MeV}~ Q_\gamma^{\rm dm}(E'_\gamma,z)\right]
\,\langle(1+\delta)^2\rangle \, e^{-\tau(E'_\gamma,z)}\, ,
\end{align}
where $\langle \sigma v \rangle$ is annihilation cross section, an
extra factor $\langle (1+\delta)^2 \rangle$, where the dark matter
overdensity $\delta=(\rho_{\rm dm}-\langle \rho_{\rm dm}
\rangle)/\langle \rho_{\rm dm} \rangle$ ($\rho_{\rm dm}$ is energy
density of dark matter), is needed to give gamma rays in
line-of-sight.  This factor boosts the annihilation rate, giving rise
to much larger gamma-ray intensity.  However, it is accompanied with
an uncertainty related to how one simulates clustering of (dark)
matter and extrapolates the results down to sub-grid scales.  We note
that recent theoretical studies ({\it e.g.}, ref.\,\cite{BA}) discuss
how to reduce the uncertainty to have a better handle on the
clustering properties. (See discussion in Sec.\,\ref{sec:res_DMonly}.)

\section{Extragalactic gamma-ray background from astrophysical sources}
\label{sec:EGRB_astro}
\setcounter{equation}{0}

As in our previous paper~\cite{Ando:2015qda}, we considered two
astrophysical sources as a potential contaminating background for dark
matter searches; blazars and star-forming galaxies.

Blazars are the dominant gamma-ray source, thousands of which have
been detected with Fermi-LAT~\cite{Acero:2015hja}.  Their number
densities per unit luminosity range ({\it i.e.}, the luminosity
function $dn_\gamma/dL_\gamma$) have been constructed by using the
luminosity-dependent density evolution model~\cite{Ajello:2015mfa}.
They are characterized by a double-power-law function, where the break
luminosity depends on redshift.  The window function $W_\gamma(z)$
is then computed as
\begin{equation}
 W_\gamma^X(z) = \chi^2 \int dL_\gamma \frac{dn_\gamma^X
  (L_\gamma,z)}{dL_\gamma} F_\gamma (L_\gamma,z),
  \label{eq:window_astro}
\end{equation}
where the superscript $X$ represents astrophysical sources: blazars and
star-forming galaxies, $F_\gamma = L_\gamma/[4\pi (1+z) \chi^2]$ is the
number flux of the gamma-ray photons from a source with the luminosity
$L_\gamma$ and at the redshift $z$.
With the most recent luminosity function, ref.\,\cite{Ajello:2015mfa}
showed that the blazar component could explain about 50\% of the
gamma-ray background above 100~MeV, while most above 100~GeV.

At low energies, there are softer astrophysical components playing a
major role in the gamma-ray intensity.
One such class is the star-forming and starburst galaxies, where cosmic
rays supplied by supernovae produce gamma rays from interactions with
interstellar medium.
There is an established correlation between the gamma-ray luminosity and
infrared luminosity, which is a proxy for star-formation
activity~\cite{Ackermann:2012vca}.
Combining this with recent measurements of the infrared luminosity
function~\cite{Gruppioni:2013jna}, ref.\,\cite{Tamborra:2014xia} obtained
the gamma-ray luminosity function for both the star-forming and
starburst galaxies.
One can then compute the window function for these galaxies by using
eq.~(\ref{eq:window_astro}).
They can give significant contribution to the gamma-ray background,
particularly at low energies.
A similar contribution is also expected from another gamma-ray source
class: misaligned active galactic nuclei~\cite{Inoue:2011bm,
DiMauro:2013xta}.
We do not, however, include this population in our analysis, as their
contribution to the gamma-ray background is similar to that from the
star-forming galaxies, but with larger uncertainties. Including it will
further constrain the parameter spaces of dark matter.

The astrophysical contributions from all these sources can accommodate
most (if not all) of the measured energy spectrum of the EGRB.  See
Fig.~9 of ref.~\cite{Fornasa:2015qua} for a summary plot.  Therefore,
one can obtain stringent constraints on both dark matter
decay~\cite{Ando:2015qda} and annihilation~\cite{Ajello:2015mfa,
  DiMauro:2015tfa} from the energy spectrum alone.  One can further
tighten these limits by investigating clustering properties of the
gamma-ray data, {\it e.g.}, by cross-correlating with galaxy
distributions, as we shall discuss in the following sections.

\section{Cross correlation between the extragalactic gamma-ray background with galaxy catalogs}
\label{sec:crosscorrelation}
\setcounter{equation}{0}

Galaxies trace underlying dark matter distribution, and therefore, it
is expected that the distribution of the gamma-ray photons that come
from dark matter decay or annihilation are spatially correlated with
those of galaxies.  This cross correlation between gamma-ray emitters
($\gamma$) and galaxies (g) is quantified by the angular cross-power
spectrum as
\begin{equation}
 C_\ell^{\gamma {\rm g}} = \int\frac{d\chi}{\chi^2}W_\gamma(z) W_{\rm
  g}(z) P_{\gamma {\rm g}}\left(k=\frac{\ell}{\chi},z\right),
  \label{eq:angular cross-power}
\end{equation}
where $W_{\rm g}(z)$ is related to the redshift distribution of
galaxies in a catalog through $W_{\rm g} = (d\ln N_{\rm
  g}/dz)(dz/d\chi)$.  $P_{\gamma {\rm g}}(k,z)$ is the cross-power
spectrum between the gamma-ray sources and galaxies at wave number $k$
and redshift $z$.  For discussions of decaying dark matter and
astrophysical sources, we assume that it is well approximated by a
matter power spectrum with a constant bias parameter: $P_{\gamma {\rm
    g}}(k,z) \approx b_\gamma b_{\rm g} P_{\rm m}(k,z)$ ($P_{\rm m}$
is matter power spectrum), and $b_{\rm dm} = 1$ for the dark matter
component. (See later discussion for $b_g$.) For annihilating dark
matter, on the other hand, since the rate of annihilation depends on
the density squared, one has to evaluate the cross-power spectrum
between the density squared and density, $P_{\delta^2\delta}(k,z)$,
and we assume $P_{\gamma {\rm g}}(k,z) \approx b_{\rm g}
P_{\delta^2\delta}(k,z)$.  For computing $P_{\delta^2\delta}(k,z)$, we
follow an analytic halo-model prescription introduced in
ref.\,\cite{Ando:2013xwa}.

A great advantage of taking cross correlation over analyzing the
energy spectrum of the gamma-ray background is that one can filter
gamma-ray emission from a preferred redshift range.  This can be seen
from the fact that the integrand of eq.~(\ref{eq:angular cross-power})
depends on $W_\gamma(z) W_{\rm g}(z)$, because distribution of
gamma-ray sources are uncorrelated with galaxy distribution at
different redshifts.  Since larger contributions from dark matter
annihilation and decay come from lower redshift, while it is opposite
for ordinary astrophysical sources~\cite{Ando:2014aoa}, one can
efficiently remove the astrophysical backgrounds by choosing galaxy
catalogs.  Several theoretical studies showed that the sensitivity for
dark matter from the cross correlation would be much tighter than that
from the energy spectrum~\cite{Ando:2013xwa, Ando:2014aoa,
  Fornengo:2013rga}.

Recently, ref.\,\cite{Xia:2015wka} measured cross correlations between
the gamma-ray background measured with Fermi-LAT with several galaxy
catalogs.  The catalogs that the authors used for the analyses are
SDSS QSO, 2MASS, NVSS, SDSS MG, and SDSS LRG, and they found positive
signatures at greater than 3.5$\sigma$ for the first three catalogs,
and about 3$\sigma$ for the last two.  Particularly, the 2MASS catalog
is for the lowest-redshift galaxies, which peak around $z \sim 0.1$,
which is the most suitable for dark matter searches.  The
cross-correlation measurements are then interpreted and used to put
constraints on dark matter properties as well as astrophysical models.
By conservatively including dark matter alone,
ref.\,\cite{Regis:2015zka} showed that the lower limits on, {\it
  e.g.}, decay lifetime of dark matter, were quite stringent, being
improved by about one order of magnitude compared with the previous
limits obtained with the spectral analysis~\cite{Ando:2015qda}.
Reference\,\cite{Cuoco:2015rfa} extended the analyses to include a few
astrophysical components, and showed that the limits further improved
by factors of several.

In this paper, we consider both decaying (or annihilating) dark matter
component and two classes of astrophysical sources (blazars and
star-forming galaxies).  For the cross-power spectrum $P_{\gamma {\rm
    g}}$, we assume that biases of the astrophysical sources
($b_\gamma$) are both 1.  This is conservative because typical
astrophysical sources are considered and that they are found to be
positively biased ({\it i.e.}, $b_\gamma >
1$)~\cite{Allevato:2014qga}.  For the bias parameter of galaxies in
the chosen catalogs ($b_{\rm g}$), we adopt the values found in
ref.\,\cite{Xia:2015wka}.  In addition to these `source' terms for
$C_\ell^{\gamma {\rm g}}$, we also include a noise term, constant as a
function of $\ell$.  This is to accommodate the shot-noise term that
comes from the fact that the astrophysical sources are point-like
sources, and hence, it yields such a scale-independent term in the
power spectrum~\cite{Ando:2014aoa}.  It is also meant to correct for
uncertainties of the power spectrum at small angular scales, as also
mentioned in ref.\,\cite{Cuoco:2015rfa}.  For these correction terms,
we follow the prescription in ref.\,\cite{Cuoco:2015rfa}, where
$C_{\rm corr}^{\gamma {\rm g}} = A_{\gamma {\rm g}}$, and they are
function of both gamma-ray energy ($\gamma$) and galaxy catalog (${\rm
  g}$).  But we assume that the energy spectrum is represented as
$E^{-2.3}$ that is the measured EGRB spectrum, which leaves five
independent parameters $A_{\rm g}$ for $> 1$~GeV, corresponding to
five galaxy catalogs (${\rm g} = \{{\rm QSO, 2MASS, NVSS, MG,
  LRG}\}$).

\section{Results}
\label{sec:results}
\setcounter{equation}{0}

\subsection{Analysis including dark matter component alone}
\label{sec:res_DMonly}

First we perform a conservative analysis by taking dark matter component
only into account.
We use the data for the angular cross-power spectrum $C_\ell^{\gamma
{\rm g}}$ for gamma-ray data in three energy bins ($>500$~MeV, $>1$~GeV,
and $>10$~GeV) and five different galaxy catalogs, as obtained in
ref.\,\cite{Xia:2015wka}.
For a given dark matter mass and a final state, we compute a $\chi^2$
statistic:
\begin{equation}
 \chi^2 = \sum_{\gamma,{\rm g}} \sum_{\ell,\ell^\prime}
  \left(C_{\rm dat}^{\gamma {\rm g}}-C_{\rm th}^{\gamma {\rm g}}\right)_\ell
  \left(\mbox{Cov}^{-1}\right)_{\ell\ell^\prime}
  \left(C_{\rm dat}^{\gamma {\rm g}}-C_{\rm th}^{\gamma {\rm
   g}}\right)_{\ell'},
  \label{eq:chi2}
\end{equation}
in order to obtain a constraint on $\tau_{\rm dm}$ ($\langle \sigma v
\rangle$) for decay (annihilation) scenario.\footnote{This $\chi^2$
  has nothing to do with the comoving distance, although the same
  symbol $\chi$ is used.}  Here `dat' and `th' represent data and
theoretical value, respectively, `Cov' is the covariance matrix for
the cross-correlation measurements, and $\gamma$ and g run for three
energy and five galaxy bins, respectively.  After obtaining the
minimum $\chi^2$ by solving $\partial \chi^2 / \partial \tau_{\rm dm}
= 0$ or $\partial \chi^2 / \partial \langle \sigma v\rangle = 0$, we
solve $\Delta \chi^2 \equiv \chi^2 - \chi_{\rm min}^2 = 2.71$ in order
to obtain the 95\% confidence level limit on $\tau_{\rm dm}$ or
$\langle \sigma v \rangle$.

\begin{figure}
 \begin{center}
  \includegraphics[width=7cm]{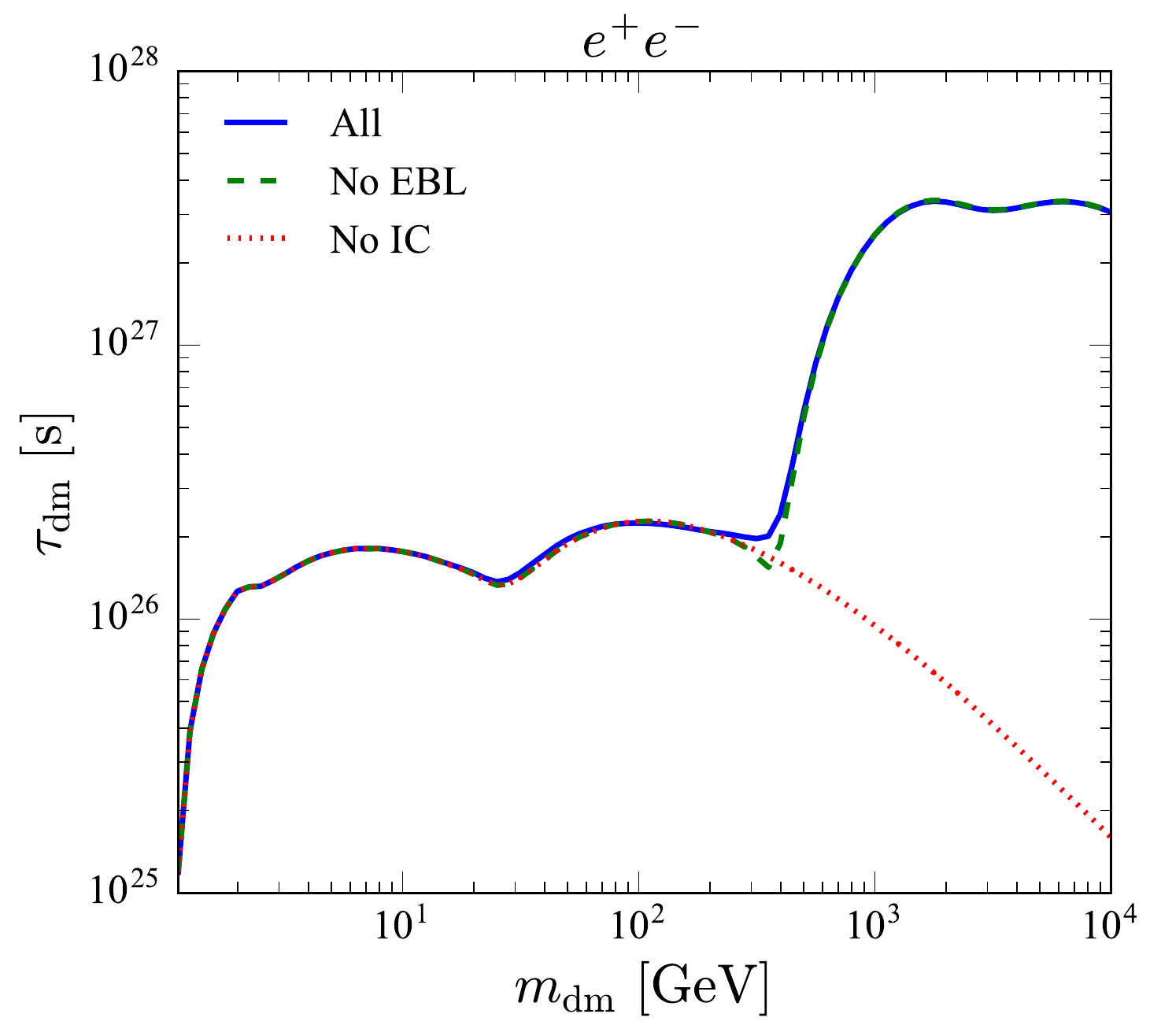}
  \includegraphics[width=7cm]{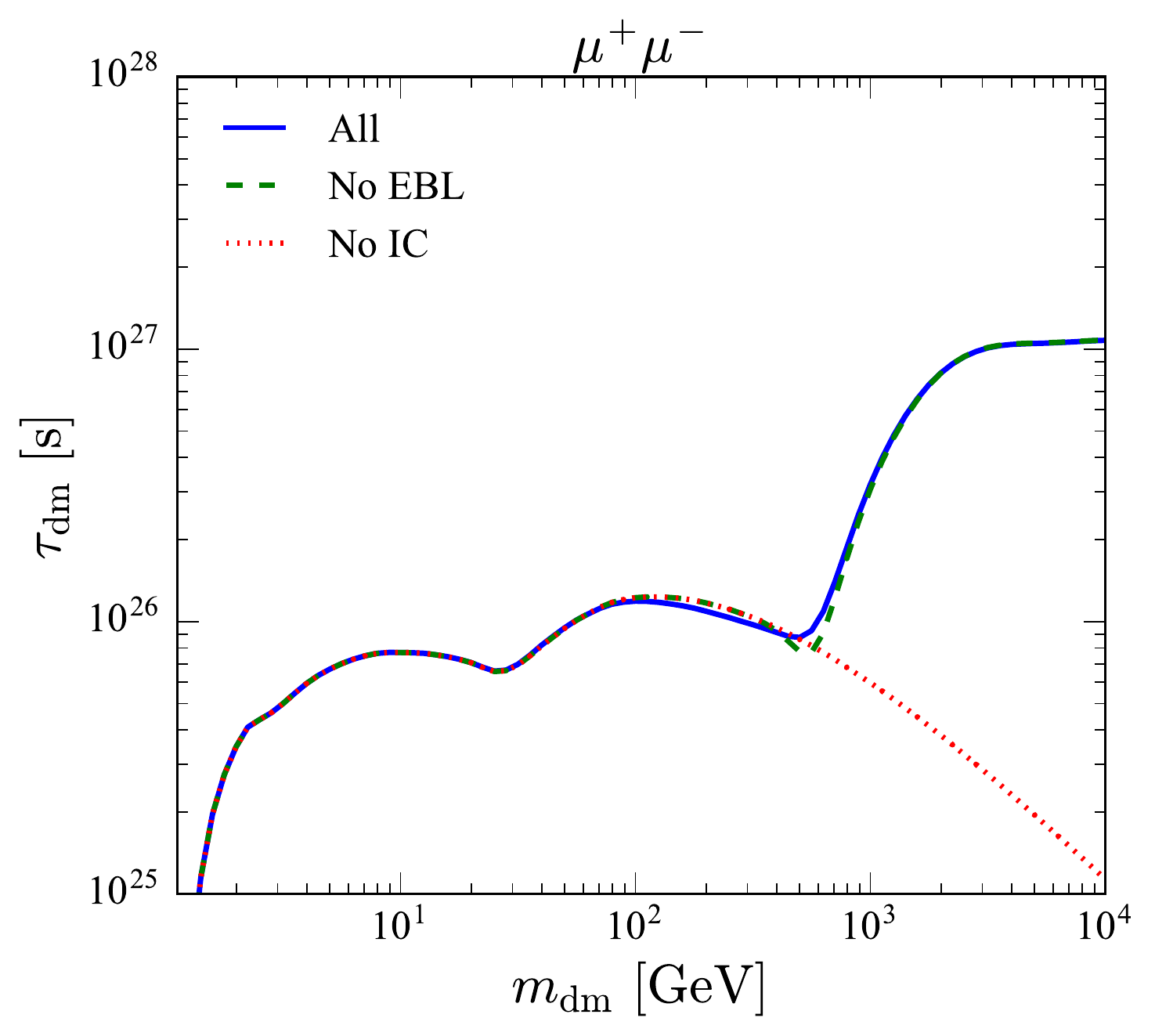}
  \includegraphics[width=7cm]{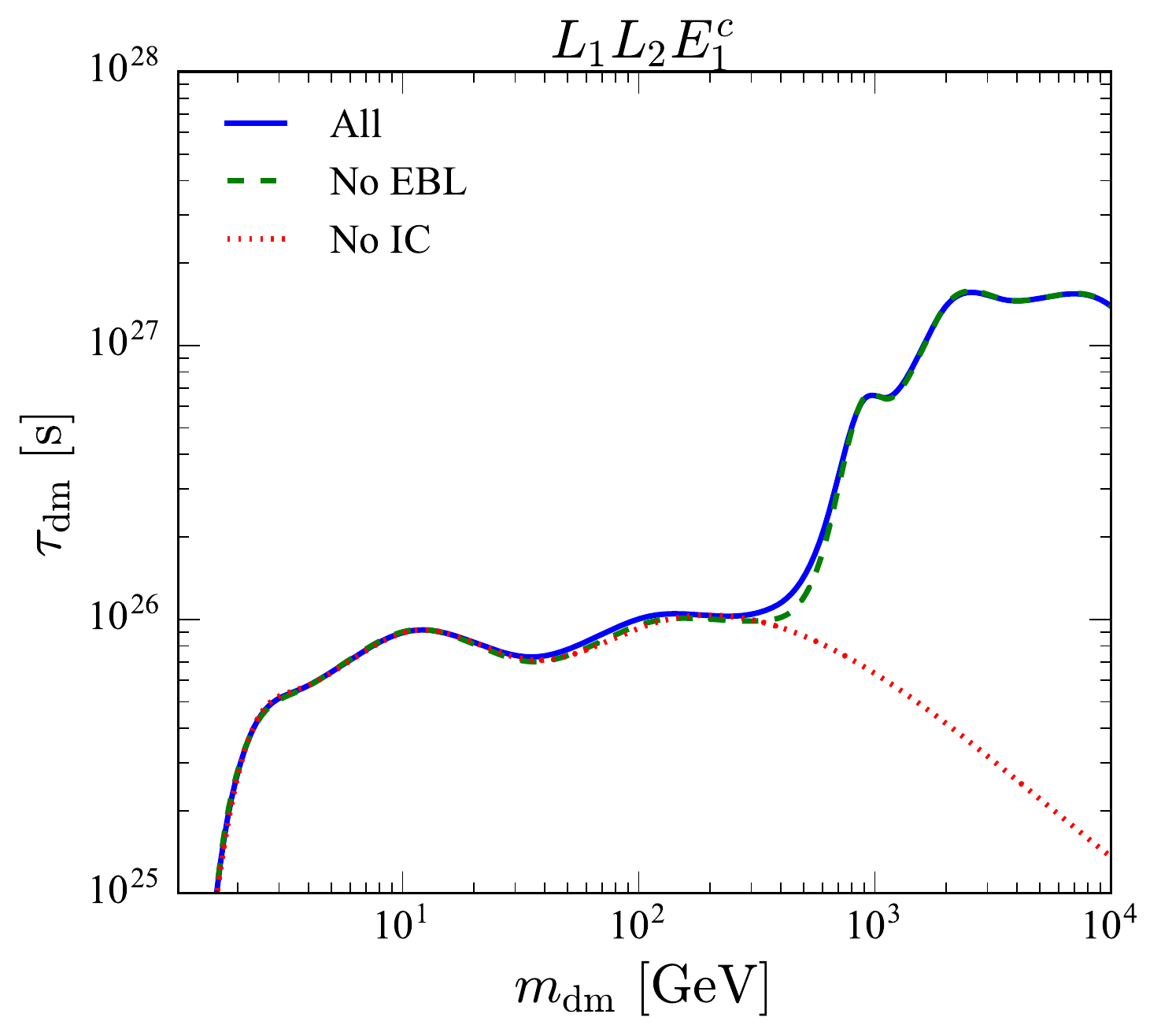}
  \includegraphics[width=7cm]{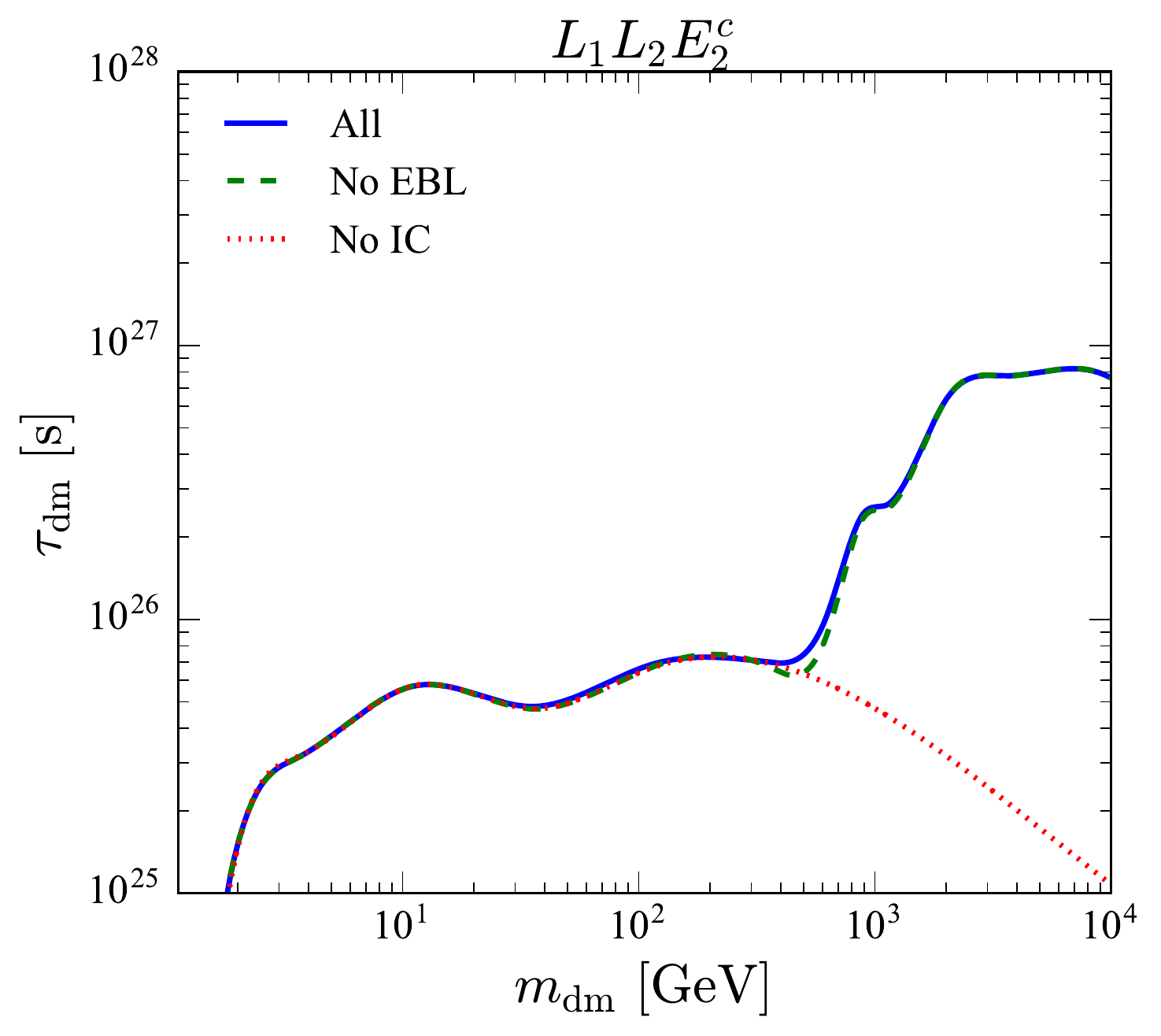}
  \caption{Lower limits on dark matter lifetime for leptonic channels
    due to cross correlation between the gamma-ray background and five
    galaxy catalogs. In this analysis, only dark matter component is
    included. IC scattering off both the EBL and CMB photons are
    included in solid, only CMB in dashed, and no IC effect is
    included in dotted.}
  \label{fig:limit95_DMonly_leptonic}
 \end{center}
\end{figure}

\begin{figure}
 \begin{center}
  \includegraphics[width=7cm]{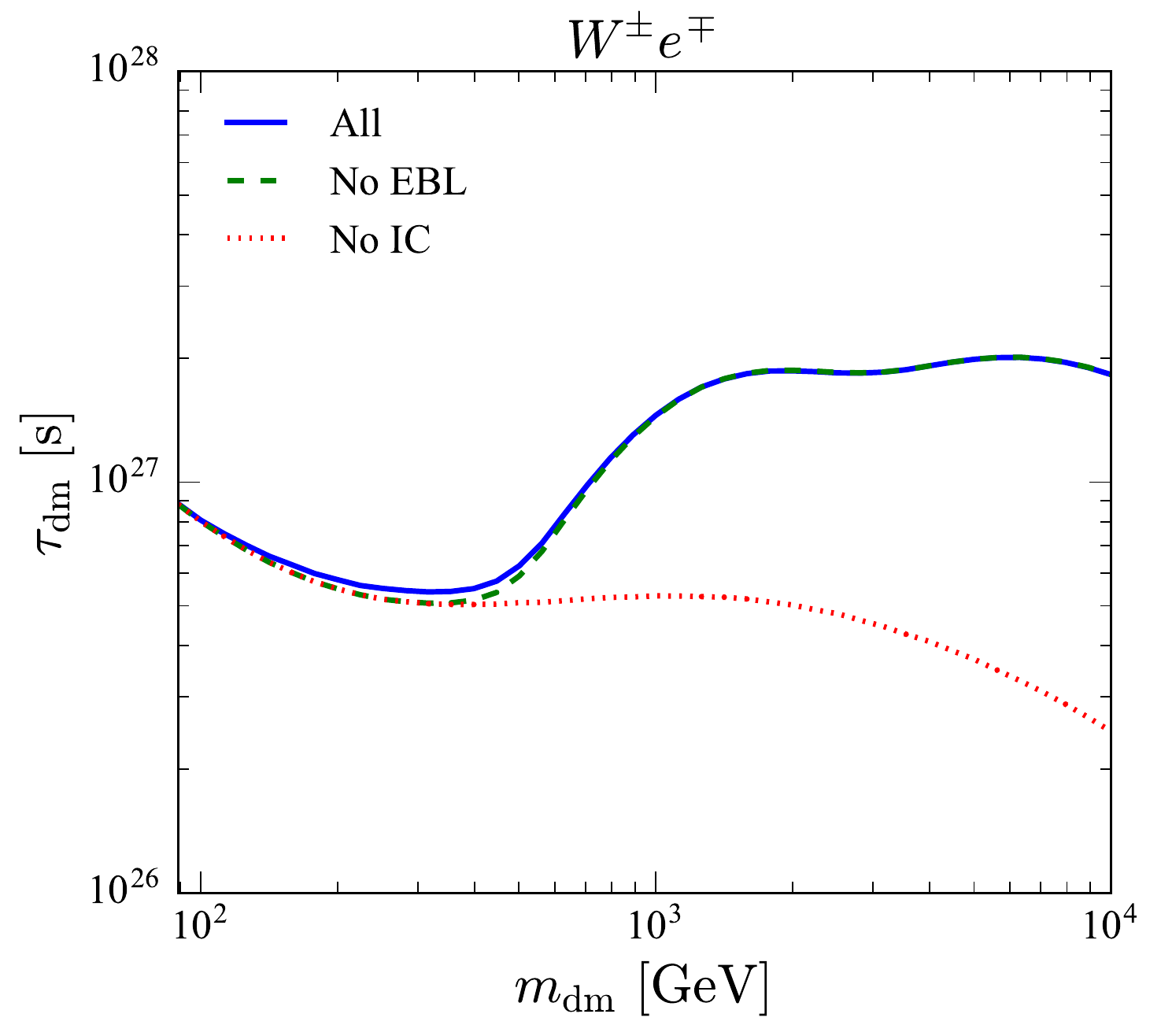}
  \includegraphics[width=7cm]{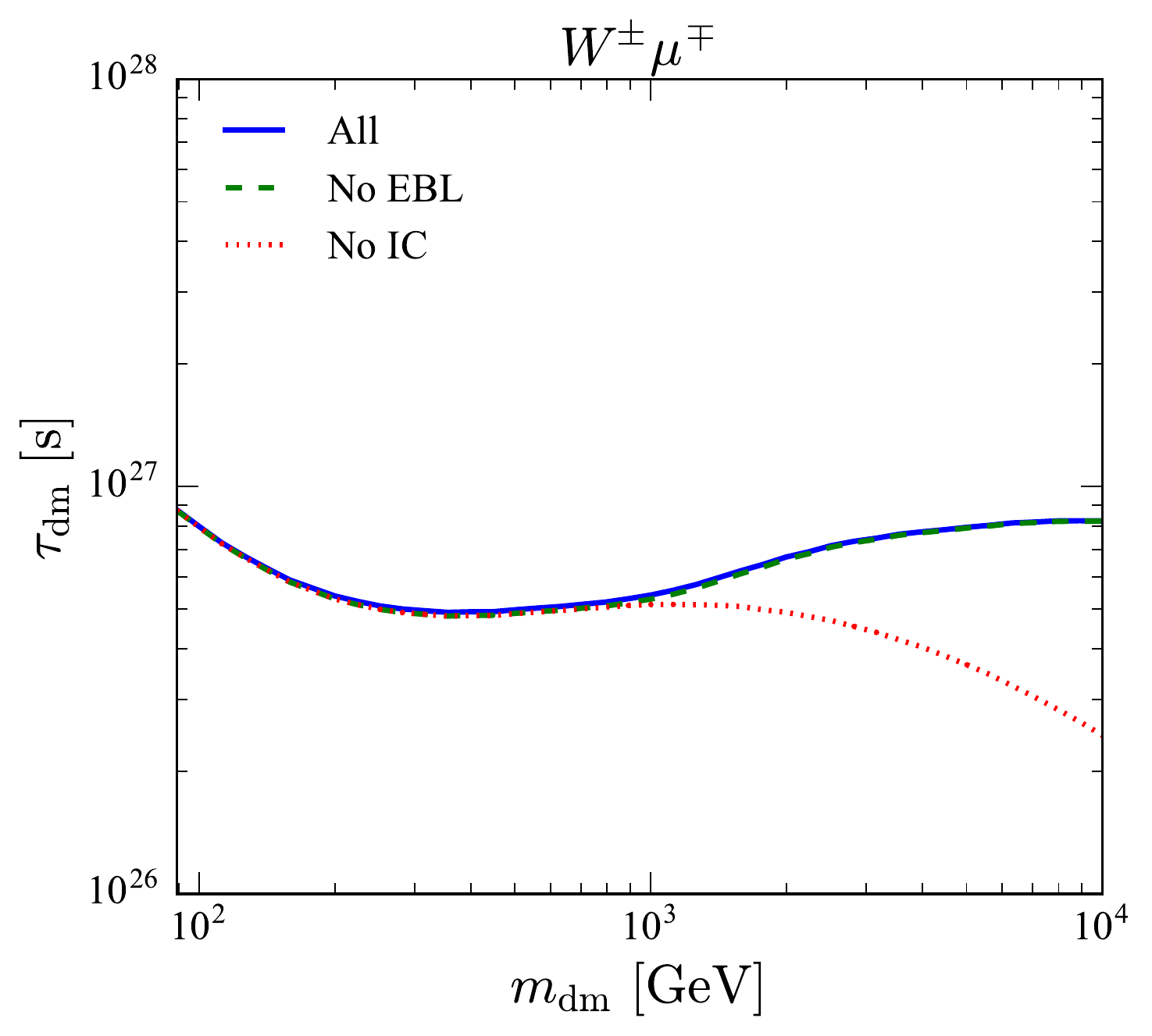}
  \includegraphics[width=7cm]{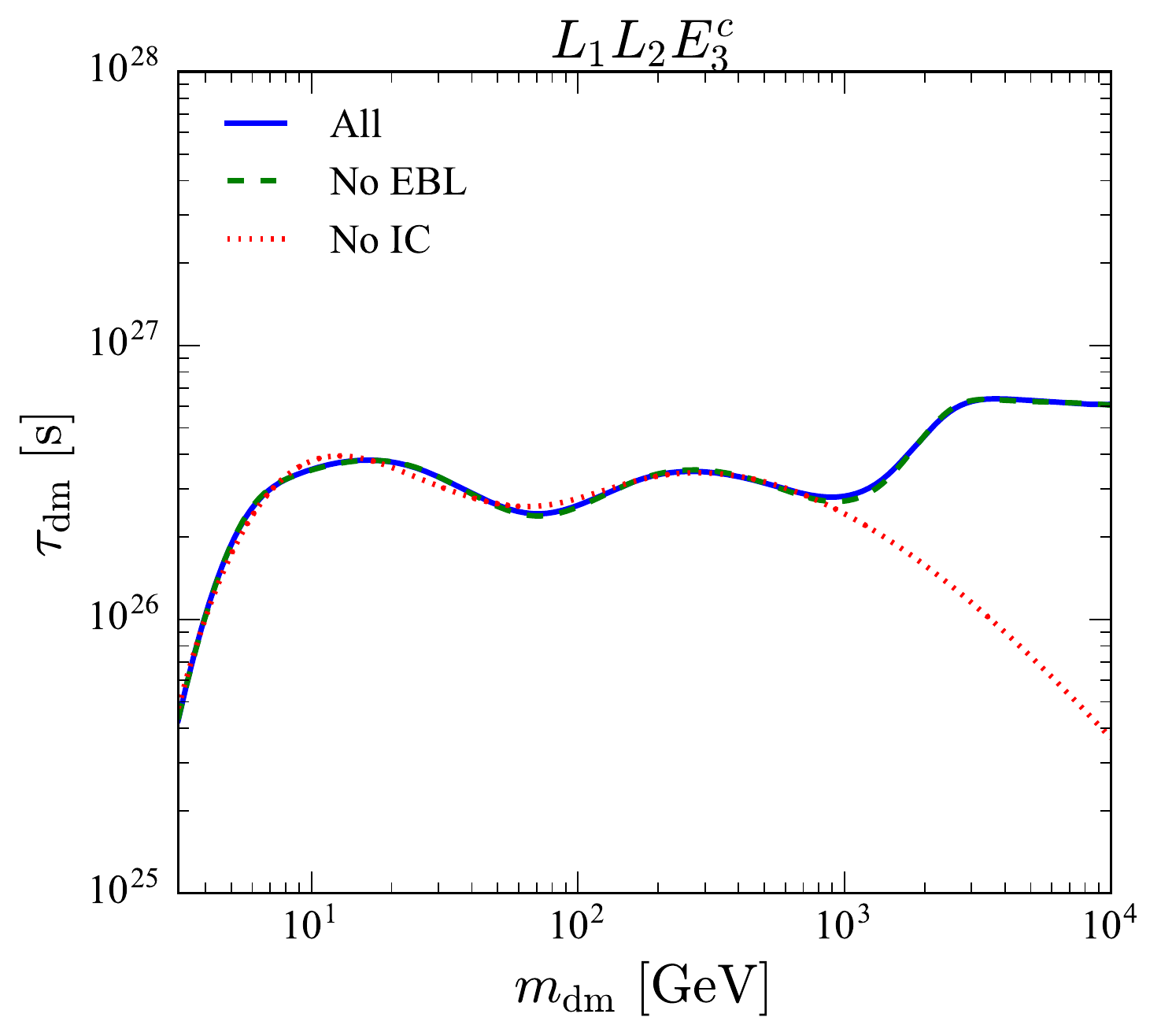}
  \includegraphics[width=7cm]{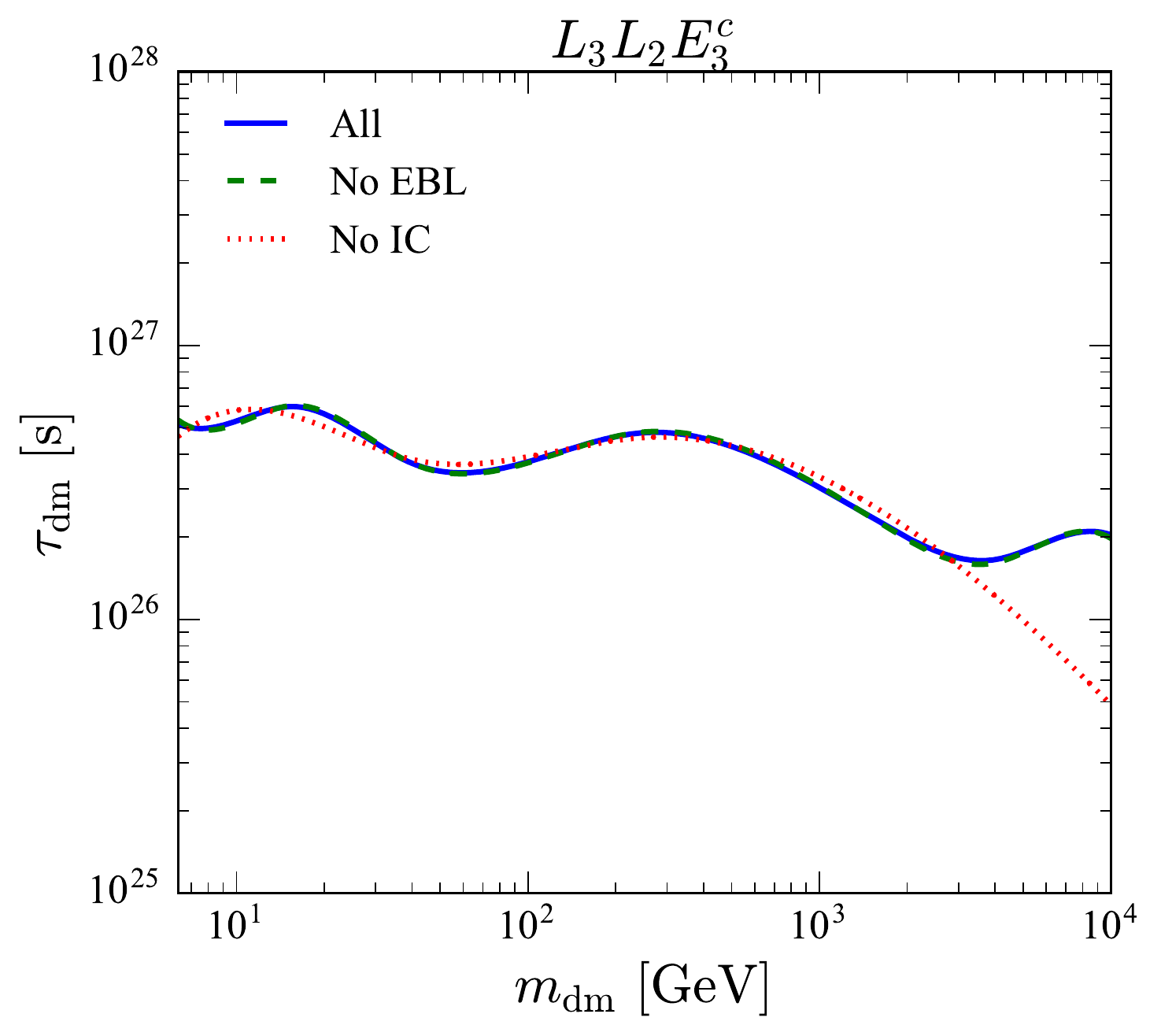}
  \caption{The same as Fig.~\ref{fig:limit95_DMonly_leptonic} but for
    hadroleptonic channels.}
  \label{fig:limit95_DMonly_hadro-leptonic}
 \end{center}
\end{figure}

\begin{figure}
 \begin{center}
  \includegraphics[width=7cm]{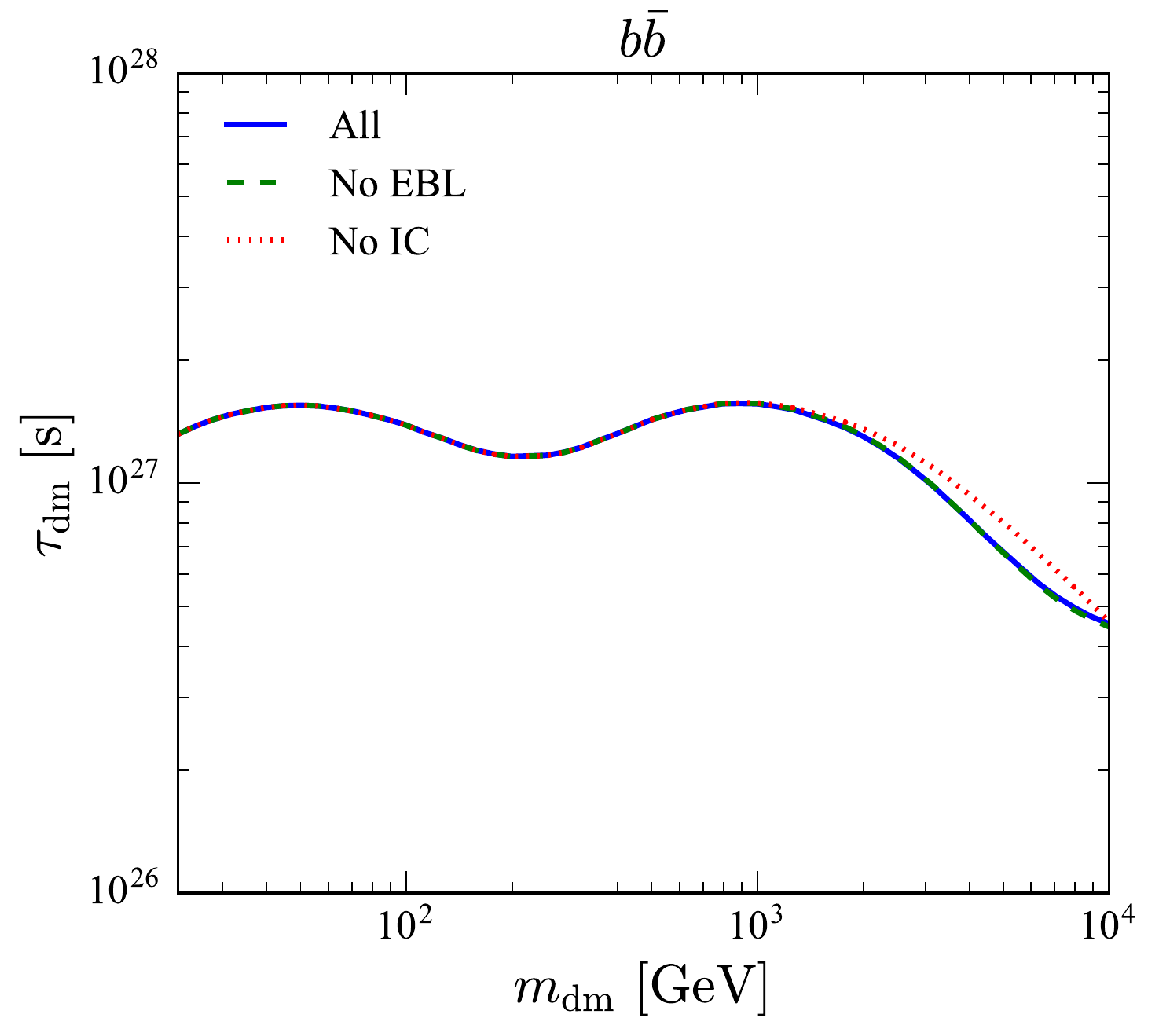}
  \includegraphics[width=7cm]{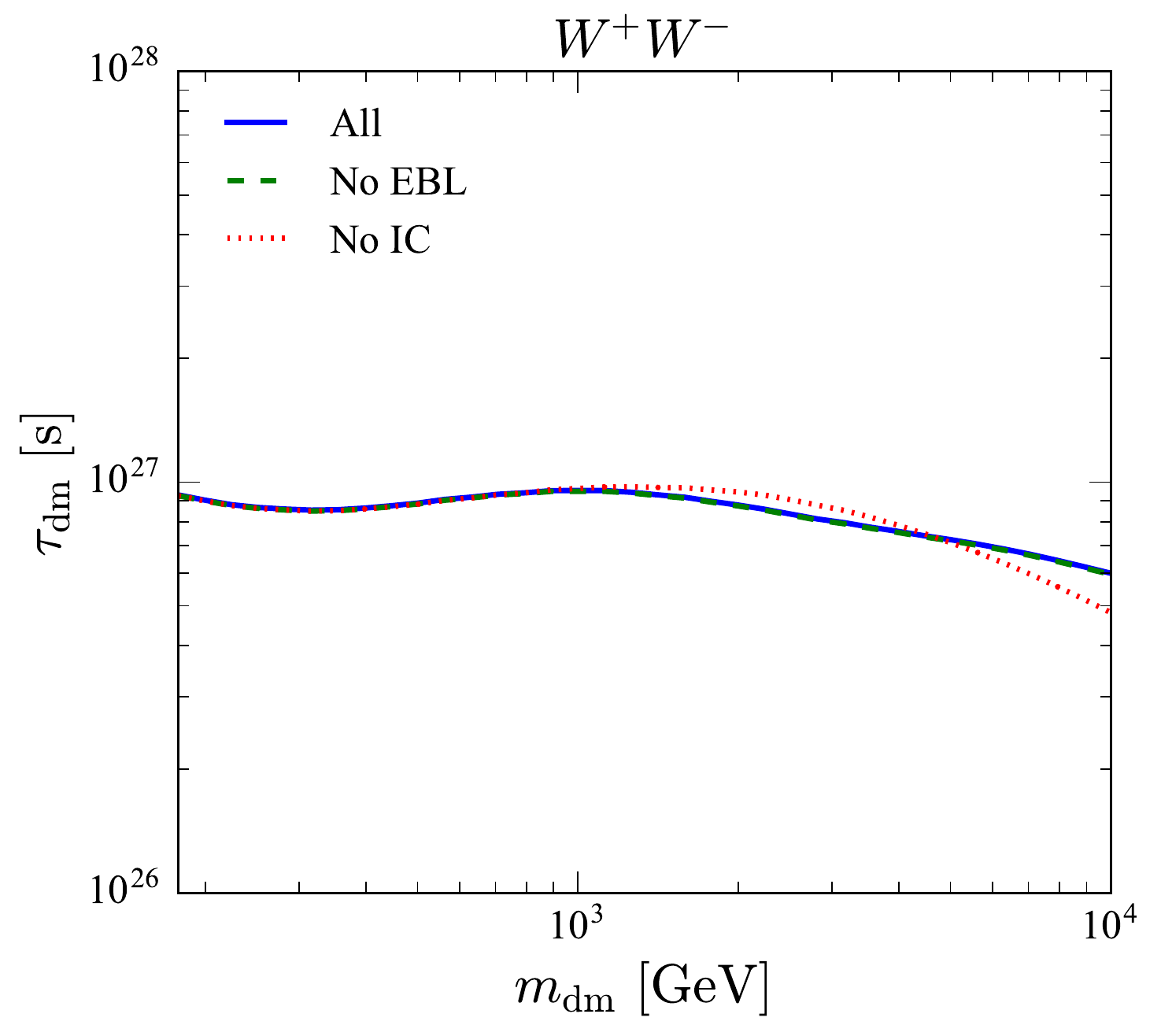}
  \includegraphics[width=7cm]{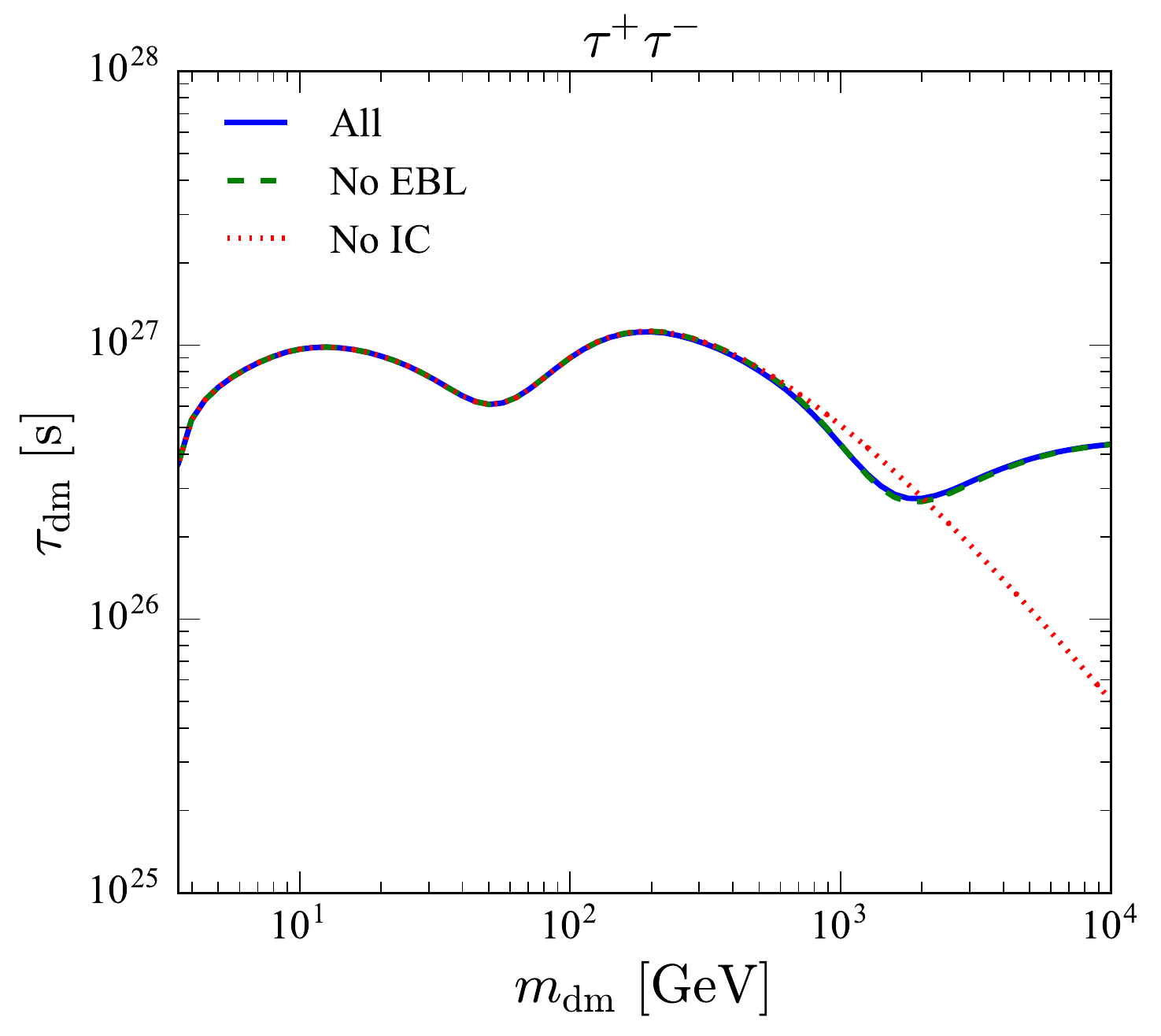}
  \includegraphics[width=7cm]{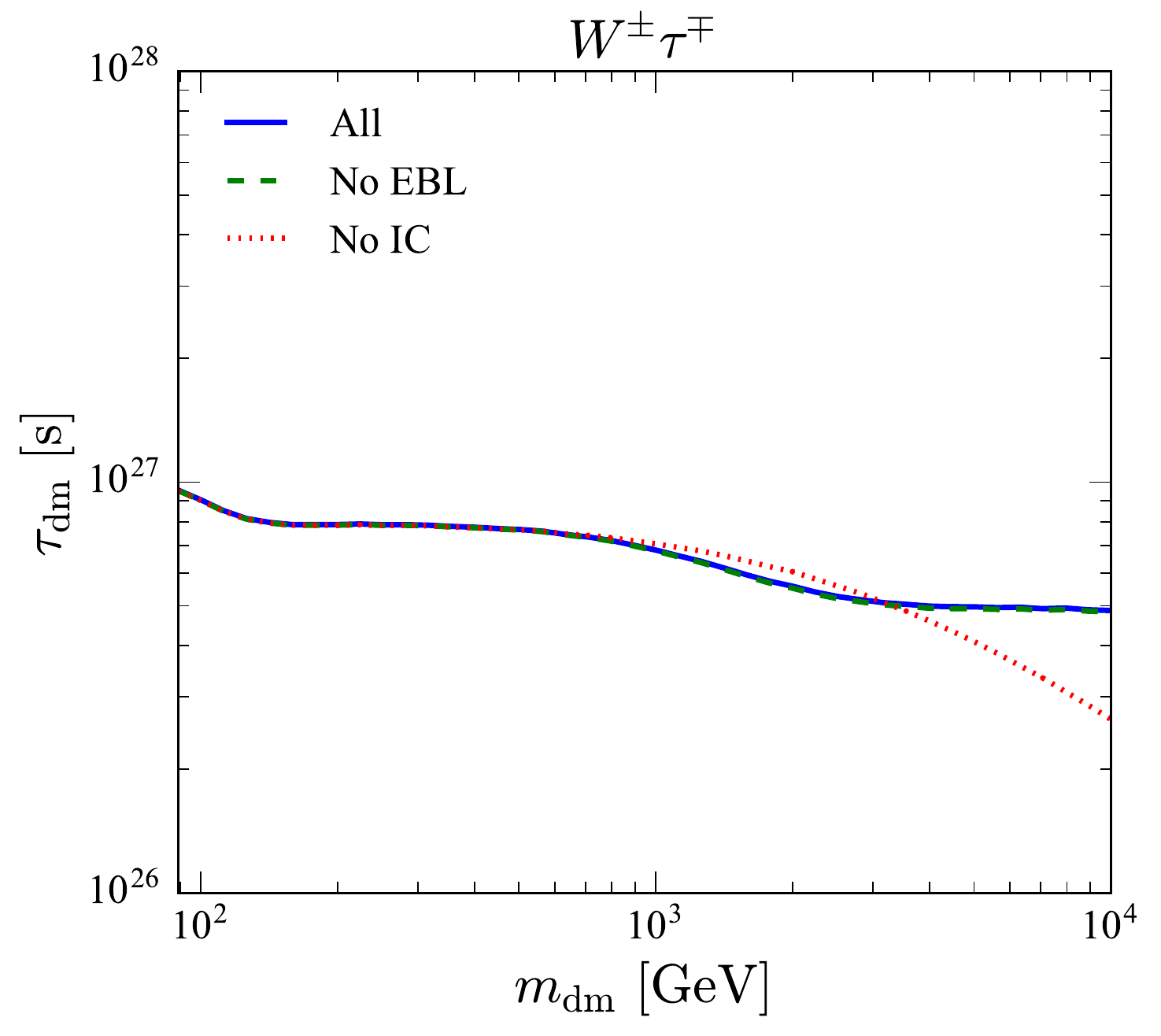}
  \caption{The same as Fig.~\ref{fig:limit95_DMonly_leptonic} but for
  the hadronic channels.}
  \label{fig:limit95_DMonly_hadronic}
 \end{center}
\end{figure}

In
Figs.\,\ref{fig:limit95_DMonly_leptonic}--\ref{fig:limit95_DMonly_hadronic},
we show 95\% confidence level lower limits on the decay lifetime for
various decay modes.  We also show the effect of including the IC
scattering off both the CMB and EBL photons in the analysis.  In fact
it is found that the IC scattering is very important for the decay
channels that involve leptons in the final state and for dark matter
heavier than $\sim$1~TeV.  This is clearly seen in
Figs.\,\ref{fig:limit95_DMonly_leptonic} and
\ref{fig:limit95_DMonly_hadro-leptonic}.  The constraints due to the
IC process have improved the lower limits by more than one order of
magnitude for TeV region. The limits are better than those given from
clusters or dwarf galaxies. For example, in $\mu^+\mu^-$ channel, our
result shows that $\tau_{\rm dm}\lesssim$ (3--10)$\times 10^{26}\,{\rm
  sec}$ for $m_{\rm dm}=1$--10 TeV. This is more stringent limit by a
factor of a few compared to the latest constraints from dwarf
galaxies~\cite{Baring:2015sza}. (See also related past
works~\cite{Dugger:2010ys,Huang:2011xr}.)  We note that our results
for no-IC case well agree with those of ref.\,\cite{Regis:2015zka}.

Let us see the scenarios motivated to explain the AMS-02 positron
excess.  As discussed in Sec.~\ref{sec:scenario_DM}, three-body
leptonic decay gives a good fit to explain the positron excess.  The
numerical results for `$L_iL_jE^c_k$' in
Figs.~\ref{fig:limit95_DMonly_leptonic} and
\ref{fig:limit95_DMonly_hadro-leptonic} correspond to the final state
$\nu_i e_j^{\pm}e_k^{\mp}$\&$\nu_je_i^{\pm}e_k^{\mp}$. It is seen that
$\tau_{\rm dm}\lesssim 10^{27}\,{\rm s}$ is excluded for $m_{\rm
  dm}\gtrsim 1~{\rm TeV}$ in cases of $L_1L_2E^c_1$ and $L_1L_2E^c_2$.
These constraints exclude part of preferred region to explain the
positron excess~\cite{Ibe:2014qya}.  Similarly, the preferred
parameter spaces are partly excluded for $L_1L_2E^c_3$ and
$L_3L_2E^c_3$.

We have drawn a similar conclusion for the scenario to explain the
AMS-02 anti-proton excess, {\it i.e.}, the preferred region in
$W^{\pm}\mu^{\mp}$ scenario~\cite{Hamaguchi:2015wga} is partly
excluded. It is well-known that there is a huge uncertainty in the
calculation of cosmic-ray (anti-)protons.  This uncertainty can be
estimated by considering three propagation models, MIN, MED and
MAX~\cite{Donato:2003xg}. Reading from the result in
ref.\,\cite{Hamaguchi:2015wga}, one can explain the excess while
satisfying the present constraint if MAX or MED models for
(anti-)proton propagation are adopted. It will be shown in the next
subsection, however, that the remaining parameter regions for
$L_iL_jE^c_k$ and $W^{\pm}\mu^{\mp}$ final states are excluded when
the astrophysical components are taken into account.

In Figs.~\ref{fig:limit95_DMonly_leptonic} and
\ref{fig:limit95_DMonly_hadro-leptonic}, we also show the results for
the decay channels which are not suitable for the explanation of these
excesses. Those are for readers who are interested in the decay
channels in different context.
Figure~\ref{fig:limit95_DMonly_hadronic}, which shows the results for
hadronic channels, is for the same purpose. For hadronic channels the
impact of the IC process is small. This is expected since there are
fewer energetic electrons and positrons in the cascading products.

Two more remarks are in order. First, including the EBL photons has
little impact on the total IC results, since their energy density is
much smaller than that of the CMB photons.  This is expected from the
discussion given in Sec.\,\ref{sec:gamma_DM}. Therefore, we conclude
that our results presented in these figures are robust constraints on
decaying dark matter. Second, the obtained constraints are much
stronger than those in the previous studies, especially in the TeV
mass region for leptonic and hadroleptonic channels. In
ref.\,\cite{Ando:2015qda}, the gamma-ray background {\it spectrum} is
used to constrain the decaying dark matter. (Here the IC process is
considered.) Figure~6 in the reference, which is obtained by including
only dark matter contribution, shows that $\tau_{\rm dm}\lesssim
10^{26}\,{\rm s}$ is excluded for $m_{\rm dm}\gtrsim 1~{\rm
  TeV}$. Similar constraints are obtained by
ref.\,\cite{Regis:2015zka} using the {\it cross-correlation} technique
but the IC process is ignored.  Thus, these two facts show that
analysis of the angular cross-correlation of the gamma-ray background
by taking the IC effect into account is important for the
investigation on dark matter scenarios.

\begin{figure}
 \begin{center}
  \includegraphics[width=7cm]{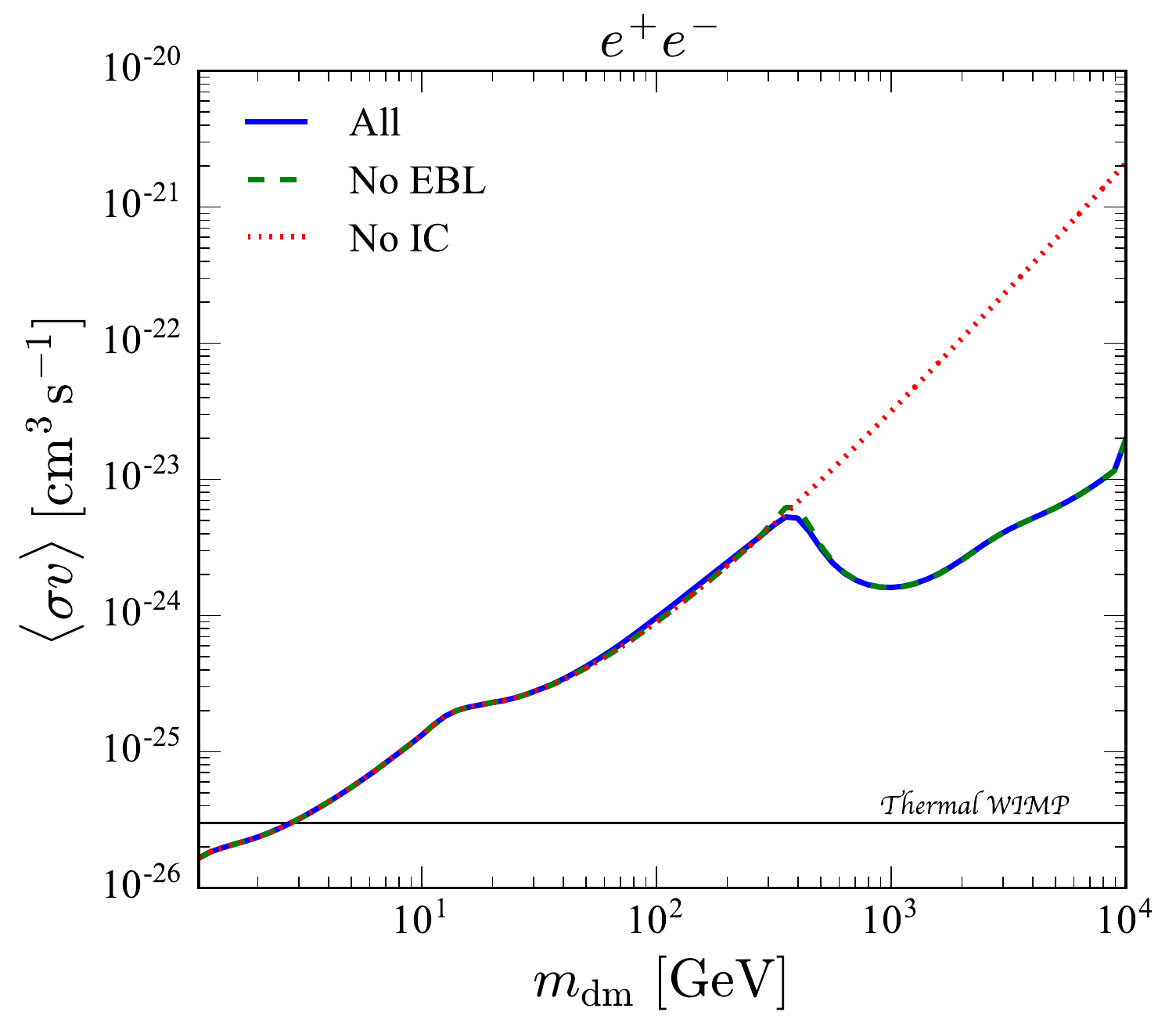}
  \includegraphics[width=7cm]{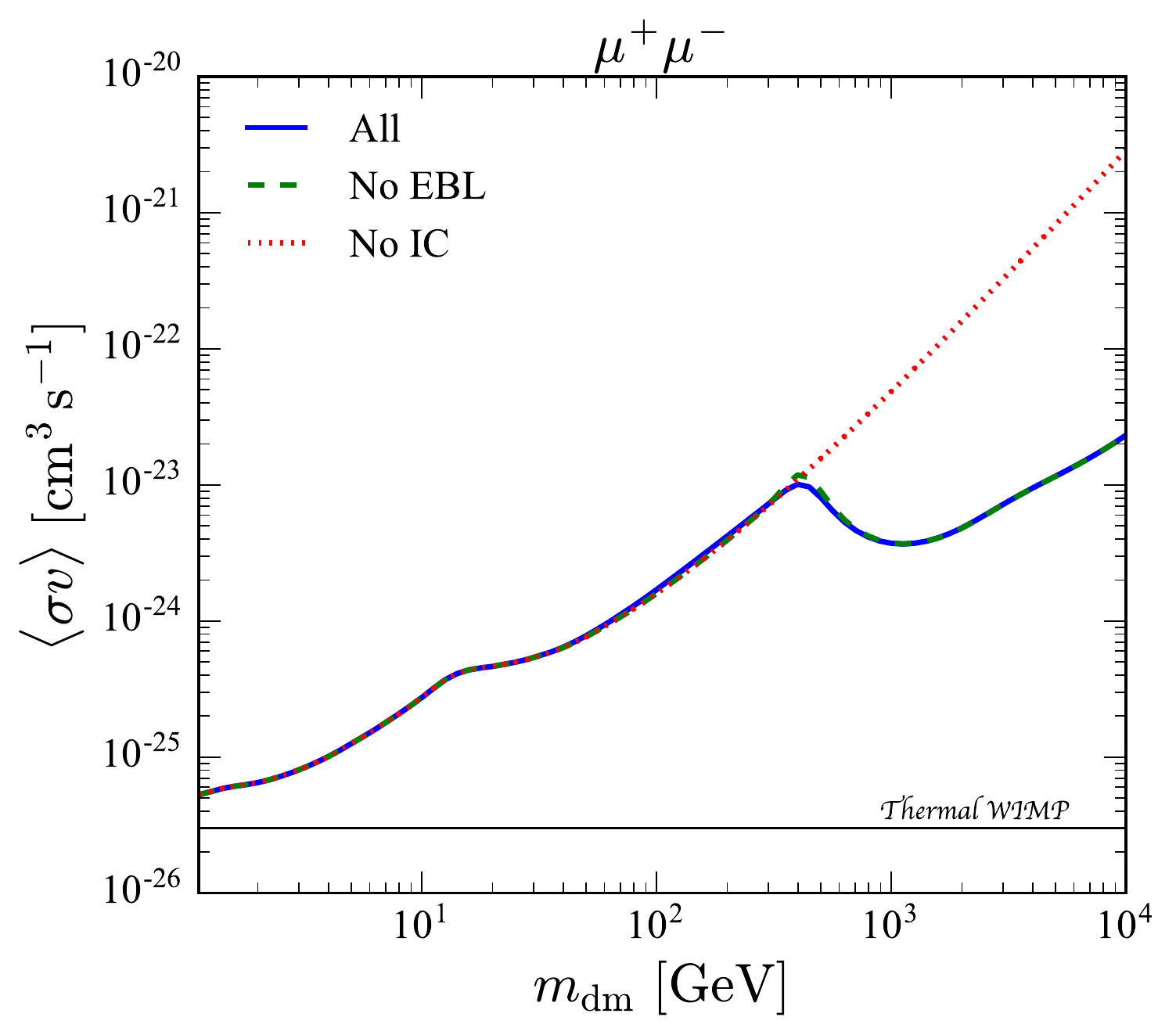}
  \includegraphics[width=7cm]{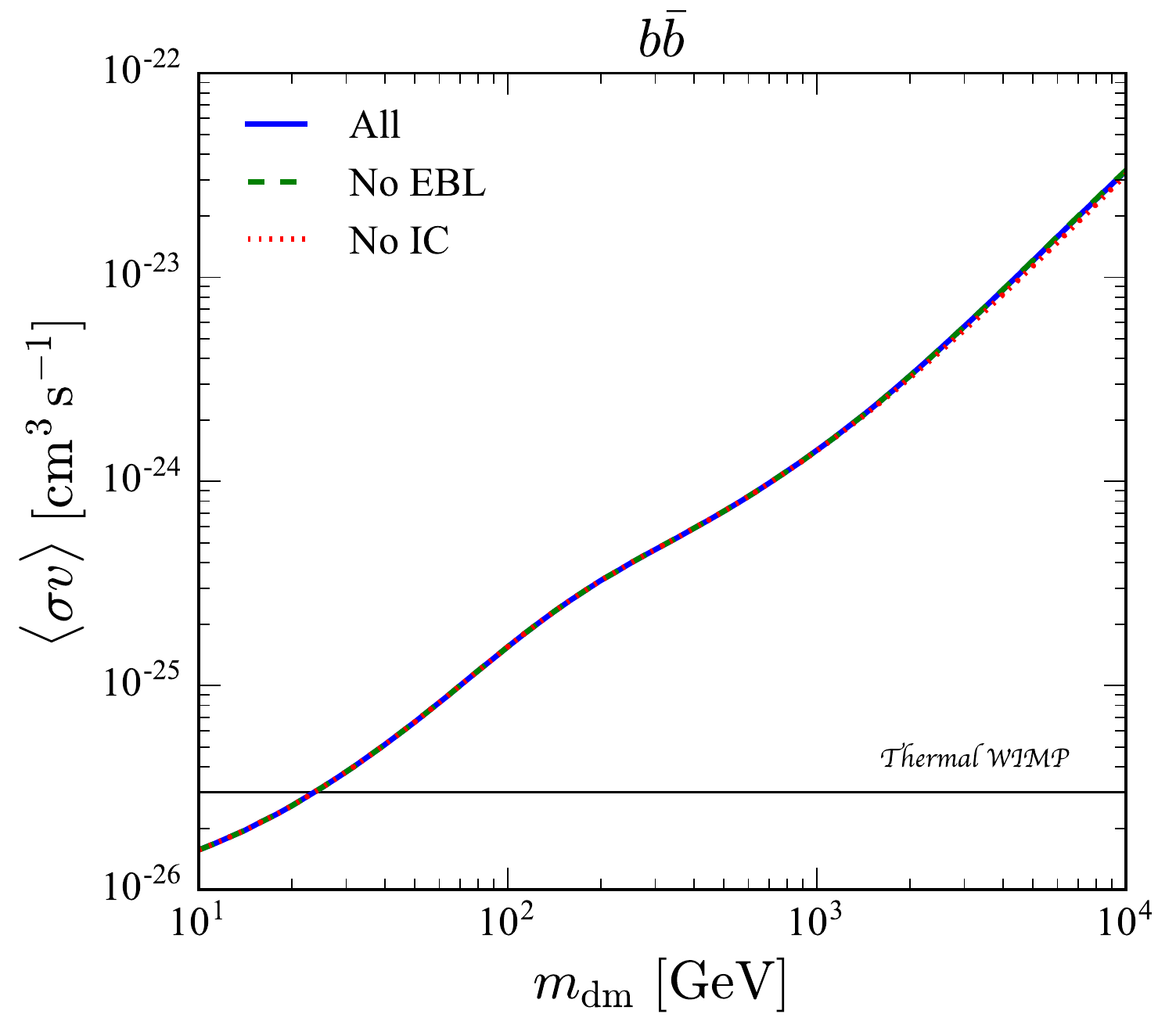}
  \includegraphics[width=7cm]{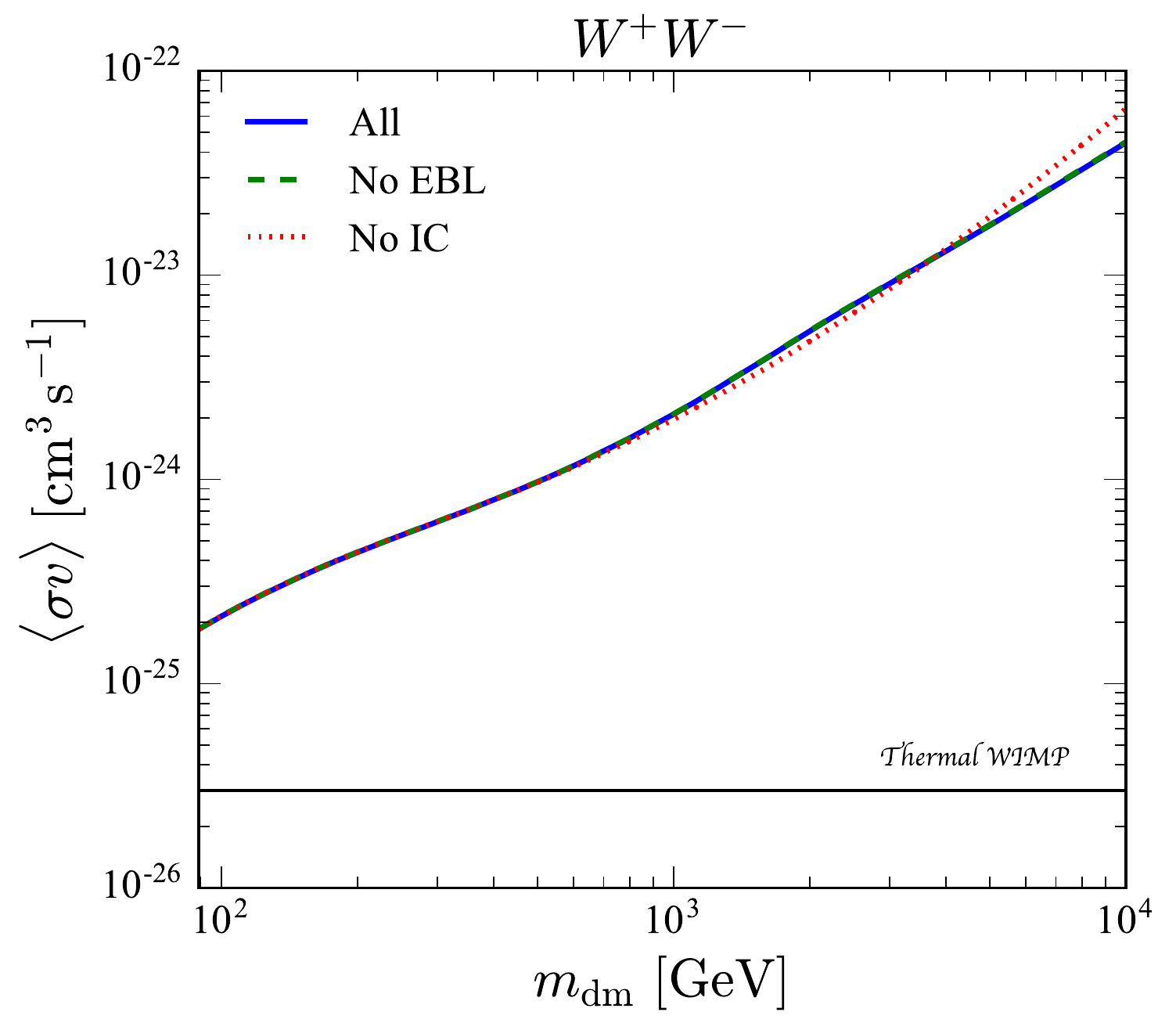}
  \includegraphics[width=7cm]{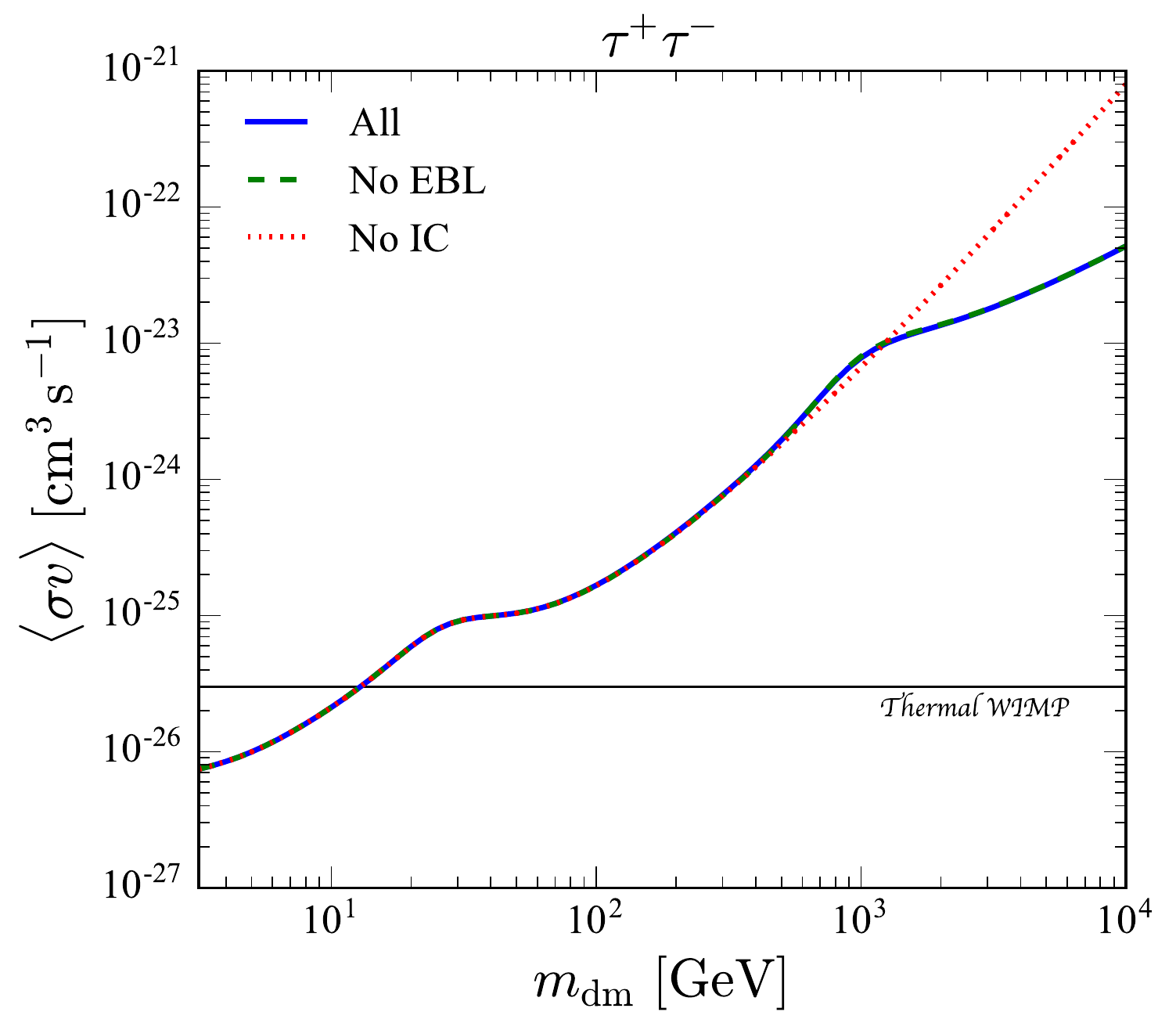}
  \caption{Upper limit on dark matter annihilation cross section
    $\langle \sigma v\rangle$ at 95\% confidence level due to cross
    correlation between the gamma-ray background and five galaxy
    catalogs. In this analysis, only dark matter component is
    included. IC scattering off both the EBL and CMB photons are
    included in solid, only CMB in dashed, and no IC effect is
    included in dotted.  `Thermal WIMP' line shows canonical value
    of the cross section which is suggested in thermal WIMP scenario.}
  \label{fig:limit95_DMonly_ann}
 \end{center}
\end{figure}

Finally we did the same analysis for annihilating scenarios, $e^+e^-$,
$\mu^+\mu^-$, $b\bar{b}$, $W^+W^-$, and $\tau^+\tau^-$.  Upper limits
on $\langle \sigma v \rangle$ at 95\% confidence level are shown in
Fig.\,\ref{fig:limit95_DMonly_ann}. Compared to the decaying case, the
computation of $P_{\delta^2\delta}(k,z)$ involves uncertainty.  In the
current analysis we largely followed ref.\,\cite{Ando:2013xwa}, except
for adopting the substructure boost factor given in ref.\,\cite{BA}.
This substructure model is based on the the latest development of the
observation and simulation by taking tidal stripping and dynamical
friction into account, and consequently it predicts substructure boost
in between the optimistic and conservative models in the previous
literature~\cite{Ando:2013xwa}.  Indeed, the obtained constraints
reflects this fact; `No IC' lines in
Fig.\,\ref{fig:limit95_DMonly_ann} shows a bit tighter constraints
compared to `annLOW' lines in Fig.\,3 of ref.\,\cite{Regis:2015zka},
but it is weaker compared to `annHIGH' line in the same figure.  The
effect of IC, on the other hand, is clearly seen in the annihilation
cases too.  Especially for $e^+e^-$, $\mu^+\mu^-$, the constraints get
stronger by one to two orders of magnitude in $m_{\rm dm}\gtrsim
1~{\rm TeV}$.

\subsection{Analysis including astrophysical sources and dark matter
  component}

Realistically, it is expected that the astrophysical sources such as
blazars and star-forming galaxies also contribute to the
measured cross correlations.
Indeed, both the shape and the amplitude of the measurements can be well
explained by these astrophysical sources~\cite{Xia:2015wka}.
Including them in the analysis will not only be realistic but also
tighten the constraints on the dark matter component significantly.

Here we adopt the Bayesian statistics, where the posterior distribution
of theoretical parameters $\bm\vartheta$ given data $\bm d$ is obtained
through the Bayes theorem:
\begin{equation}
 P(\bm\vartheta|\bm d) \propto P(\bm\vartheta) \mathcal L(\bm d|\bm\vartheta),
\end{equation}
where $\mathcal L(\bm d|\bm\vartheta)$ is the likelihood function and
$P(\bm\vartheta)$ is the prior distribution.
The likelihood function is characterized as Gaussian: $\mathcal L =
\exp(-\chi^2/2)$ with $\chi^2$ defined in eq.~(\ref{eq:chi2}).
This time, $C_{\rm th, \ell}^{\gamma {\rm}}$ in $\chi^2$ depends on all
the theoretical parameters $\bm\vartheta$.

We adopt $\bm\vartheta = (\tau_{\rm dm}, A_{\rm SFG}, A_{\rm blazar},
A_{\rm QSO}, A_{\rm 2MASS}, A_{\rm NVSS}, A_{\rm MG}, A_{\rm LRG})$ as
eight free parameters for each case with a fixed $m_{\rm dm}$ and
decay channel. ($\tau_{\rm dm}$ should be replaced with $\langle
\sigma v \rangle$ for the annihilation case.) Here $A_{\rm SFG}$ and
$A_{\rm blazar}$ are the amplitudes of the angular cross-power
spectrum $C_{\ell}^{\gamma {\rm g}}$ for these astrophysical sources,
and they are normalized to 1 for our model with the canonical choice
of relevant parameters as described in Sec.~\ref{sec:EGRB_astro}.  The
other parameters $\{A_{\rm QSO}, A_{\rm 2MASS}, A_{\rm NVSS}, A_{\rm
  MG}, A_{\rm LRG}\}$ are introduced to correct for shot-noise terms
that are expected for point-like astrophysical sources as well as
uncertain behaviors of $C_\ell^{\gamma {\rm g}}$ at small angular
scales (Sec.~\ref{sec:crosscorrelation}).  We choose flat priors in
logarithmic space for $\tau_{\rm dm}$ and $\langle \sigma v \rangle$,
and linear space for the rest, and the ranges are summarized in
Table~\ref{table:priors}.

\begin{table}
 \begin{center}
  \begin{tabular}{|c|c|} \hline
   $\log (\tau_{\rm dm}/{\rm s})$ & $(25,35)$\\
   $\log (\langle\sigma v \rangle/\mathrm{cm^3\,s^{-1}})$ &
       $(-30,-20)$\\
   $A_{\rm SFG}$ & $(0,3)$\\
   $A_{\rm blazar}$ & $(0,3)$\\
   $A_{\rm QSO}$ & $(-0.5,0.5)\times 10^{-12}$~cm$^{-2}$~s$^{-1}$\\
   $A_{\rm 2MASS}$ & $(-0.5,0.5)\times 10^{-12}$~cm$^{-2}$~s$^{-1}$\\
   $A_{\rm NVSS}$ & $(-0.5,0.5)\times 10^{-11}$~cm$^{-2}$~s$^{-1}$\\
   $A_{\rm MG}$ & $(-0.5,0.5)\times 10^{-12}$~cm$^{-2}$~s$^{-1}$\\
   $A_{\rm LRG}$ & $(-0.5,0.5)\times 10^{-12}$~cm$^{-2}$~s$^{-1}$\\
   \hline
  \end{tabular}
  \caption{Ranges of flat priors for the parameters studied in MCMC.}
  \label{table:priors}
 \end{center}
\end{table}

\begin{figure}
 \begin{center}
  \includegraphics[width=7cm]{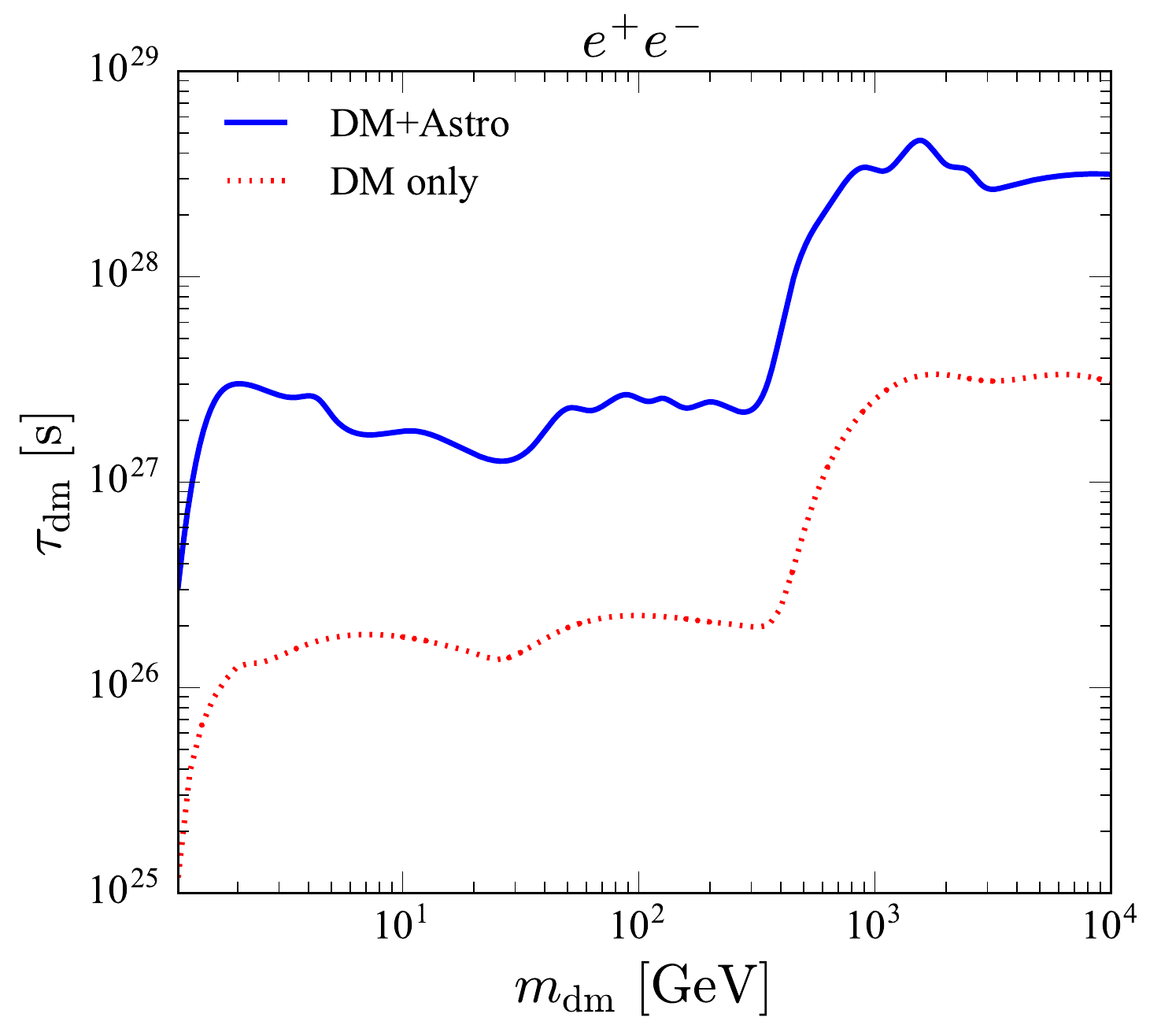}
  \includegraphics[width=7cm]{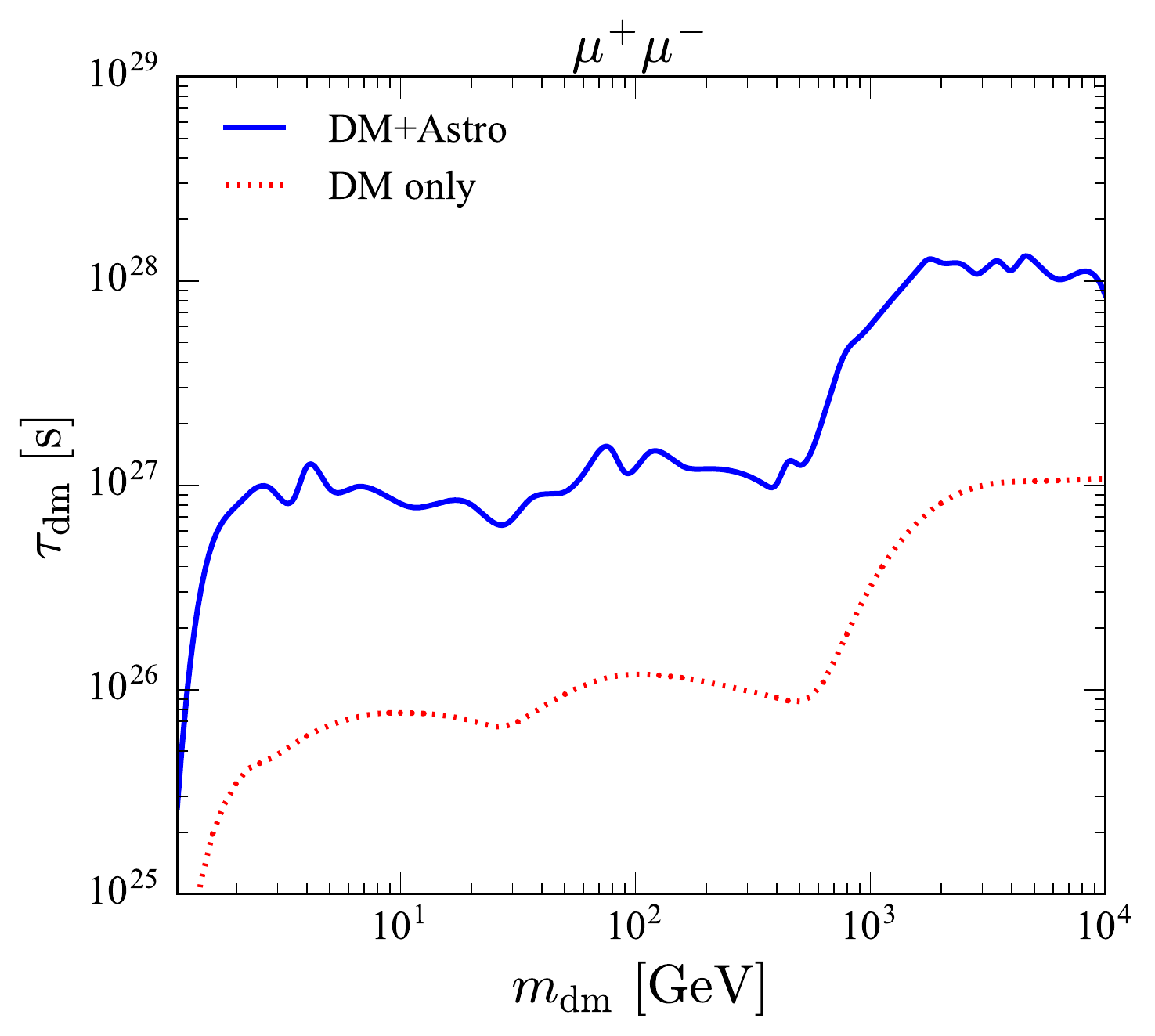}
  \includegraphics[width=7cm]{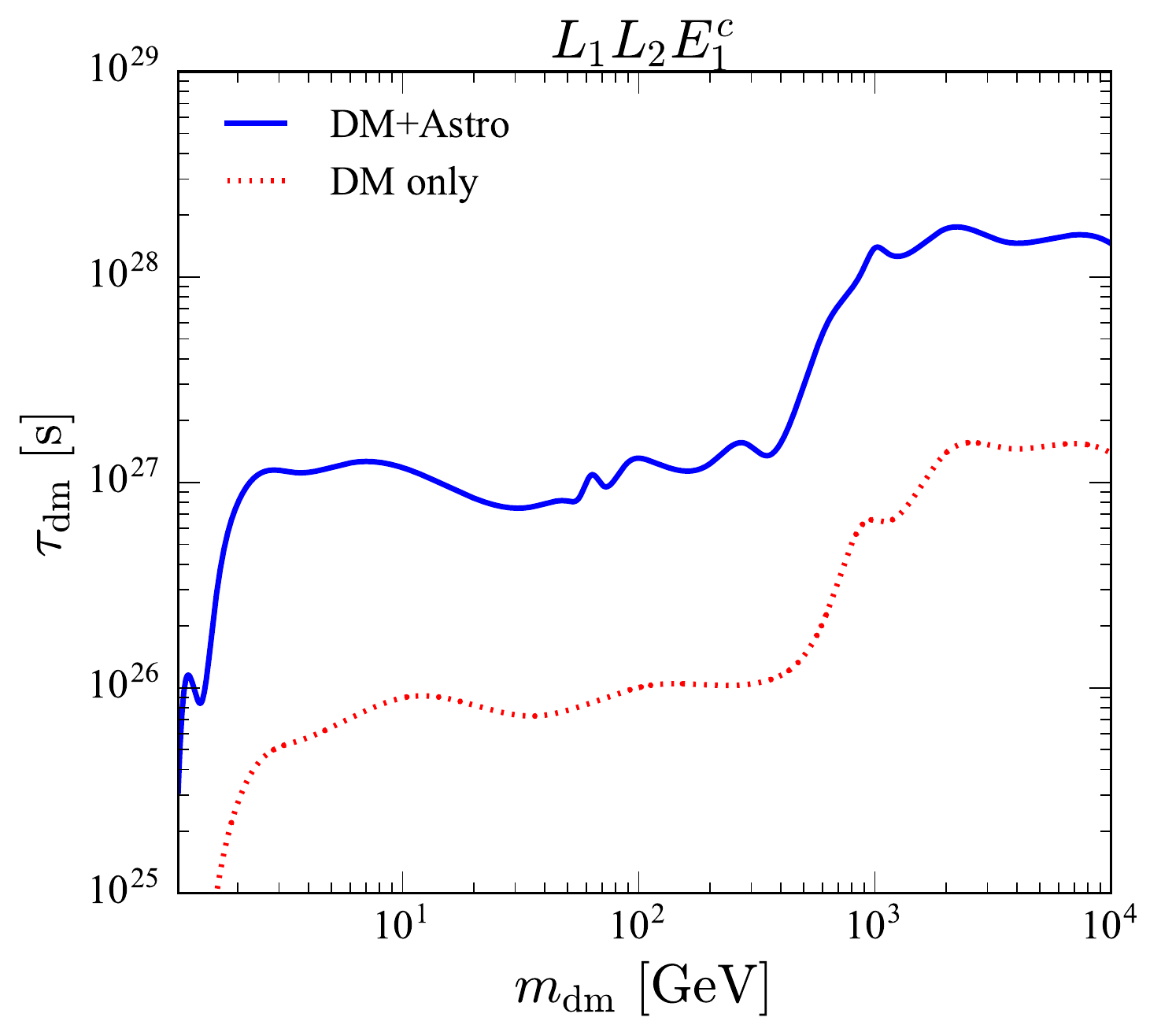}
  \includegraphics[width=7cm]{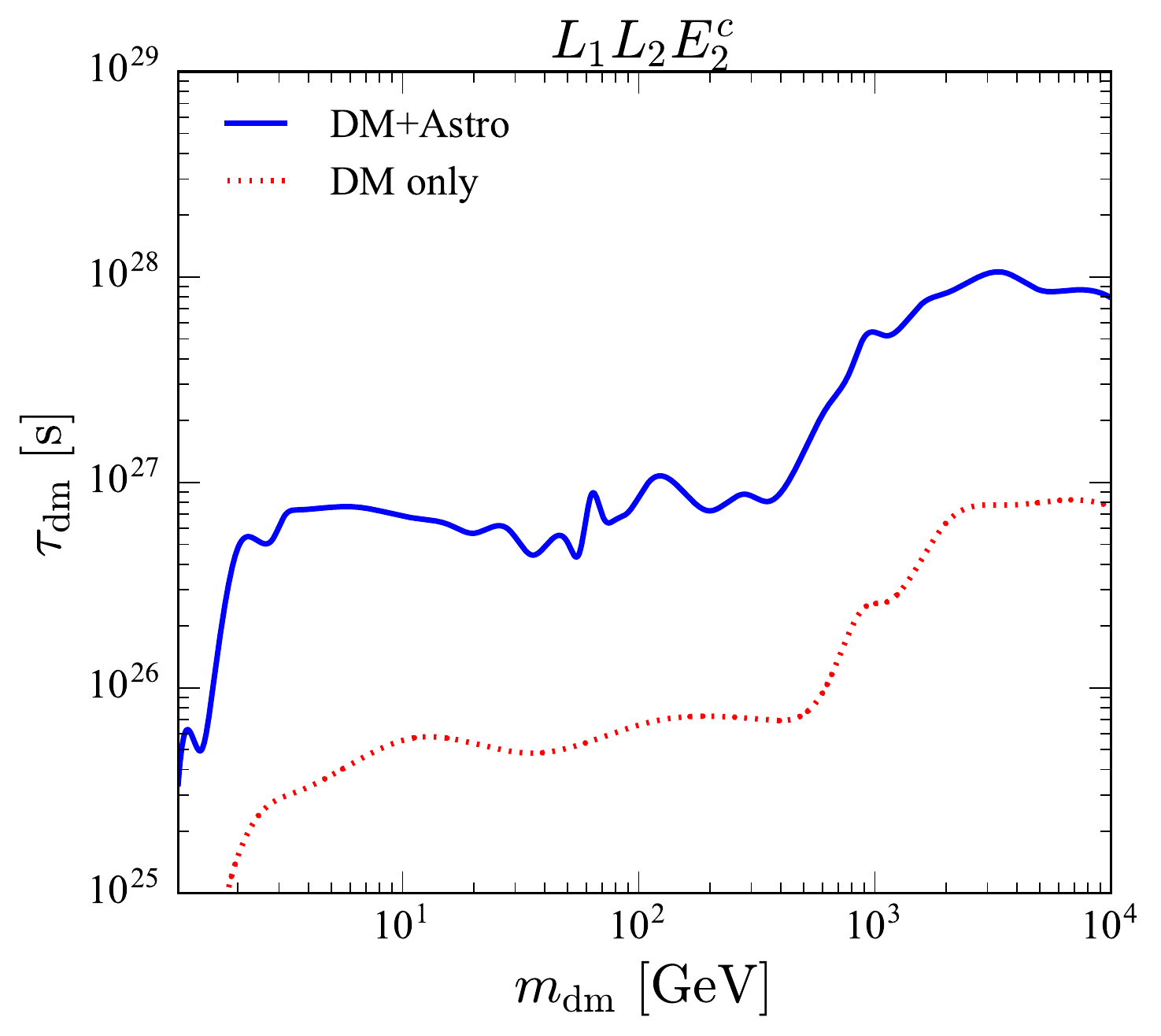}
  \caption{Lower limits on dark matter lifetime for leptonic channels
    due to cross correlation between the EGRB and five galaxy
    catalogs. The solid curves are the limits as the result of
    analysis that takes into account astrophysical contributions,
    while the dotted curves are the dark matter only analysis as shown
    in
    Figs.~\ref{fig:limit95_DMonly_leptonic}--\ref{fig:limit95_DMonly_hadronic}.
    For example, for $\mu^+\mu^-$ channel, the obtained results are
    better than the latest limit~\cite{Baring:2015sza} given from
    dwarf galaxies by a few factors.}
  \label{fig:limit95_leptonic}
 \end{center}
\end{figure}

\begin{figure}
 \begin{center}
  \includegraphics[width=7cm]{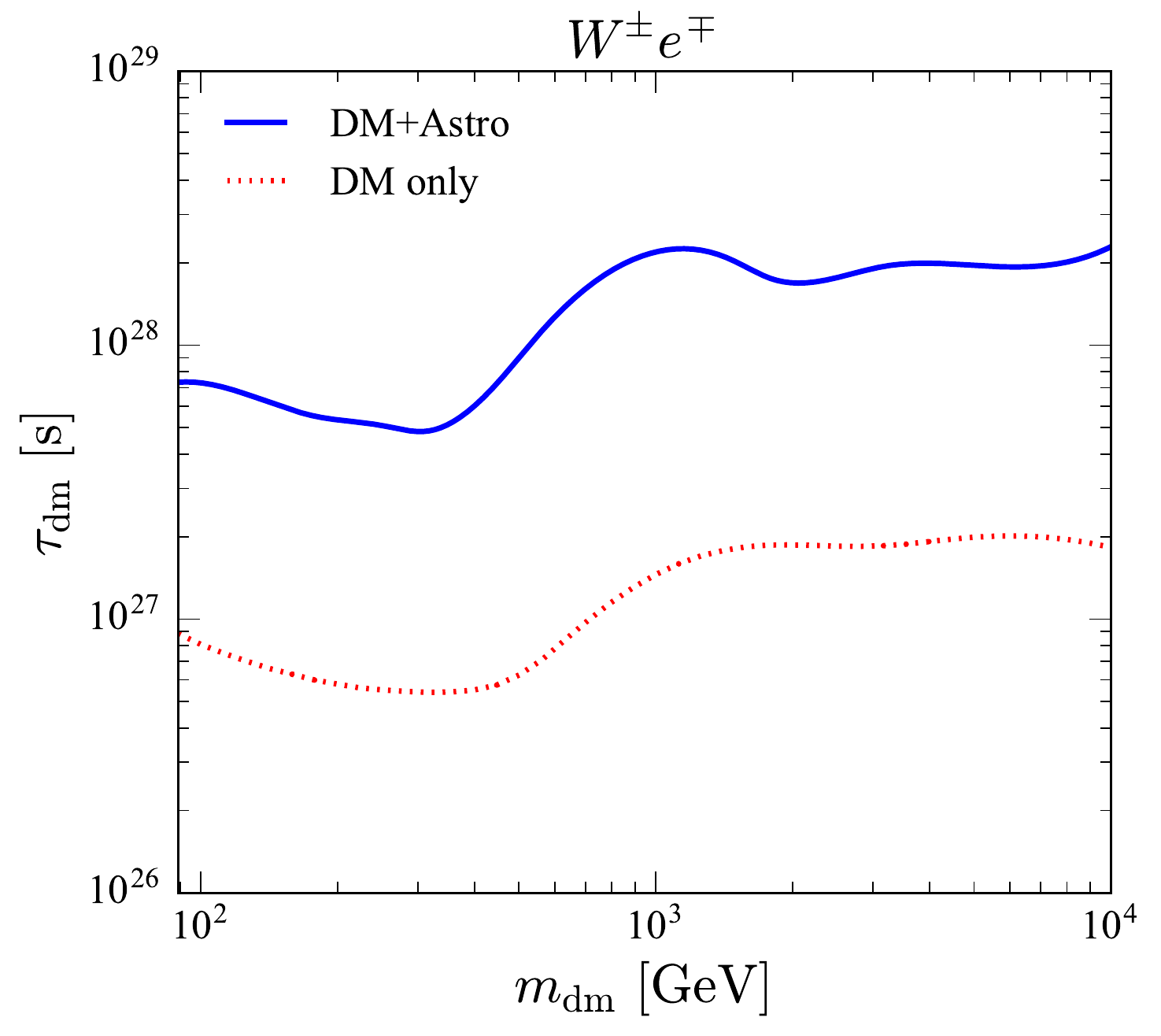}
  \includegraphics[width=7cm]{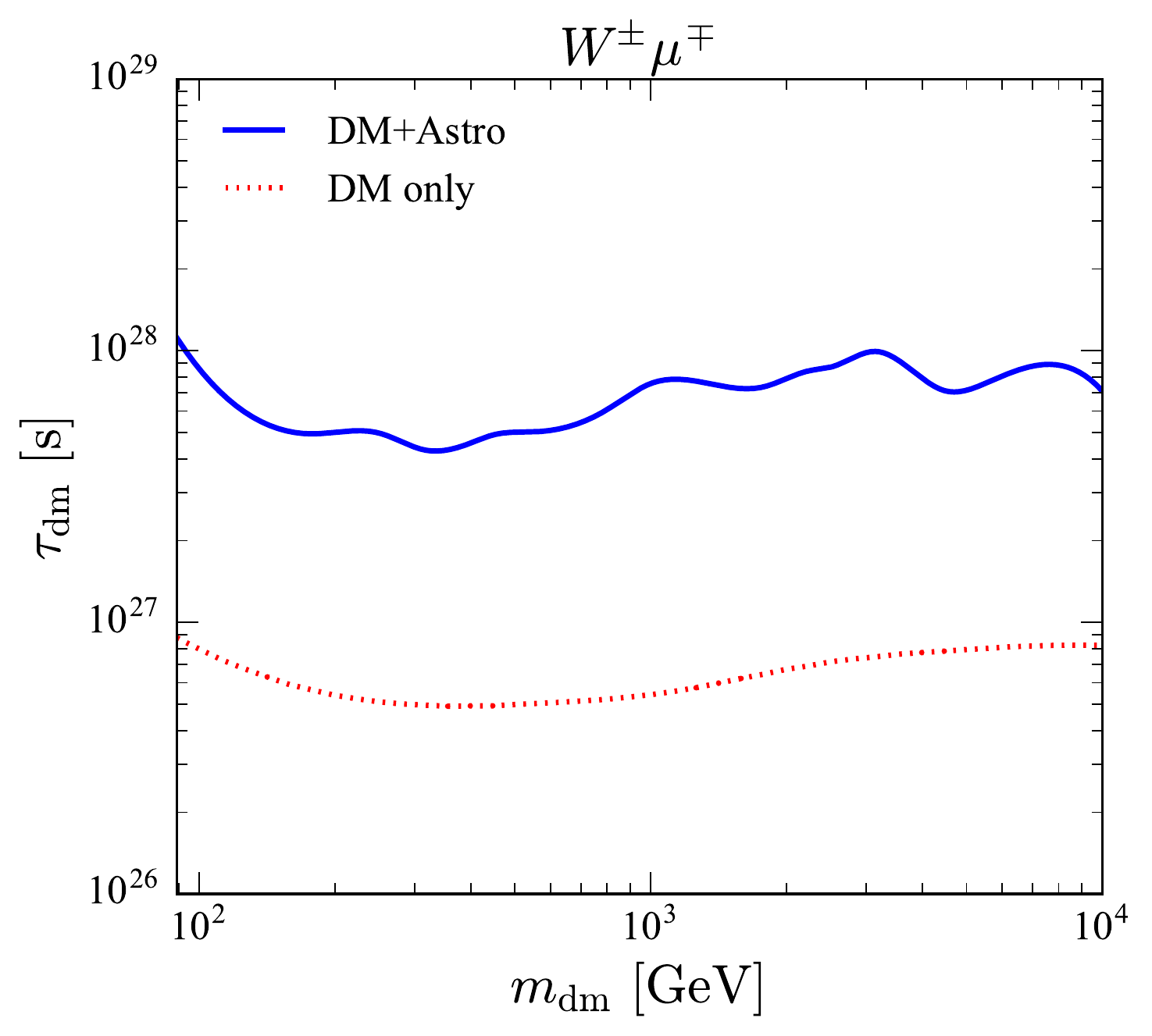}
  \includegraphics[width=7cm]{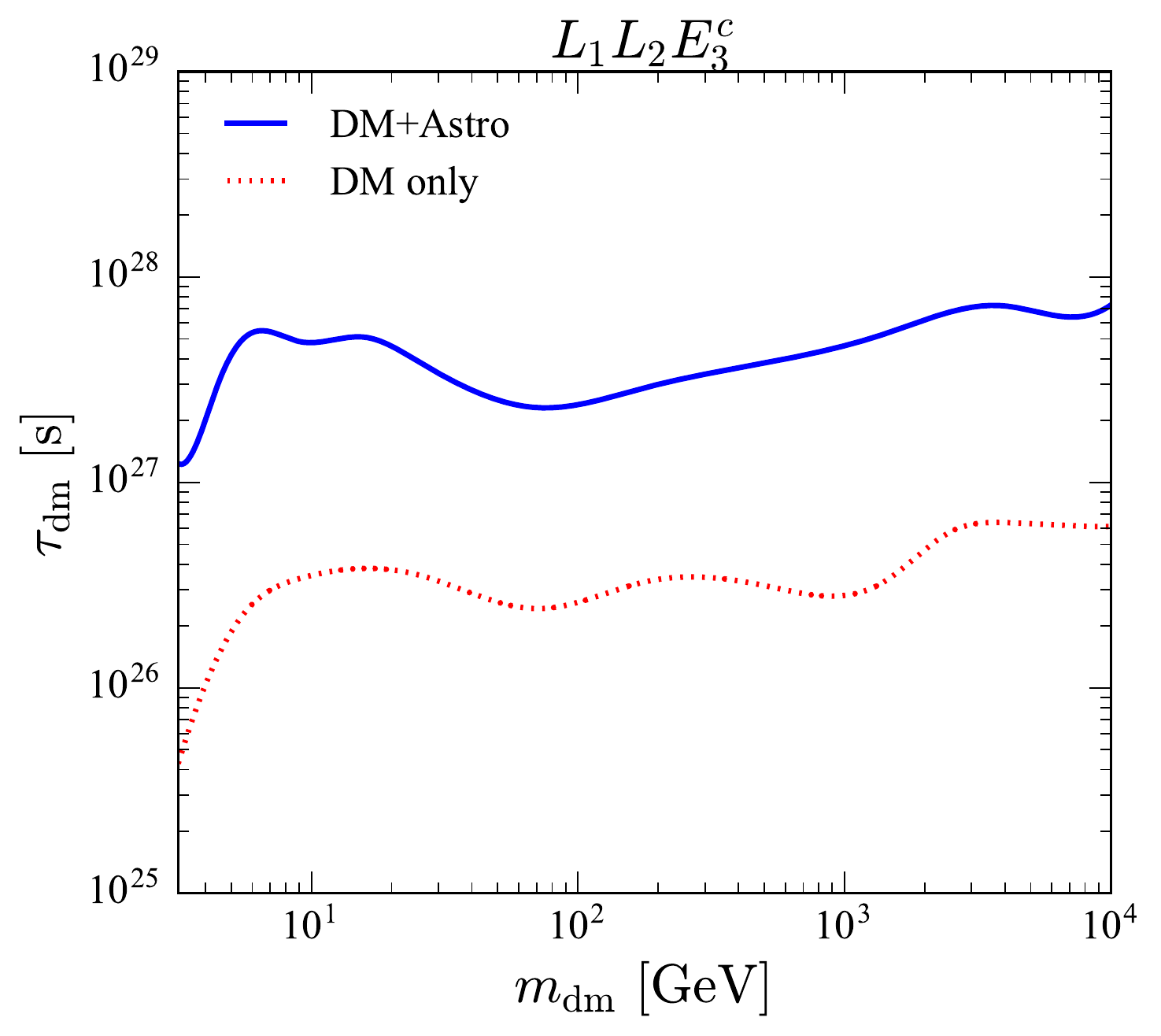}
  \includegraphics[width=7cm]{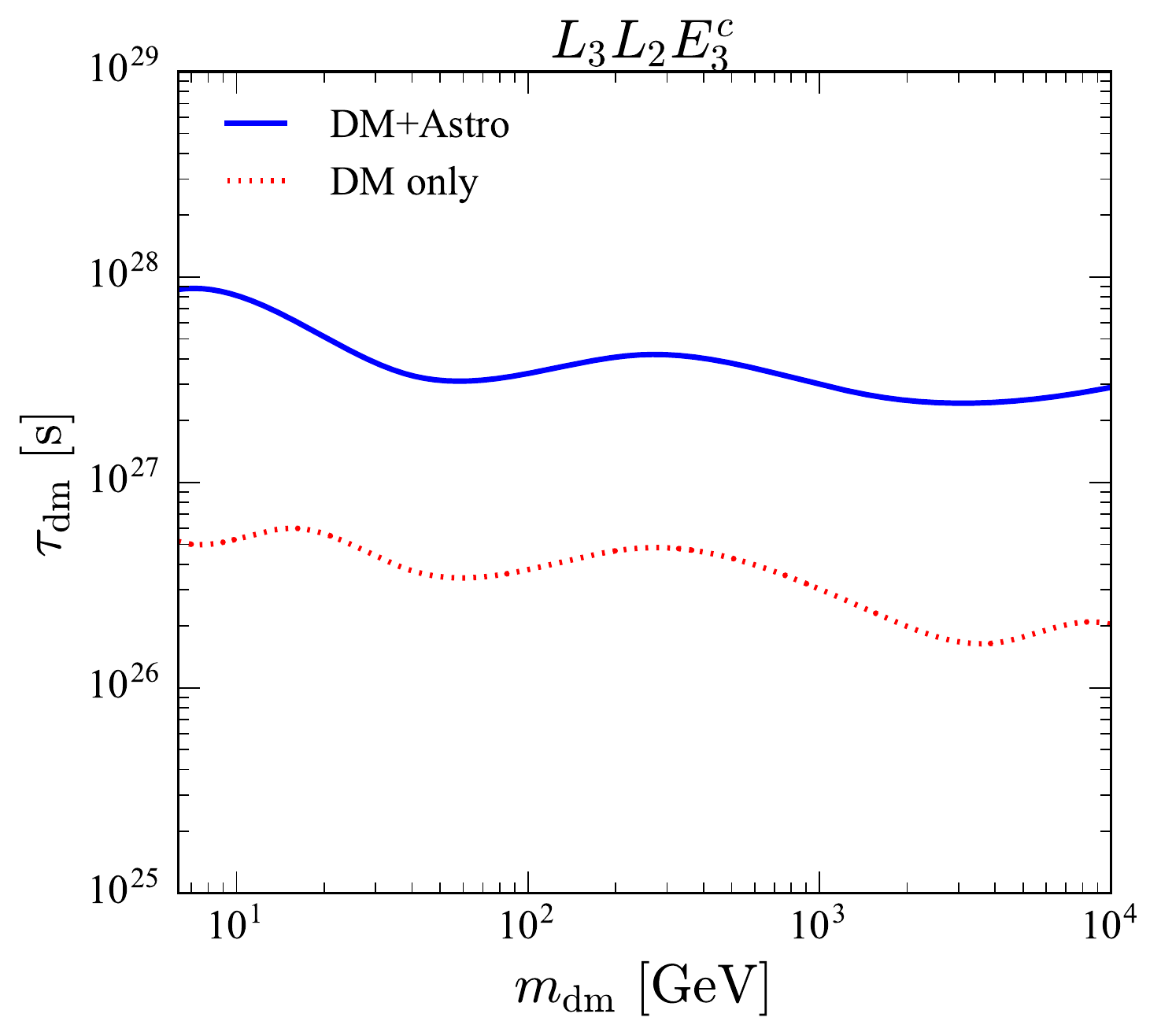}
  \caption{The same as Fig.~\ref{fig:limit95_leptonic} but for
  the hadroleptonic channels.}
  \label{fig:limit95_hadroleptonic}
 \end{center}
\end{figure}

\begin{figure}
 \begin{center}
  \includegraphics[width=7cm]{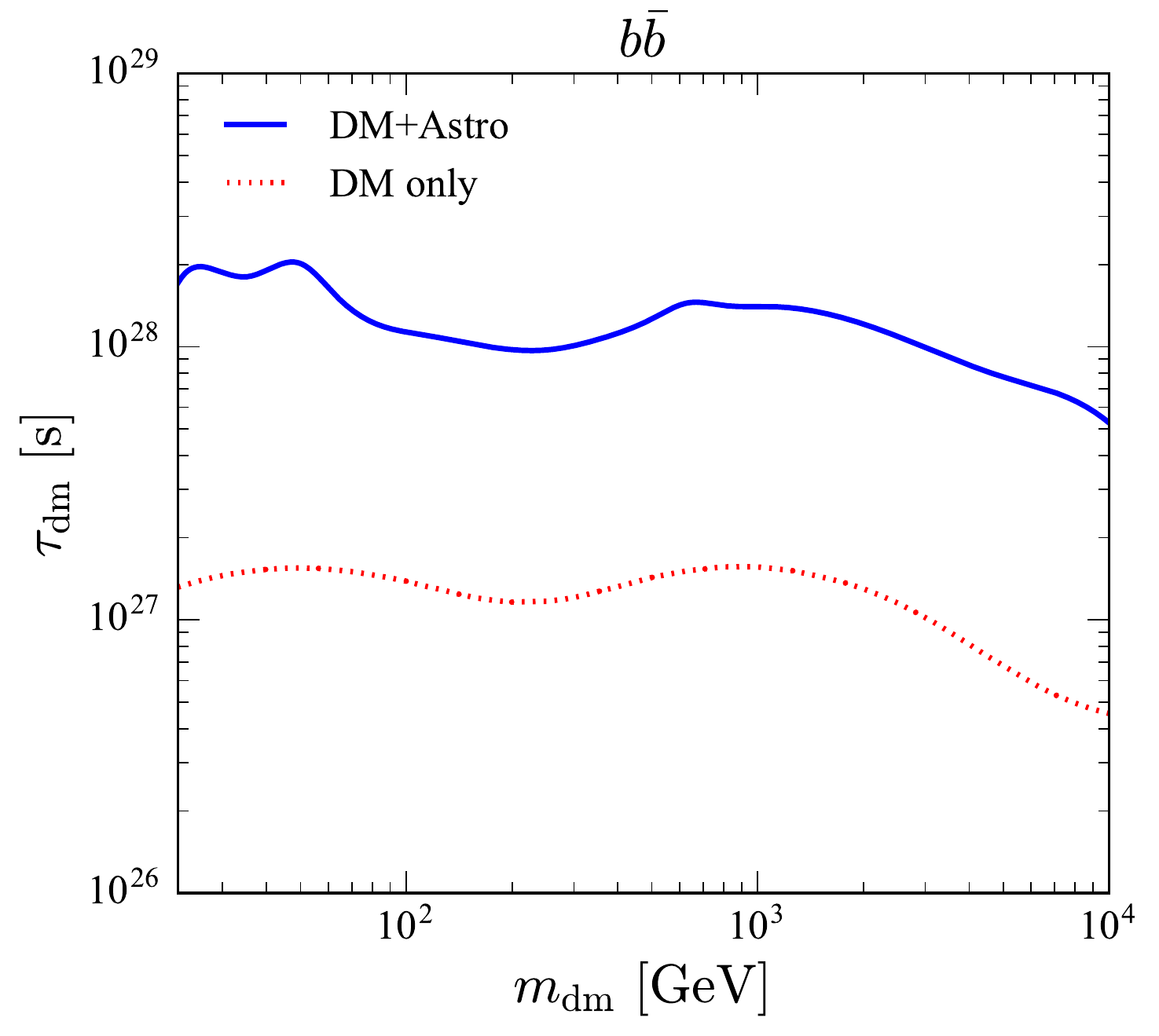}
  \includegraphics[width=7cm]{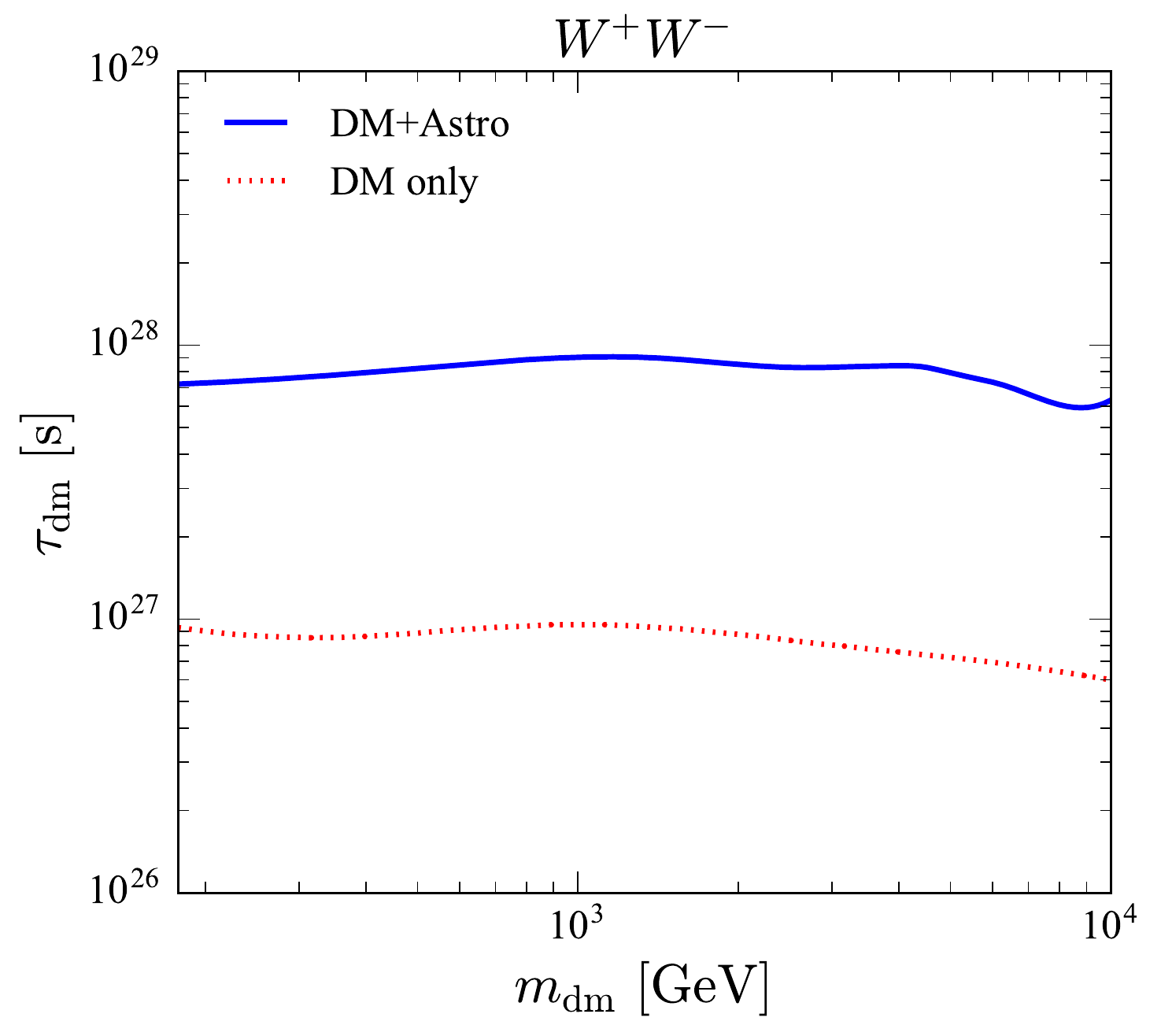}
  \includegraphics[width=7cm]{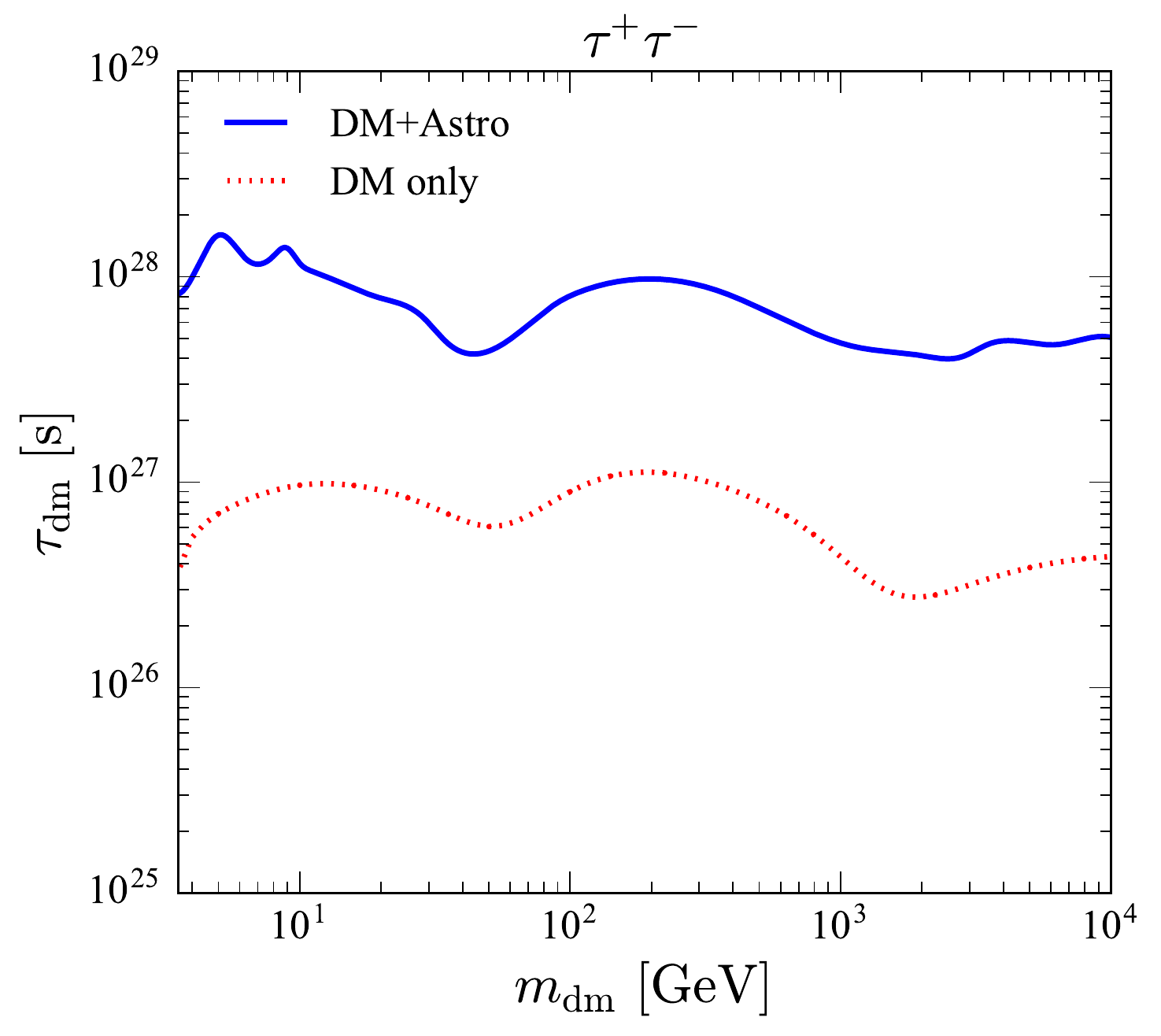}
  \includegraphics[width=7cm]{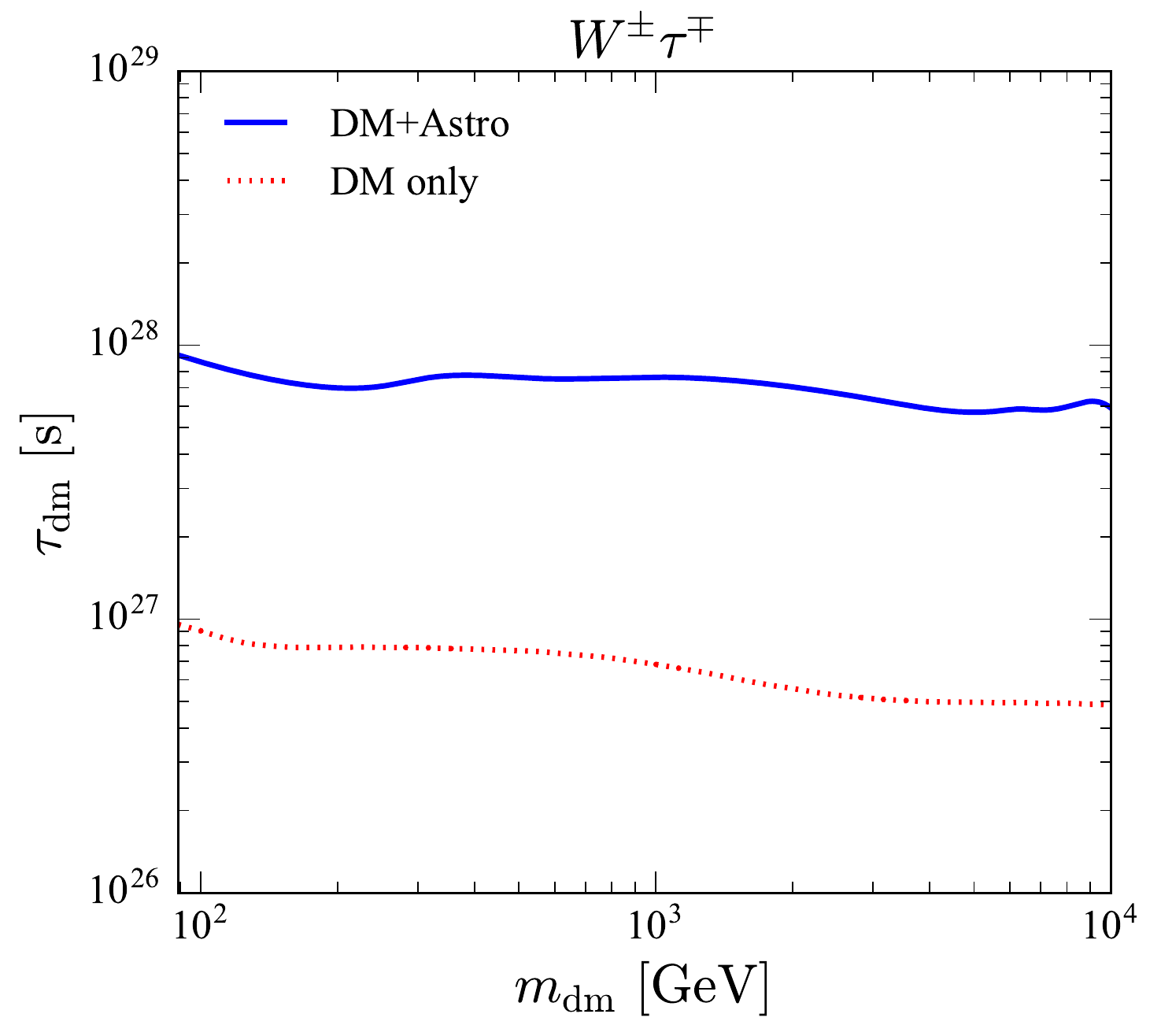}
  \caption{The same as Fig.~\ref{fig:limit95_leptonic} but for
    the hadronic channels.}
  \label{fig:limit95_hadronic}
 \end{center}
\end{figure}

With these parameters and priors, we perform the Markov-Chain Monte
Carlo (MCMC) simulations using the {\tt MultiNest}
package~\cite{Feroz:2008xx, Feroz:2007kg, Feroz:2013hea}, and show the
results in
Figs.\,\ref{fig:limit95_leptonic}--\ref{fig:limit95_hadronic}.  They
corresponds to the results shown in
Figs.\,\ref{fig:limit95_DMonly_leptonic}--\ref{fig:limit95_DMonly_hadronic},
respectively, but including the contribution from the astrophysical
sources.  In all the cases, the constraints get stronger by an order
of magnitude compared to
Figs.\,\ref{fig:limit95_DMonly_leptonic}--\ref{fig:limit95_DMonly_hadronic}.
The results for $b \bar{b}$ and $W^+W^-$ well agree with
ref.\,\cite{Cuoco:2015rfa}. The constraints on $\mu^+\mu^-$ and
$\tau^+\tau^-$ are also consistent for $m_{\rm dm}\lesssim 1~{\rm
  TeV}$. However, more stringent constraints have been obtained in
$m_{\rm dm}\gtrsim 1~{\rm TeV}$ due to the IC process.  For example,
leptonic channels are constrained to $\tau_{\rm dm}\lesssim
10^{28}\,{\rm sec}$ for $m_{\rm dm}\gtrsim 1~{\rm TeV}$.
Consequently, the parameter regions to explain the positron excess in
$L_iL_jE^c_k$ scenario or anti-proton excess in $W^{\pm}\mu^{\mp}$
final state are excluded. For anti-proton, there is uncertainty in the
computation of the cosmic-ray anti-proton as mentioned in the previous
subsection.  Our present result for $W^{\pm}\mu^{\mp}$ final state
excludes possibility to explain the anti-proton excess even in MAX or
MED models.\footnote{It might be possible to find loophole for the
  constraint, {\it e.g.}, considering unconventional propagation model
  or spectrum for astrophysical proton source. Our discussion is based
  on conventional cases.} As stressed before, including the impact of
the IC gamma rays is crucial to get this conclusion.

\begin{figure}
 \begin{center}
  \includegraphics[width=7cm]{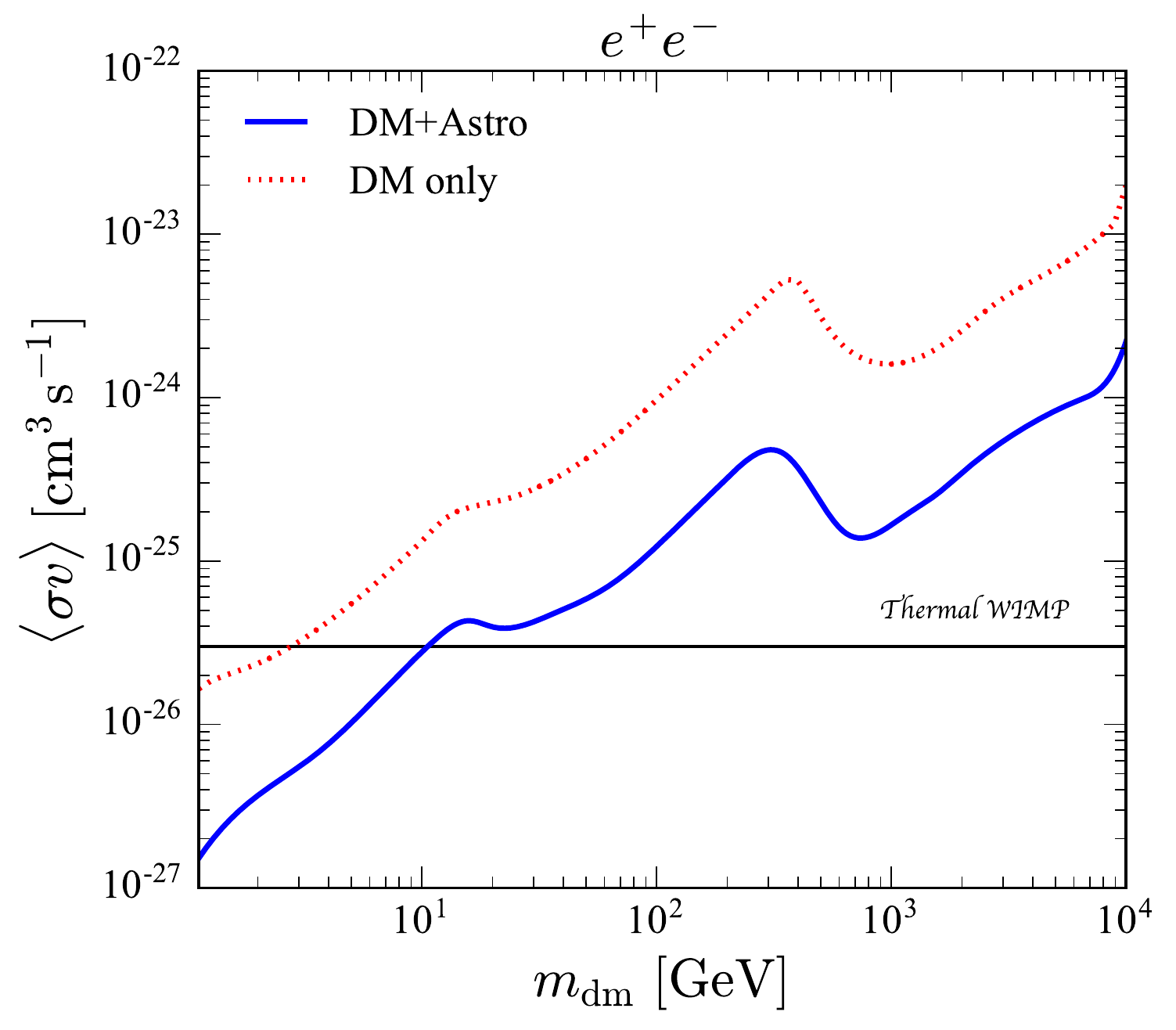}
  \includegraphics[width=7cm]{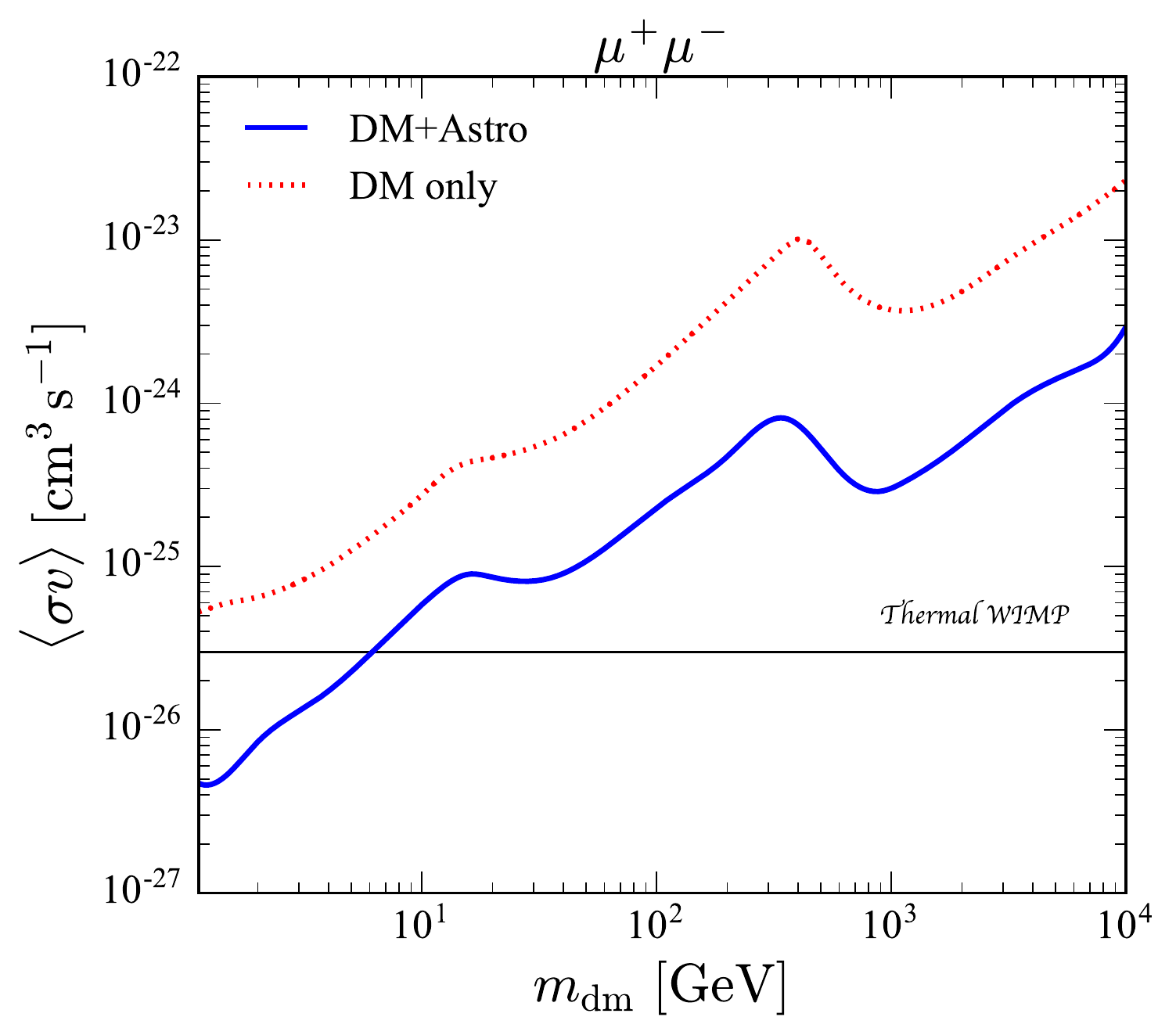}
  \includegraphics[width=7cm]{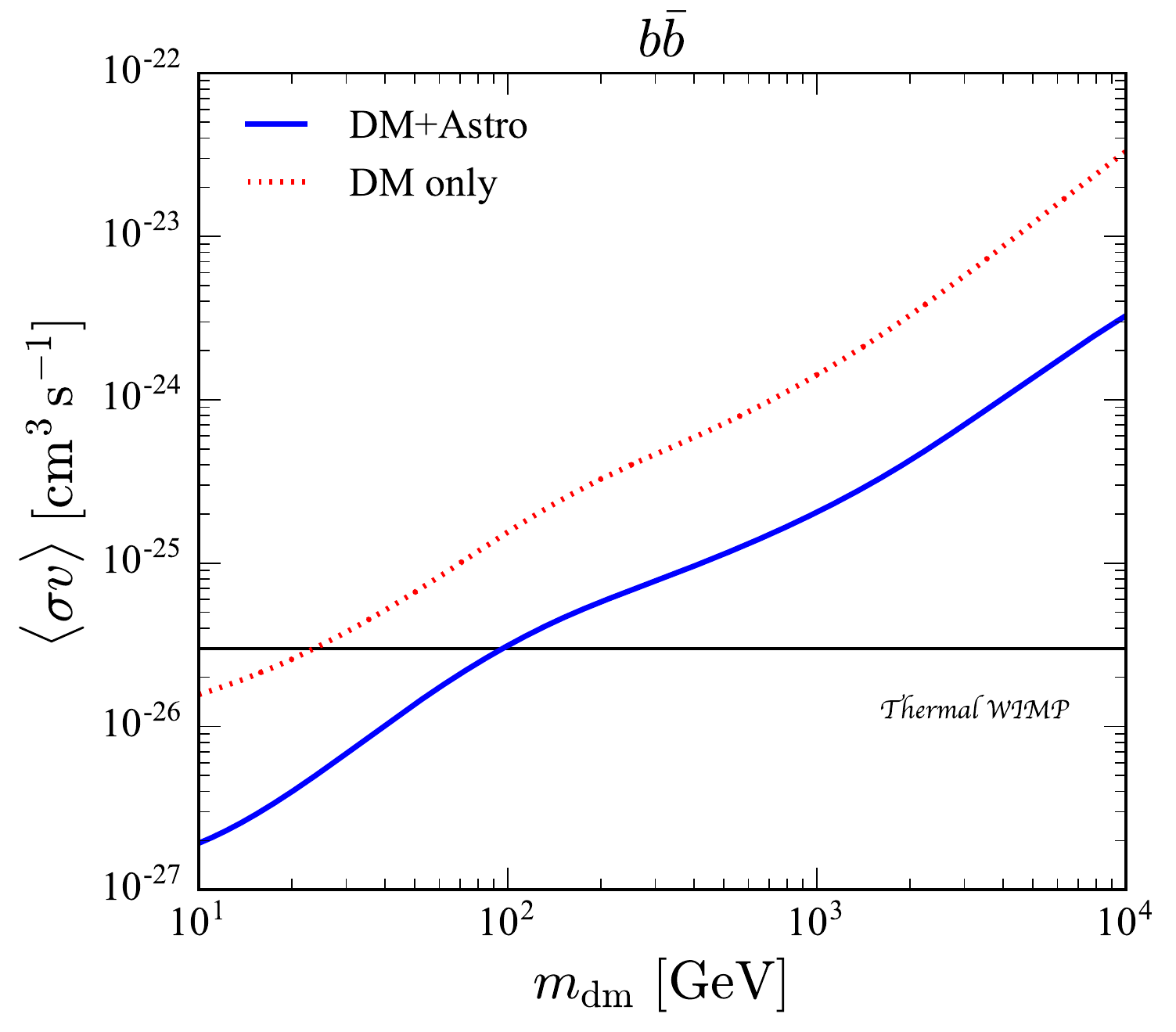}
  \includegraphics[width=7cm]{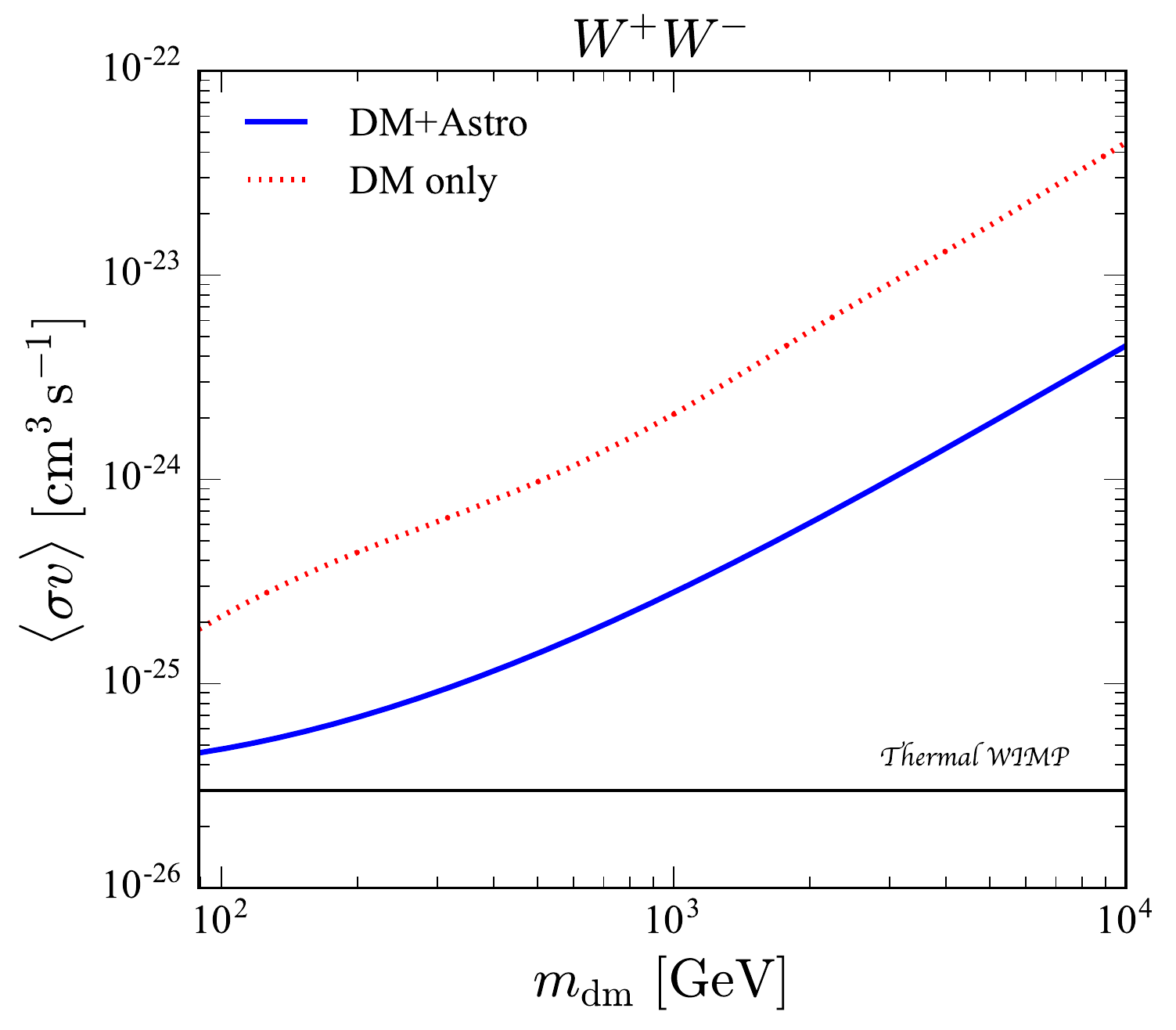}
  \includegraphics[width=7cm]{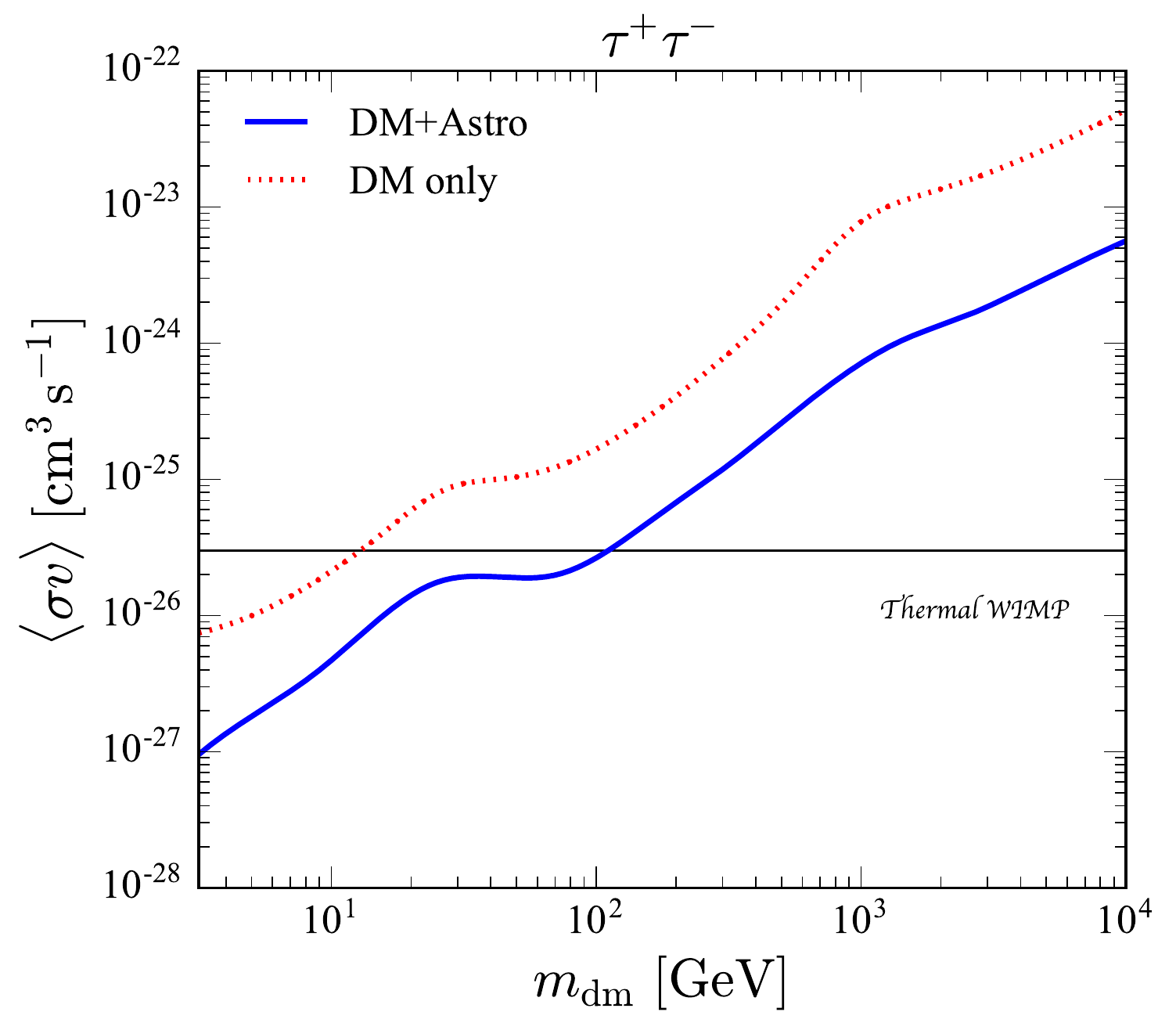}
  \caption{Upper limit on dark matter annihilation cross section due
    to cross correlation between the gamma-ray background and five
    galaxy catalogs. Line contents are the same as the decaying cases,
    {\it i.e.},
    Figs.\,\ref{fig:limit95_leptonic}--\ref{fig:limit95_hadronic}.
    Additionally, `Thermal WIMP' line is included as in
    Fig.\,\ref{fig:limit95_DMonly_ann}.}
  \label{fig:limit95_ann}
 \end{center}
\end{figure}

Finally we give results for annihilation
cases. Figure\,\ref{fig:limit95_ann} shows the same result as
Fig.\,\ref{fig:limit95_DMonly_ann} but taking astrophysical components
into account. Similar to decaying cases given in
Figs.\,\ref{fig:limit95_leptonic}--\ref{fig:limit95_hadronic}, the
constraints become more stringent by an order of magnitude compared to
those by taking only dark matter component. For a reference, a line
$\langle \sigma v \rangle =3\times 10^{-26}\,{\rm cm}^3\,{\rm
  s}^{-1}$, which is required value for the conventional thermal WIMP
production scenario, is also shown in the plot. It is found that the
mass region $m_{\rm dm}\lesssim 100~{\rm GeV}$ is excluded for the
thermal WIMP mainly annihilating to $b {\bar b}$ or
$\tau^+\tau^-$. Namely the cross-correlating analysis is beginning to
investigate the weak-scale mass region of the WIMP dark matter.

\section{Conclusions}
\label{sec:conclusion}
\setcounter{equation}{0}

Indirect detection of dark matter via cosmic rays is a promising way
for the probe of the nature of dark matter. Anomalous fluxes in
cosmic-ray positron and anti-proton recently reported by AMS-02
collaboration indicate the signal of decaying or annihilating dark
matter with a mass of 1--10 TeV. On the other hand, there is an
alternative solution for the anomalies, such as nearby pulsars for the
positron flux or known inner-galactic activities for anti-proton.
Extragalactic gamma-ray background has a potential to clarify the
situation.  In the dark matter scenarios suggested to explain the
anti-proton excess, lots of gamma rays are produced due to the cascade
decay of the final state products. Even in the leptonic final states
motivated by the positron excess, gamma rays are produced via final
state radiation or inverse-Compton process. On observational side,
gamma-ray data has been accumulated, which enables us to study their
spatial distribution with better accuracy. In addition, the
astrophysical sources for extragalactic gamma rays, such as blazars,
star-forming galaxies, etc., have been identified due to the recent
development of the observations.

In the article we compute the cross correlation of gamma rays from
dark matter with local galaxy catalogs for the study of dark
matter. In the previous studies which use the same technique, gamma
rays from dark matter in inverse-Compton scattering were neglected,
which we have included in the present work. We have found that the
inverse-Compton scattering due to the CMB photon is crucial to
constrain TeV scale dark matter, and consequently most stringent
constrains has been obtained, especially on leptonic channels of the
decay or annihilation of dark matter.  We have done two types of
analysis; computing the angular cross correlation of the gamma rays
from i) only dark matter or from ii) both dark matter and
astrophysical sources. To do comprehensive analysis, we have
considered variety of final states, $l^+_il^-_i$, $\nu
l^{\pm}_jl^{\mp}_k$, $W^{\pm}l^{\mp}_i$, $b\bar{b}$, $W^+W^-$ for dark
matter contribution. On the leptonic channels of decaying scenarios,
lifetime of dark matter is constrained as $\tau_{\rm dm}\gtrsim
10^{27}\,{\rm s}$ and $\tau_{\rm dm}\gtrsim 10^{28}\,{\rm s}$ in
analysis i) and ii), respectively, for $m_{\rm dm}\gtrsim 1~{\rm
  TeV}$. Thus, the decaying scenarios which are suggested to explain
the positron or anti-proton excesses are excluded. This conclusion is
robust since there is no uncertainty in the calculation of gamma rays
from dark matter once decay channel is specified and that the we have
based our discussion on the data of the observed astrophysical
sources. We have checked consistency with the previous
papers~\cite{Regis:2015zka,Cuoco:2015rfa}, {\it i.e.}, the similar
results are obtained for hadronic channels ($b\bar{b}$, $W^+W^-$,
$\tau^+\tau^-$) and on leptonic channels ($\mu^+\mu^-$) in $m_{\rm
  dm}\lesssim 1~{\rm TeV}$ region. For annihilating scenarios,
similarly, the constraints get one to two orders of magnitude more
stringent than the previous ones~\cite{Regis:2015zka,Cuoco:2015rfa} in
TeV mass region. Besides, we have found that $m_{\rm dm} \lesssim
100~{\rm GeV}$ is excluded for a WIMP dark matter which mainly
annihilates into $b\bar{b}$ or $\tau^+\tau^-$, which is comparable
constraint to that obtained by using dwarf spheroidal
galaxies~\cite{Ackermann:2015zua}.  Those results are obtained by
adopting the latest updates in the subhalo model~\cite{BA}, where the
effects of tidal stripping are better treated compared to the past
studies.  Note, however, that the estimates are still subject to other
uncertainties intrinsic to dark matter clustering down to extremely
small scales (on the order of Earth mass), effects of baryons, etc.,
although future studies will address these issues. There is also a
huge uncertainty in the computation of (anti-)proton flux in the
galaxy. Thus, the constraint obtained here cannot exclude, for
example, annihilating scenario to $W^+W^-$ to explain the anti-proton
excess. However, it is expected that more cosmic ray data by AMS-02
will reduce the uncertainty in the cosmic-ray propagation model, which
will make it possible to test dark matter hypothesis for the origin of
the anomalous anti-proton flux.

\appendix
\section{Energy loss rate due to inverse-Compton scattering}
\label{sec:bic}
\setcounter{equation}{0}

\begin{figure}
 \begin{center}
  \includegraphics[width=7cm]{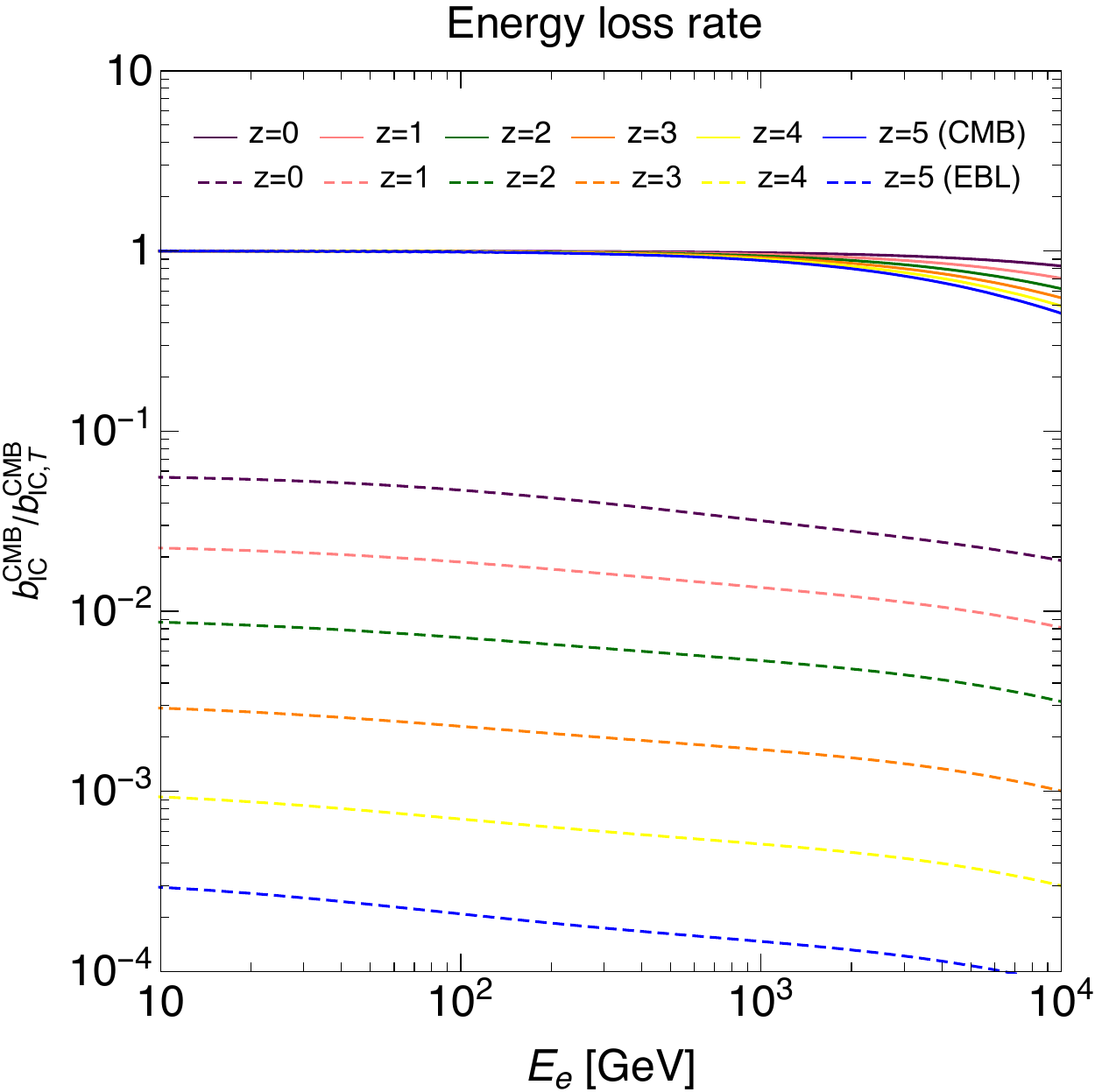}
  \caption{Energy loss rate of $e^{\pm}$ as function of $e^{\pm}$
    energy, which is normalized by the analytic formula in Thomson
    limit. The energy loss rates due to the CMB (EBL) are given in
    solid (dotted) lines for $z=0$, $1$, $2$, $3$, $4$, $5$.  }
  \label{fig:bic}
 \end{center}
\end{figure}

In Fig.\,\ref{fig:bic} we plot $b_{\rm IC}^{\rm CMB}(E_e,z)$ and
$b_{\rm IC}^{\rm EBL}(E_e,z)$ as function of $E_e$ normalized by the
analytic expression in Thomson limit, $b_{\rm IC,T}^{\rm CMB}(E_e,z)$.
For the energy loss under the CMB, it is seen that the analytic
expression in the Thomson limit well agrees with the numerical
result. However, in larger energy range $E_e\gtrsim 1~{\rm TeV}$,
which we are interested in the current study, the numerical result
deviates from $b_{\rm IC,T}^{\rm CMB}(E_e,z)$ especially larger
$z$. In the EBL, on the other hand, Thomson limit can not be applied.
However, the energy loss due to the EBL is less than 5\% compared to
one due to the CMB. Thus the EBL itself merely affects our numerical
study. The numerical results given in
Figs.\,\ref{fig:limit95_DMonly_leptonic}--\ref{fig:limit95_DMonly_ann}
(see ``All'' and ``No EBL'' in the figures) are consistent with this
fact.

\section*{Acknowledgments}

The authors thank M.~Regis for useful discussions.  This work was
supported by the Netherlands Organization for Scientific Research
(NWO) through Vidi grant (SA).



\end{document}